\documentclass[manuscript]{emulateapj}

\usepackage{apjfonts}
\usepackage{graphics}

\newcommand{\cm}{{\rm cm}}
\newcommand{\s}{{\rm s}}
\newcommand{\kms}{{\rm km}\,{\rm s}^{-1}}
\newcommand{\K}{{\rm K}}

\newcommand{\kpc}{{\rm kpc}}
\newcommand{\mpc}{{\rm Mpc}}

\newcommand{\HI}{\ion{H}{1}} 
 
\newcommand{\HeII}{\ion{He}{2}} 

\newcommand{\CIII}{\ion{C}{3}}
\newcommand{\CIV}{\ion{C}{4}}

\newcommand{\NV}{\ion{N}{5}}
\newcommand{\OVI}{\ion{O}{6}}
\newcommand{\SiIII}{\ion{Si}{3}}
\newcommand{\SiIV}{\ion{Si}{4}}
\newcommand{\FeII}{\ion{Fe}{2}}

\newcommand{\lya}{Ly$\alpha$} 
\newcommand{\lyb}{Ly$\beta$}

\shorttitle{Distribution of Metals as Traced by \CIV}
\shortauthors{Schaye et al.}

\begin{document}

\slugcomment{Accepted for publication in the Astrophysical Journal}

\title{Metallicity of the Intergalactic Medium Using Pixel
Statistics.\\ II. The Distribution of Metals as Traced by
\CIV\altaffilmark{1}} 
\altaffiltext{1}{Based on
public data obtained from the ESO archive of observations from the UVES
spectrograph at the VLT, Paranal, Chile and on data obtained at the W. M. Keck 
Observatory, which is operated as a scientific partnership among the
California Institute of Technology, the University of California, and
the National Aeronautics and Space Administration. The W. M. Keck
Observatory was made possible by the generous financial support of the
W. M. Keck Foundation.}

\author{Joop~Schaye\altaffilmark{2}, Anthony Aguirre\altaffilmark{2},\\
Tae-Sun Kim\altaffilmark{3,4}, Tom Theuns\altaffilmark{5,6}, Michael
Rauch\altaffilmark{7}, Wallace L. W. Sargent\altaffilmark{8}}
\altaffiltext{2}{School of Natural Sciences, Institute for Advanced
Study, Einstein Drive, Princeton, NJ 08540; schaye@ias.edu; aguirre@ias.edu.}
\altaffiltext{3}{European Southern Observatory,
Karl-Schwarzschild-Strasse 2, D-85748 Garching bei M\"unchen, Germany.}
\altaffiltext{4}{Institute of Astronomy, Madingley Road, Cambridge CB3
0HA, UK.} 
\altaffiltext{5}{Institute for Computational Cosmology, Department of
Physics, University of Durham, South Road, Durham, DH1 3LE, UK.}
\altaffiltext{6}{University of Antwerp, Universiteits plein 1,
B-2610 Antwerpen, Belgium.}
\altaffiltext{7}{Carnegie Observatories, 813 Santa Barbara Street,
Pasadena, CA 91101.} 
\altaffiltext{8}{Department of Astronomy, California Institute of
Technology, Pasadena, CA 91125.}  

\begin{abstract}
We measure the distribution of carbon in the intergalactic medium as a
function of redshift $z$ and overdensity $\delta$. Using a
hydrodynamical simulation to link the \HI\ absorption to the density
and temperature of the absorbing gas, and a model for the UV
background radiation, we convert ratios of \CIV\ to \HI\ pixel optical
depths into carbon abundances. For the median metallicity this
technique was described and tested in Paper I of this series. Here we
generalize it to reconstruct the full probability distribution of the
carbon abundance and apply it to 19 high-quality quasar absorption
spectra.  We find that the carbon abundance is spatially highly
inhomogeneous and is well-described by a lognormal distribution for
fixed $\delta$ and $z$. Using data in the range $\log\delta = -0.5 -
1.8$ and $z=1.8-4.1$, and a renormalized version of the Haardt \&
Madau (2001) model for the UV background radiation from galaxies and
quasars, we measure a median metallicity of $[{\rm C}/{\rm H}] =
-3.47_{-0.06}^{+0.07} + 0.08_{-0.10}^{+0.09}(z-3) +
0.65_{-0.14}^{+0.10}(\log\delta-0.5)$ and a lognormal scatter of
$\sigma([{\rm C}/{\rm H}]) = 0.76_{-0.08}^{+0.05} +
0.02_{-0.12}^{+0.08}(z-3) -
0.23_{-0.07}^{+0.09}(\log\delta-0.5)$. Thus, we find significant trends
with overdensity, but no evidence for evolution. These measurements imply that
gas in this density range accounts for a cosmic carbon abundance of
$[{\rm C}/{\rm H}]=-2.80 \pm 0.13$ ($\Omega_{\rm C} \approx 2\times
10^{-7}$), with no evidence for evolution. The dominant source of
systematic error is the spectral shape of the UV background, with
harder spectra yielding higher carbon abundances. While the
systematic errors due to uncertainties in the spectral hardness may
exceed the quoted statistical errors for $\delta < 10$, we stress that
UV backgrounds that differ significantly from
our fiducial model give unphysical results. The measured lognormal
scatter is strictly independent of the spectral shape, provided the
background radiation is uniform.  We also present measurements of the
\CIII/\CIV\ ratio (which rule out temperatures high enough for
collisional ionization to be important for the observed \CIV) and
of the evolution of the effective \lya\ optical depth.  \end{abstract}


\section{Introduction}
\label{sec:intro}

The enrichment of the intergalactic medium (IGM) with heavy elements
provides us with a fossil record of past star formation, as well as a unique
laboratory to study the effects of galactic winds and early
generations of stars. With the advent of the HIRES spectrograph on the
Keck telescope it rapidly  
became clear that a substantial fraction of the high column density
($N_{\rm HI} \ga 10^{14.5}~\cm^{-2}$) \lya\ absorption lines seen at
redshift $z\sim 3$ in the spectra of distant quasars 
have detectable associated absorption by \CIV\ (Cowie et al.\ 1995;
Ellison et al.\ 2000) and \SiIV\ (Songaila \& Cowie 1996). More
recently, observations with the UVES instrument on the Very Large
Telescope (VLT), as well as with the Keck Telescope, have
revealed that the same is true for \OVI\ at $z 
\sim 2$ (Carswell, Schaye, \& Kim 2002; Simcoe et al.\ 2002;
Bergeron et al.\ 2002, see also Telfer et al.\ 2002).
Both simple photoionization models (e.g., Cowie et al.\ 1995; Songaila \&
Cowie 1996; Carswell et al.\ 2002; Bergeron et al.\ 2002) and
numerical simulations assuming a uniform metallicity
(Haehnelt et al.\ 1996; Rauch et al.\ 1997a; Hellsten et al.\ 1997;
Dav\'e et al.\ 1998) indicate that the gas has a typical metallicity
of $10^{-3}$ to $10^{-2}$ solar.

The gas in which metals can be detected directly is thought to be
significantly overdense [$\delta \equiv \rho/\left<\rho\right > \sim
10~ (N_{\rm HI} / 
10^{15}~\cm^{-2})^{2/3}[(1+z)/4]^{-3}$; Schaye 2001] 
and consequently fills only a small fraction of the volume. The typical
metallicity of the low-density IGM (away from local 
sources of metals) remains largely unknown, although statistical
analyses based on pixel optical depths do indicate that there is \CIV\
(Cowie \& Songaila 1998; Ellison et al.\ 2000) and \OVI\ (Schaye et
al.\ 2000a) associated with the low-column density \lya\ forest.

Despite the recent progress, many questions remain regarding the
distribution of metals in the IGM: Are metals generally present at
$\delta \ll 10$? Does the metallicity vary with overdensity? Does the
metallicity vary with redshift? How much scatter is there in the
metallicity for a fixed density and redshift? Does this scatter change
with density or redshift?  Are the metals photoionized or does
collisional ionization dominate?  How does the metallicity change with
the assumed spectral shape of the UV-background? What are the relative
abundances of the different elements? Except for the last,
which we will address in future publications, this paper
will address all of these questions.

We measure the distribution of carbon in the IGM by 
applying an extension of the pixel optical depth
method, which we developed and tested in Aguirre, Schaye, \& Theuns
(2002, hereafter Paper I) ---
which was itself an extension of the earlier work by Cowie \& Songaila
(1998) (see also Songaila 1998, Ellison et al.\ 2000, and Schaye et
al.\ 2000a) --- to a 
set of high-quality quasar spectra taken with the Keck and VLT
telescopes. The basic idea is to measure the distribution of \CIV\
pixel optical 
depths as a function of the corresponding \HI\ optical depth,
corrected for noise and absorption by other transitions, and to
convert this correlation into an 
estimate of the metallicity as a function of the
density. Our method differs in two respects from that
proposed in Paper I: (1) we measure the full distribution 
of metals instead of just the median metallicity and (2) we correct for
noise and contamination using the data themselves instead of
simulations. Furthermore, we use the \CIII/\CIV\ ratio to set an 
upper limit on the temperature of the enriched gas, ruling
out collisional ionization as the dominant ionization mechanism.

This paper is organized as follows. In \S\ref{sec:obs} we describe our
sample of quasar spectra, and in \S\ref{sec:taueff} we use this sample
to measure the evolution of the mean absorption, which we need to
normalize the intensity of our UV background models. Before describing
these models in \S\ref{sec:uvmodels}, we briefly describe our
hydrodynamical simulation in \S\ref{sec:sim}. The simulation is used
to compute the gas density and temperature as a function of the \HI\
optical depth, which are needed to compute the ionization
correction factors, as discussed in detail in
\S\ref{sec:ionizcorr}. In \S\ref{sec:method} we provide a step-by-step
description of our method for measuring the median metallicity
(\S\ref{sec:method:median}) and the distribution of metals at a fixed
density (\S\ref{sec:method:scatter}). To
make the reading more 
interesting, we illustrate the method by presenting results for one of
our best quasar spectra. Section \ref{sec:ciii} contains our measurements
of the \CIII/\CIV\ ratio, which support our assumption that the
temperature is low enough for collisional ionization of \CIV\ to be
unimportant. In \S\ref{sec:results} we present our measurements of the
distribution of carbon as a function of overdensity and redshift. In
\S\ref{sec:varyinguv} we show how the 
results change if we vary the UV background. In \S\ref{sec:discussion}
we discuss the results, compute mean metallicities and filling factors,
compare with previous work, and estimate the size of systematic
errors. Finally, we summarize our conclusions in \S\ref{sec:conclusions}.

How should this paper be read? Given the length of the paper, we
encourage reading the conclusions (\S\ref{sec:conclusions})
first. Readers who are not 
interested in the details of the method would in addition only need to
read \S\S \ref{sec:results} and \ref{sec:discussion}. Those who would
like to understand the method, but are not interested in knowing all of
the details, would benefit from also reading \S\S
\ref{sec:uvmodels}, 
\ref{sec:ionizcorr}, and \ref{sec:method}. Readers who would like 
to know more about the mean absorption or the 
constraints on the ionization mechanism may want to read \S\S
\ref{sec:taueff} and \ref{sec:ciii}, respectively. 

\section{Observations}
\label{sec:obs}
We analyzed spectra of the 19 quasars listed in the first column of
Table~\ref{tbl:sample}. Fourteen spectra were taken with the UV-Visual
Echelle Spectrograph (UVES; D'Odorico et al.\ 2000) on the VLT and
five were taken with the High Resolution Echelle Spectrograph (HIRES;
Vogt et al.\ 1994) on the Keck telescope (see col.\ [7] of
Table~\ref{tbl:sample}). The UVES spectra were taken from the ESO
archive. Only spectra that have a signal-to-noise ratio (S/N) $\ga$ 40 in
the \lya\ region and that were publicly available as of 2003 January 31
were used. The UVES spectra were reduced with the
ESO-maintained MIDAS ECHELLE package (see Kim, Cristiani, \& D'Odorico
2001 and Kim et al.\ 2003 for details on the data
reduction). The reduction procedures for the HIRES spectra are
described in Barlow \& Sargent 
(1997). The spectra have a nominal velocity resolution of $6.6~\kms$
(FWHM) and a pixel size of 0.04 and 0.05 \AA\ for the HIRES and
UVES data, respectively.

For a quasar at redshift $z_{\rm em}$ (the emission redshift), we
analyze data in the redshift range
$(1+z_{\rm em})\lambda_{{\rm Ly}\beta}/\lambda_{{\rm Ly}\alpha}-1 < z
< z_{\rm em} - (1+z_{\rm em})\Delta v/c$, where $\Delta v =
\max(4000,8\,\mpc\,H(z)/h)~\kms $ and $H(z)$ is the Hubble
parameter at 
redshift $z$ extrapolated from its present value ($H_0 \equiv
100h~\kms\,\mpc^{-1}$) assuming $(\Omega_m,\Omega_\Lambda) =
(0.3,0.7)$. The lower limit ensures that \lya\ falls redwards of the
quasar's \lyb\ emission line, thereby avoiding confusion with the
\lyb\ forest. The region close to the quasar is excluded to avoid
proximity effects. 
The minimum, 
median, and maximum absorption redshifts considered are listed in
columns (3), (4), and (5) of Table~\ref{tbl:sample}. 

Regions thought to be contaminated by absorption features that are not
present in our simulated spectra were excluded. Examples are
absorption by metal lines other than \CIII, \CIV, \SiIII, \SiIV, \OVI,
\NV, and \FeII, and atmospheric lines. Seven spectra
(Q1101-264, Q1107+485, Q0420-388, Q1425+604, Q2126-158, Q1055+461, and
Q2237-061) contain \lya\ lines with damping wings (i.e.,
$N({\rm HI})\ga 10^{19}~\cm^{-2}$). Regions contaminated by these
lines or their corresponding higher order Lyman series lines were also
excluded. For the case of damped \lya\ lines the excluded regions can
be large (up to several tens of angstroms). Contaminating metal lines
were identified by eye by searching for absorption features corresponding to the
redshifts of strong \HI\ lines and to the redshifts of the absorption
systems redward of the quasar's \lya\ emission line (trying a large
number of possible identifications). Due to the robustness of our
method, which uses the \CIV\ doublet to correct for contamination and
is based on nonparametric statistics, our results are nearly
identical if we do not remove any of the contaminating metal lines.


\begin{deluxetable*}{llccccll} 
\tablecolumns{8} 
\tablewidth{0pc} 
\tablecaption{Observed quasars \label{tbl:sample}}
\tablehead{ 
\colhead{QSO} & \colhead{$z_{\rm em}$} & \colhead{$z_{\rm min}$} &
\colhead{med($z$)} & \colhead{$z_{\rm max}$} & \colhead{$\lambda_{\rm
min}$ (\AA)} & \colhead{Instrument} & \colhead{Reference}}
\startdata 
Q1101-264   & 2.145 & 1.654 & 1.878 & 2.103 & 3050.00 & UVES & 1 \\  
Q0122-380   & 2.190 & 1.691 & 1.920 & 2.147 & 3062.00 & UVES & 2 \\  
J2233-606   & 2.238 & 1.732 & 1.963 & 2.195 & 3055.00 & UVES & 3 \\  
HE1122-1648 & 2.400 & 1.869 & 2.112 & 2.355 & 3055.00 & UVES & 1 \\  
Q0109-3518  & 2.406 & 1.874 & 2.117 & 2.361 & 3050.00 & UVES & 2 \\  
HE2217-2818 & 2.406 & 1.874 & 2.117 & 2.361 & 3050.00 & UVES & 3 \\  
Q0329-385   & 2.423 & 1.888 & 2.133 & 2.377 & 3062.00 & UVES & 2 \\  
HE1347-2457 & 2.534 & 1.982 & 2.234 & 2.487 & 3050.00 & UVES & 1,2\\  
PKS0329-255 & 2.685 & 2.109 & 2.373 & 2.636 & 3150.00 & UVES & 2 \\  
Q0002-422   & 2.76  & 2.173 & 2.441 & 2.710 & 3055.00 & UVES & 2 \\  
HE2347-4342 & 2.90  & 2.291 & 2.569 & 2.848 & 3428.00 & UVES & 2 \\  
Q1107+485   & 3.00  & 2.375 & 2.661 & 2.947 & 3644.36 & HIRES & 4 \\ 
Q0420-388   & 3.123 & 2.479 & 2.774 & 3.068 & 3760.00 & UVES & 2 \\  
Q1425+604   & 3.20  & 2.544 & 2.844 & 3.144 & 3736.20 & HIRES & 4 \\ 
Q2126-158   & 3.268 & 2.601 & 2.906 & 3.211 & 3400.00 & UVES & 2 \\  
Q1422+230   & 3.62  & 2.898 & 3.225 & 3.552 & 3645.24 & HIRES & 4 \\ 
Q0055-269   & 3.655 & 2.928 & 3.257 & 3.586 & 3423.00 & UVES & 1 \\  
Q1055+461   & 4.12  & 3.320 & 3.676 & 4.033 & 4586.36 & HIRES & 5\\ 
Q2237-061   & 4.558 & 3.690 & 4.070 & 4.451 & 4933.68 & HIRES & 4   
\enddata 
\tablerefs{(1) Kim et al.\ 2002; (2) Kim et al.\ 2003; (3)
  Kim, Cristiani, \& D'Odorico 2001; (4) Rauch et al.\ 1997; (5)
  Boksenberg, Sargent, \& Rauch 2003}
\end{deluxetable*} 


\section{The observed evolution of the mean \lya\ absorption}
\label{sec:taueff}
Our measurements of the effective optical depth,
$\tau_{\rm eff} \equiv \left < -\ln F\right >$, where $F$ is the
normalized flux in the \lya\ forest, are plotted as a
function of redshift in Fig.~\ref{fig:taueff} and are listed in
Table~\ref{tbl:taueff} in Appendix B. Besides the quasars listed in
Table~\ref{tbl:sample}, we measured $\tau_{\rm eff}$ for two
additional HIRES quasar spectra from the Rauch et al.\ (1997) sample:
Q1442+293 ($z_{\rm em} = 2.67$) and Q0000-262 ($_{\rm em} =
4.11$). These spectra are not part of our main sample because they
do not have sufficient coverage to be useful for 
our metal analysis. Each \lya\ forest spectrum was divided in two,
giving 42 data points in total. The left-hand panel shows the results
obtained using all
pixels in the \lya\ forest range; the right-hand panel shows the results
obtained after removing all pixels that are thought to be contaminated by
metal lines, \HI\ lines with damping wings, and
atmospheric lines. Comparison of the two panels shows that removing
the contamination generally reduces both the mean absorption and the
scatter in the absorption at a given redshift, particularly at lower
redshifts. 

The dotted lines show least-squares power-law fits to the data
points. Note that the scatter is much larger than expected given the
error bars: the $\chi^2/$dof is about 5.2 and 2.4 for the left-
and right-hand panels respectively (for 40 dof). This probably indicates
that most of the remaining scatter is due to cosmic variance (see also
Kim et al.\ 2002).

The dashed curve in Figure~\ref{fig:taueff} indicates the evolution of
the \lya\ optical depth measured by Bernardi et al.\ (2003), who
applied a novel $\chi^2$ minimization technique to 1061
low-resolution QSO spectra (their S/N >
4 sample) drawn from the Sloan Digital Sky Survey database. The
Bernardi et al.\ fit gives absorption systematically higher by about
0.1~dex than we obtain from our high-resolution spectra. 
For $z\ga 3$ our local continuum fits may underestimate the absorption
and the Bernardi et al.\ results may be more reliable. However, for lower
redshifts the continuum is well-defined and our results may be more
accurate than those of Bernardi et al., whose spectra have
insufficient resolution to allow for the identification and removal of
metal lines. 


\begin{figure*}[t]
\epsscale{1.15}
\plottwo{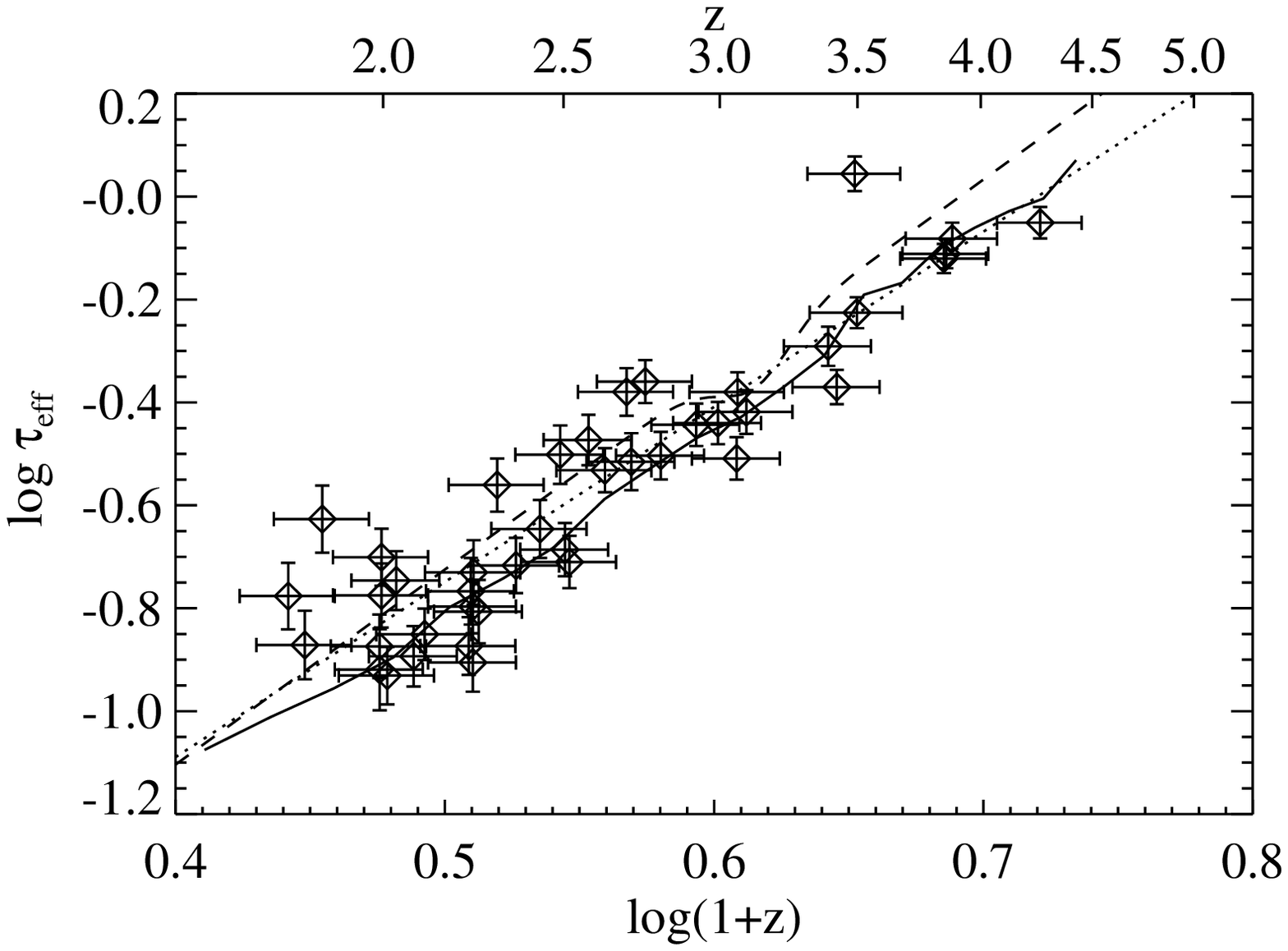}{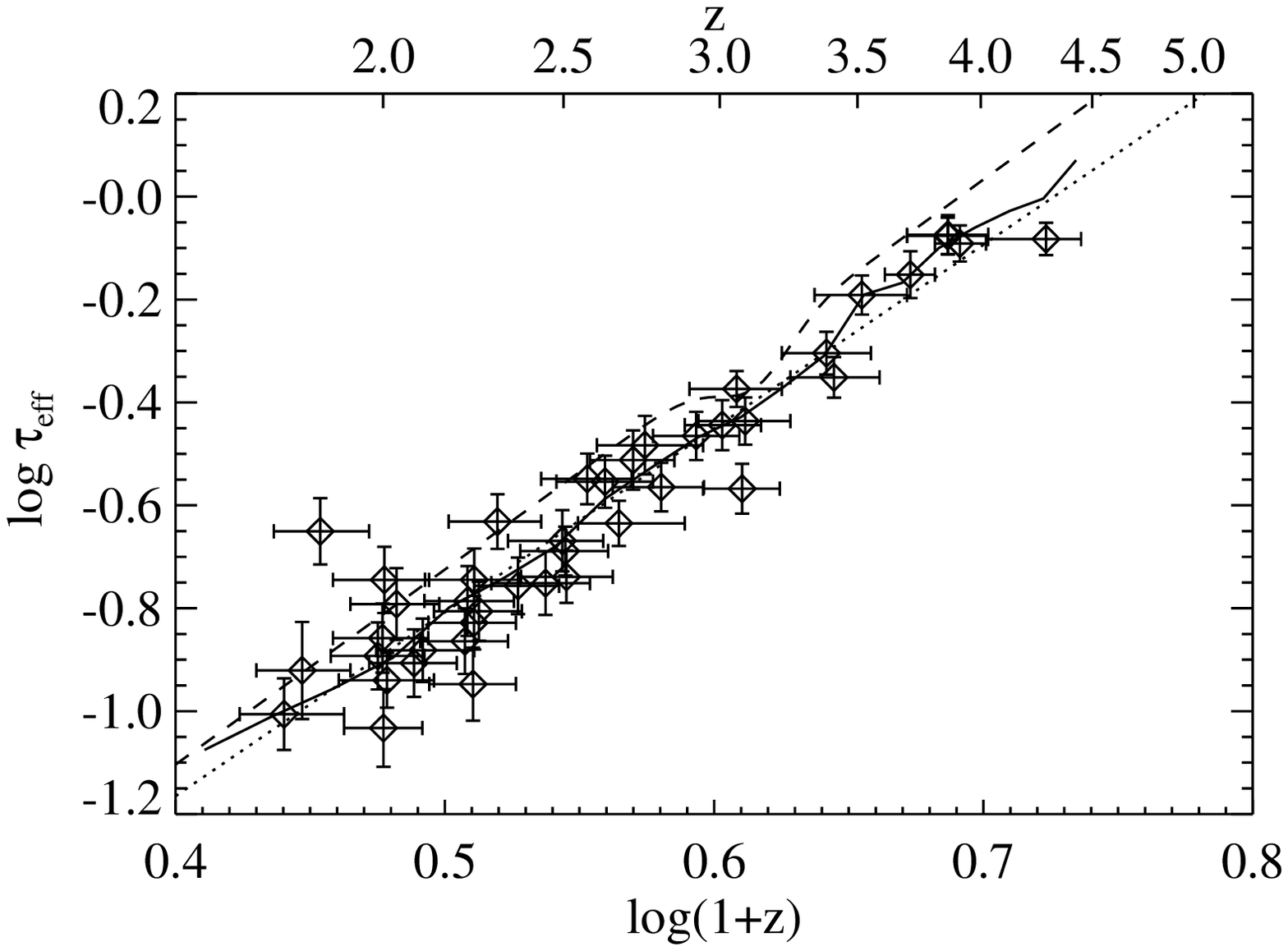}
\caption{Observed effective optical depth as a
function of redshift for all pixels (\emph{left-hand panel}) and after
removal of 
pixels contaminated by metal lines, Ly$\alpha$ lines with damping
wings, or atmospheric lines (\emph{right-hand panel}). Horizontal error bars
indicate redshift ranges 
used, vertical error bars are 1$\sigma$ errors measured by bootstrap
resampling the spectra using chunks of 100 pixels. The solid curve
shows the average effective optical depth in the simulated spectra;
the dotted lines are power-law least-squares fits to the data points: 
$\log \tau_{\rm eff} = \log \tau_0 + \alpha \log[(1+z)/4]$.
The best-fit values are $(\log \tau_0,\alpha) = (-0.40 \pm 0.02,3.40
\pm 0.23)$ and $(-0.44 \pm 0.01,3.57 \pm 0.20)$
for the left- and right-hand
panels, respectively. The errors on the parameters of the fit were
estimated by bootstrap resampling the data points.
The dashed curve indicates the 
evolution of the \lya\ optical depth measured by Bernardi et al.\
(2003) using a $\chi^2$ minimization technique on QSO spectra (their
S/N > 4 sample) drawn from the Sloan Digital Sky Survey
database.\label{fig:taueff}}
\end{figure*}


\section{Simulation and ionization corrections}
\label{sec:sim+ionizcorr}
Because the \CIV\ optical depth can tell us only about the density of
\CIV\ ions, we need to know the fraction of carbon that is triply
ionized to determine the density of carbon. The ionization
balance depends on the gas density, temperature, and the ambient
UV radiation field. We use (variants of) the models of Haardt \& Madau
(2001) for the UV background and a hydrodynamical simulation to compute
the density and temperature as a function of the \HI\ optical
depth. Before explaining how we convert the \CIV/\HI\ optical depth
into a metallicity in \S\ref{sec:ionizcorr}, we will briefly describe our
simulation and our method for generating synthetic spectra in
\S\ref{sec:sim} and our models for the metagalactic UV radiation in
\S\ref{sec:uvmodels}.

\subsection{Generation of synthetic spectra}
\label{sec:sim}
We use synthetic absorption spectra generated from a hydrodynamical
simulation for two purposes: (1) to determine the gas density and
temperature (which are needed to compute the ionization corrections) as a
function of the \HI\ optical depth and redshift and 
(2) to verify that
simulations using the carbon distribution measured from the observed
absorption spectra do indeed reproduce the observed optical 
depth statistics.

The simulation that we use to generate spectra is identical to the one
used in Paper I, and we refer the reader to \S3 of that paper for
details. Briefly, we use a smoothed particle hydrodynamics code to
model the evolution of a periodic, cubic region of a $(\Omega_m,
\Omega_\Lambda, \Omega_bh^2, h, \sigma_8, n, Y) = (0.3, 0.7, 0.019,
0.65, 0.9, 1.0, 0.24)$ universe of comoving size $12\,h^{-1}~\mpc$ to
redshift $z = 1.5$ using $256^3$ particles for both the cold dark
matter and the baryonic components. The gas is photoionized and
photoheated by a model of the UV background, designed to match the
temperature-density relation measured by Schaye et al.\ (2000b). Since
we recalculate the ionization balance of the gas when computing
absorption spectra, this choice of UV background only affects the
thermal state of the gas.

The software used to generate the simulated spectra is also described in
detail in Paper I (\S3). Briefly,
many sight lines through different snapshots of the simulation box are
patched together\footnote{Unlike in Paper I, we do not
cycle the short sight lines periodically as this could potentially
create spurious correlations between transitions separated by
approximately an integral number of box sizes.} to create long sight
lines spanning $z = 1.5$ to 
$z_{\rm em}$. Absorption from Ly1 ($\lambda$1216), Ly2 ($\lambda$1026),
\ldots, Ly31 ($\lambda$913), \CIII\ ($\lambda$977), \CIV\
($\lambda\lambda$1548,1551), \NV\ ($\lambda\lambda$1239,1243), \OVI\
($\lambda\lambda$1032,1038), \SiIII\ ($\lambda$1207), \SiIV\
($\lambda\lambda$1394,1403), and \FeII\ ($\lambda$1145, $\lambda$1608,
$\lambda$1063, $\lambda$1097, $\lambda$1261, $\lambda$1122, $\lambda$1082,
$\lambda$1143, $\lambda$1125) is included. The long spectra are then
processed to match the characteristics of the observed
spectrum they are compared with. First, they are convolved with a
Gaussian with FWHM of $6.6~\kms$ to mimic instrumental
broadening. Second, they are resampled onto pixels of the same size as
were used in the observations. Third, noise is added to each
pixel. The noise is assumed to be Gaussian with a variance that has the
same dependence on wavelength and flux as the noise in the
observations. 

The ionization balance of each gas particle is computed from
interpolation tables generated using the publicly available
photoionization package CLOUDY\footnote{See
\texttt{http://www.pa.uky.edu/$\sim$gary/cloudy}.} (ver.\ 94; see
Ferland et al.\ 1998 and Ferland 2000 for details), assuming the gas
to be optically 
thin and using specific models for the UV background radiation, which
we will discuss next.

\subsection{UV background models}
\label{sec:uvmodels}

We use the models of Haardt \& Madau (2001, hereafter
HM01)\footnote{The data and a description of the input parameters can
be found at 
\texttt{http://pitto.mib.infn.it/$\sim$haardt/refmodel.html}.} for the 
spectral shape of the meta-galactic UV/X-ray background
radiation. Our fiducial model, which we will refer to as model QG,
includes contributions from both quasars and galaxies, but we will
also compute some results for model Q which includes quasars only (this
model is an updated version of the widely used Haardt \& Madau 1996
model). Both models take reprocessing by the IGM into account. 

The spectra have breaks at 1 and 4~ryd as well as
peaks due to \lya\ emission. For model QG the spectral index
$-dJ_\nu/d\nu$, where $J_\nu$ is the specific intensity at frequency
$\nu$, is about 1.5 from 1 to 4 ryd, but 
above 4 ryd the spectrum hardens considerably to a spectral index
ranging from about 0.8 at $z\approx 2$ to about 0.3 at $z\approx
4$. For model QG (Q) 
the softness parameter $S\equiv \Gamma_{\rm HI}/\Gamma_{\rm 
HeII}$ where $\Gamma_i$ is the ionization rate for element $i$,
increases from 350 (175) at $z\approx 2$, to 600 (230) 
at $z\approx 3$, to 950 (280) at $z\approx 4$.

To see how the results would change if the UV background were much
softer at high redshift, as may be appropriate during and before the
reionization of \HeII, we also create model QGS, which is identical
to model QG except that the intensity above 4~ryd has been reduced by
a factor of 10.

All UV background spectra are normalized (i.e., the intensities are
multiplied by 
redshift-dependent factors)\footnote{Multiplying the QG spectra by
0.58 for $z < 2.5$ and 0.47 for $z>2.5$ gives satisfactory
results.}
so that the effective optical depth in the simulated absorption spectra 
(Fig.~\ref{fig:taueff}, \emph{solid curves}) reproduces the evolution of the
observed mean transmission. We find that our simulation agrees with
the observations if the \HI\ ionization rate $\Gamma_{\rm
  HI}/10^{-13} \approx 8.7$, 5.4, and 3.6 at $z=2$, 3, 
and 4, respectively (note that since $\Gamma \propto \Omega_b^2h^3$
[e.g., Rauch et al.\ 1997], these values would be about 27\% higher
for the currently favored cosmology [$\Omega_bh^2 = 0.0224$,
$h=0.71$; Spergel et al.\ 2003]). 

\subsection{Ionization corrections}
\label{sec:ionizcorr}
The ratio $\tau_{\rm CIV}/\tau_{\rm HI}$ is proportional
to the ratio of the \CIV\ and \HI\ number densities of the gas that is
responsible for the absorption at that redshift. Converting this into a
carbon abundance requires an estimate of the fraction of carbon
that is triply ionized and the fraction of hydrogen that is
neutral. These fractions depend on the gas density,
temperature, and the spectrum of the UV background radiation.

Both (semi-)analytic models (e.g., Bi \& Davidsen 1997; Schaye 2001)
and hydrodynamical simulations (e.g., Croft et al.\ 1997; Zhang et
al.\ 1998; Paper I) suggest that there is a tight correlation between
the \HI\ \lya\ optical depth (or column density) and the gas
density. For a given density $\tau_{\rm HI}$ depends weakly on the
temperature and is 
inversely proportional to the \HI\ ionization rate (assuming
ionization equilibrium). 

We use our hydrodynamical simulation, which
predicts a thermal evolution consistent with the measurements of the
IGM temperature of Schaye et al.\ (2000b), to compute
interpolation tables of the density and temperature as a function of
the \lya\ optical depth and redshift. As discussed in
\S\ref{sec:uvmodels}, we rescale the UV background so that the
simulated spectra reproduce the observed mean \lya\ transmission.

Because the absorption takes place in redshift space, multiple gas
elements can contribute to the optical depth of a given pixel. We
define the density/temperature in each pixel as the average over all
gas elements that contribute to the absorption in the pixel, weighted
by their \HI\ optical depths (Schaye et al.\ 1999). Therefore, \emph{the
densities we quote are effectively smoothed on the same scale as the
\lya\ forest spectra.}

Thermal broadening, the differential Hubble flow
across the absorbers, and peculiar velocity gradients all contribute
to the line widths. Simulations show that the first two are typically
most important for the low column density forest (e.g., Theuns,
Schaye, \& Haehnelt 2000). The line widths ($b$-parameters) are typically
$\sim 20-30~\kms$ for \HI\ (e.g., Carswell et al.\ 1984) and $\sim
5-15~\kms$ for (directly detectable) \CIV\ (e.g., Rauch
et al.\ 1996; Theuns et al.\ 2002b). The thermal broadening width is
$b_{\rm th} = 13~\kms (m_{\rm H}/m)^{1/2}(T/10^4~\K)^{1/2}$, 
where $m$ is the atomic weight ($m_{\rm C}\approx 12m_{\rm H}$). 
The real-space smoothing scale
is similar to the local Jeans scale, which varies from $\sim
10^2~\kpc$ for the low column density forest to $\sim 10~\kpc$ for
the rare, strong absorption lines arising in collapsed halos (e.g., Schaye
2001). For our cosmology, a 
velocity difference $\Delta v$ corresponds to a physical scale of
$34~\kpc~(\Delta v/10~\kms)$ at $z=3$. Thus, our densities are
typically smoothed on a scale of 50 - 100~kpc.  

When Hubble broadening dominates, as is the case for
all but the highest overdensities studied in this work, there is little
ambiguity in the 
relation between optical depth and gas density. However, if thermal broadening
were to dominate, then the absorption in pixels with low optical depths
could in principle arise in the thermal tails of high-density
gas. This would introduce scatter in the relation between \HI\ optical
depth and overdensity. Furthermore, because carbon is heavier
than hydrogen, it would result in discrepancies between the
densities of the gas responsible for the \CIV\ and \HI\ absorption at
a fixed redshift.\footnote{The fact that the CIV and HI fractions
scale differently with the gas density and temperature also
contributes to the differences between the $\tau_{\rm HI}$ and $\tau_{\rm
CIV}$ weighted densities. However, we find from our simulation that
the difference in 
the atomic weights is more important, except perhaps if the CIV
and HI absorption arises in different gas phases. Note that the
effects of the difference in the thermal broadening scales of HI and
CIV could in principle be partially compensated for by smoothing the
spectra on a scale greater than $b_{\rm th,CIV}$, but smaller than
$b_{\rm th,HI}$. However, we find that doing this improves the results
only marginally for CIV.}  

Fortunately, simulations indicate that there is
little scatter in the density corresponding to a fixed optical
depth. Fig.~6 of Paper I shows that the relation between
$\tau_{\rm HI}$ and $\delta$ remains tight out to the maximum
measurable optical depths, $\tau_{\rm HI} \sim 10^2$, while Fig.~7 of
Paper I shows that weighting the gas by $\tau_{\rm CIV}$ and
$\tau_{\rm HI}$ gives nearly the same densities, although the
difference does increase with overdensity.

We can understand the empirical result that the relation between gas density
and optical depth remains well-defined up to (at least) $\delta \sim 10^2$ as
follows. First, absorbers corresponding to overdensities $\delta \la 10^2$
($N_{\rm HI} \la 10^{16.5}~\cm^{-2}$ at $z=3$) are generally not
isolated, compact clouds. The absorption arises in a fairly
smooth gas distribution, which often contains multiple peaks. The
substructure is generally resolved in redshift space because
the differential Hubble
broadening across the absorption complex and/or internal peculiar velocity
gradients are not much 
smaller than the thermal broadening scale (which is in turn greater
than the instrumental resolution for \HI). More importantly, because the
column density distributions of \HI\ 
and \CIV\ are steep, most low optical depth pixels are not near high
optical depth pixels; i.e., the
contribution of ``local'' gas to the optical depth in any given pixel is
generally greater than that of the thermal tails of high column
density absorbers.

In Paper I we used \HI\ weighted quantities to compute the \HI\ fraction
and \CIV\ weighted quantities to compute the \CIV\ fraction. Since
making this distinction changes the results only marginally (and only
for the highest densities, see Fig.~7 of Paper I) and
since it is not clear what density to 
assign the resulting metallicity to, we use only \HI\ weighted
quantities here. 

\begin{figure*}
\epsscale{1.15}
\plottwo{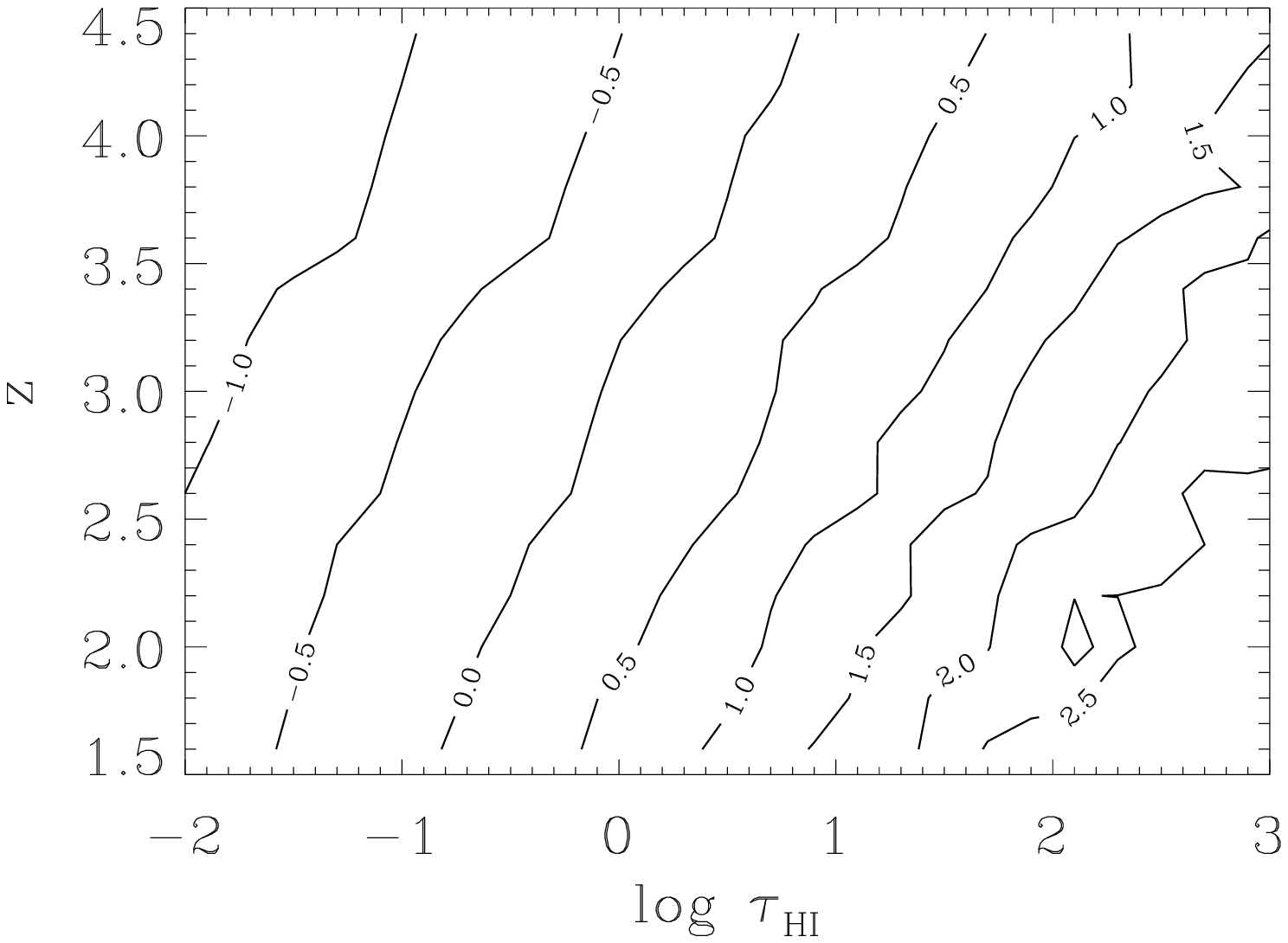}{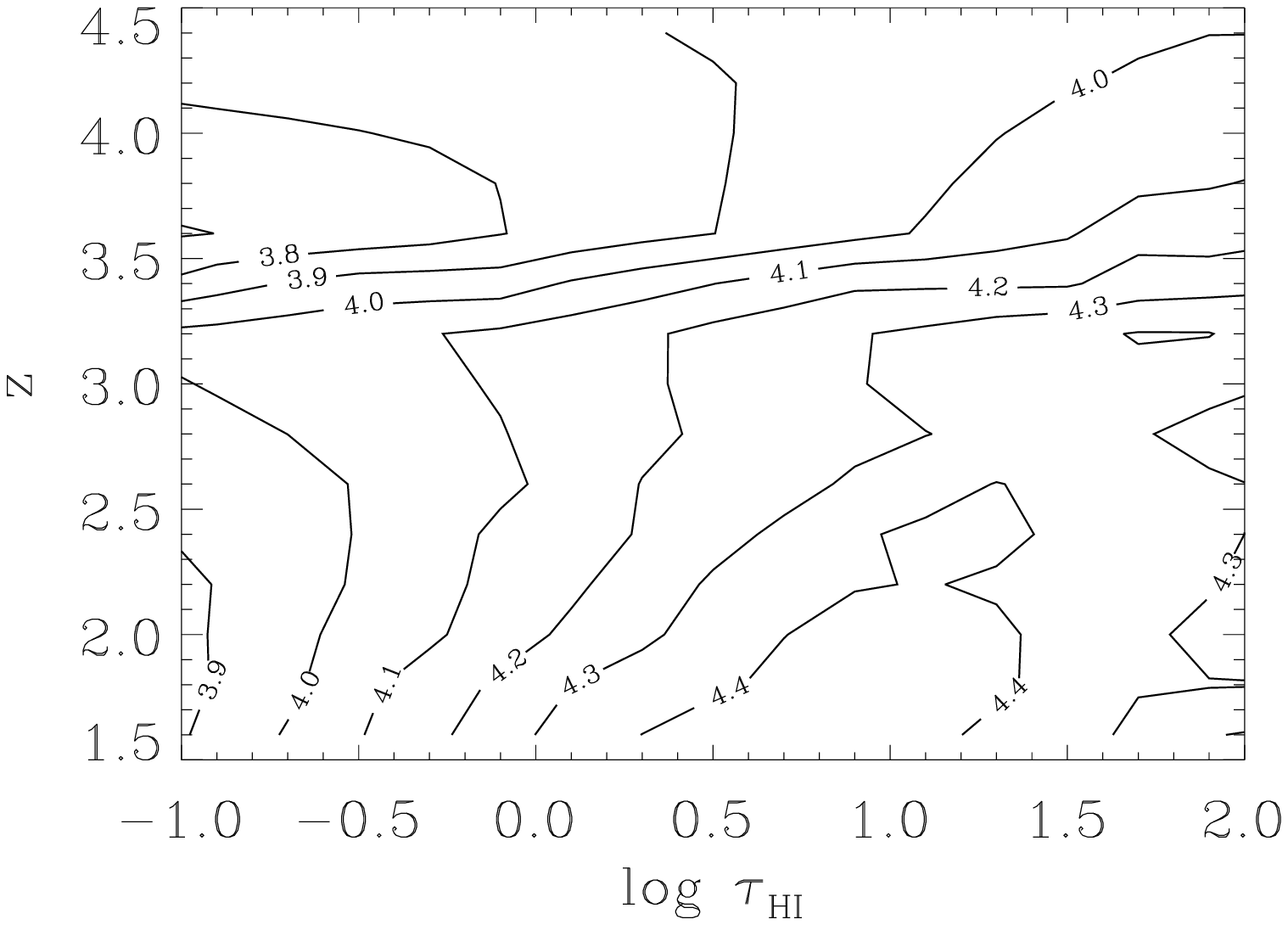}
\caption{\emph{Left:} Contour plot of $\log\delta$ as a function
of $\log\tau_{\rm HI}$ and redshift $z$. Overdensity increases with
optical depth but decreases with redshift. \emph{Right:} Contour plot
of $\log T$ as a function of $\log\tau_{\rm HI}$ and $z$. The temperature is
relatively constant over the parameter range of interest.
\label{fig:denstemp}}
\end{figure*}

Figure~\ref{fig:denstemp} shows contour plots of the overdensity (\emph{left
panel}) and temperature (\emph{right-hand panel}) as a function of
$\tau_{\rm HI}$ 
and $z$. Note that the hydrogen number density corresponding to an
overdensity $\delta$ is given by
\begin{equation}
n_{\rm H} \approx 1.04\times 10^{-5}~\cm^{-3}~\delta~\left ({1+z \over 4}
\right )^3 \left ({\Omega_bh^2 \over 0.019}\right ).
\end{equation}


\begin{figure}
\epsscale{1.2}
\plotone{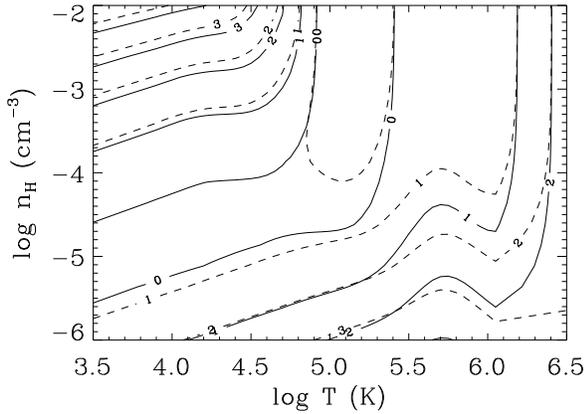}
\caption{Ionization
correction factor, $[{\rm C/H}] - \log(\tau_{\rm CIV}/\tau_{\rm HI})$,
as a function of the temperature and the hydrogen number
density. Solid (dashed) contours are for the UV background model QG
(Q). At high temperatures collisional ionization dominates and the
ionization balance becomes independent of the density. The temperature
below which photoionization becomes important (and the contours start
to curve) is higher if the density is lower. The results are only
sensitive to the hardness of the UV-background if the density is
low. In the regime of interest to us ($n_{\rm H} \sim 10^{-5}$ to
$10^{-4}~\cm^{-3}$, $T\sim 10^{4.0} - 10^{4.5}~\K$), the ionization
correction is 
near a minimum and thus relatively insensitive to the exact density
and temperature.
\label{fig:ionizcorr}}
\end{figure}

Figure~\ref{fig:ionizcorr} shows the ionization correction as a
function of temperature and density for the UV-background models QG
(\emph{solid contours}) and Q (\emph{dashed contours}), all for $z=3$.
The ionization correction is defined as the factor with
which the optical depth ratio $\tau_{\rm CIV}/\tau_{\rm HI}$ must be
multiplied to obtain the carbon abundance relative to solar (the
contours are labeled with the log of this factor). Note that the
ionization correction can be modest even when both the \HI\ and \CIV\
fractions are negligible (as is the case for $T\gg 10^5~\K$).


The shape of the contours in Fig.~\ref{fig:ionizcorr} is easily understood.
At high temperatures, collisional ionization dominates and the ionization
balance becomes independent of the density. At lower temperatures, the
change in the 
ionization correction with density can be accounted for by the degree
to which carbon is ionized. At high densities, the correction is large
because most of the carbon is less than triply ionized, and at very low
densities, it is large because most of it is more than triply
ionized. Compared with model QG, model Q predicts a lower \CIV\
fraction (and a higher ionization correction) for low-density gas
because it has more photons that can 
ionize \CIV\ (which has an ionization potential of about 4.7~ryd). 
The difference between the spectra of models QG and Q is particularly
large above 4~ryd because stars produce very few photons above the \HeII\
Lyman limit compared with quasars (recall that the models have been
scaled so that they have identical \HI\ ionization rates). 

For both models the \CIV\ fraction is near a maximum in the density
range where our data are best, $-5 \la \log n_{\rm H} \la -4$, and consequently
the ionization correction is relatively insensitive to the density over the
range that is of interest to us. The correction is also
insensitive to the temperature as long as the gas is predominantly
photoionized. In fact, for $-5 \la \log n_{\rm H} \la -4$, the temperature
dependence is weak even for temperatures as high as $10^6~\K$. 

\section{Method and results for Q1422+230}
\label{sec:method}
In this section we will discuss our method for measuring the
distribution of metals in the diffuse IGM. We will illustrate the
method by showing results for one of our best quasar spectra:
Q1422+230.

\subsection{The median metallicity as a function of density}
\label{sec:method:median}
Our method for measuring median metal abundances as a function of density
from QSO absorption spectra is presented in detail in Paper
I. Readers unfamiliar with pixel optical depth techniques may benefit
from reading \S2 of Paper I, which contains a general overview
of the method.

Our goal is to measure the abundance of carbon as a function of the
gas density and redshift, but our observable is the quasar flux as a
function of wavelength. In order to derive the metallicity we need to
do the following:

\emph{1. Normalize the spectra.} The quasar spectra are divided by a
continuum fit to obtain the normalized flux $F \equiv 
F_{\rm obs} / F_{\rm cont}$ as a function of wavelength, where 
$F_{\rm obs}$ and $F_{\rm cont}$ are the observed and
continuum flux respectively. The noise array is similarly rescaled:
$\sigma = \sigma_{\rm obs} / F_{\rm cont}$. The HIRES and UVES spectra
were continuum fitted as described in Rauch et al.\ (1997) and Kim et
al.\ (2001), respectively. 

We found that relative to their signal-to-noise ratio most UVES spectra
had larger continuum fitting errors than the HIRES spectra. All \CIV\
regions were therefore renormalized using the 
automatic procedure described in \S4.1 of Paper I: We 
divide the spectra into bins of \emph{rest-frame} size 20~\AA\ with
centers $\lambda_k$ and find 
for each bin the median flux $\bar{f_k}$. We then interpolate a spline
across the $\bar{f_k}$ and flag all pixels that are at least
$N_\sigma^{\rm cf}\sigma$ below the interpolation. The medians are
then recomputed using the unflagged pixels and the procedure is
repeated until the fit converges. Finally, the flux and error arrays
are rescaled using the fitted continuum. Tests using simulated spectra
indicate that $N_\sigma^{\rm cf}=2$ is close to optimal for the
\CIV\ region, giving errors in the continuum that are smaller than the
noise by an order of magnitude or more. When applied to the observations we
find that the adjustments in the continuum are typically smaller than
the noise by a factor of a few for the UVES spectra and by an order of
magnitude for the HIRES spectra.

\emph{2. Recover the \HI\ and \CIV\ optical depths as a function of
redshift.} The recovery $\tau = -\ln F$ is imperfect because of
contamination by other absorption lines, noise, and continuum fitting 
errors. In Paper I (\S4 and Appendix A) we describe
methods\footnote{Our method differs from that of Paper I in our
treatment of saturated metal pixels, i.e., CIV pixels with a
normalized flux $F(\lambda)< 3\sigma(\lambda)$, where
$\sigma(\lambda)$ is the normalized noise array. Instead of setting
the optical depth in these pixels to $-\ln 3\sigma(\lambda)$ as was done in
Paper I, we set it to a much larger value ($10^4$).} 
to partially correct for these effects, such as the use of higher order
Lyman series lines to estimate the \HI\ optical depth of saturated
absorption features [i.e., $F(\lambda) < 3\sigma(\lambda)$] and an
iterative procedure involving both 
components of the \CIV\ doublet to correct for
(self-)contamination. We will refer to the values resulting
from these procedures as ``recovered'' optical depths.

Pixels for which the \HI\ optical depth cannot
be accurately determined because all available Lyman series
lines are saturated are removed from the sample. Pixels that have
negative \CIV\ 
optical depth (which can happen as a result of noise and continuum fitting
errors) are given a very low positive value in order to keep the
ranking of optical depths (and thus the medians) unchanged.

\emph{3. Bin the pixels according to $\tau_{\rm HI}$ and compute the
median $\tau_{\rm CIV}$ for each bin.} 
There are two reasons for binning in $\tau_{\rm HI}$. First, binning
allows us to use nonparametric statistics, such as the median
$\tau_{\rm CIV}$ and
other percentiles, which makes the method much more robust. This is
important because our optical depth recovery is imperfect. Second, the
\HI\ optical depth is thought to be correlated with the local 
gas density (see \S\ref{sec:ionizcorr}), which is needed to compute
the ionization balance.

The errors in the medians are computed as follows.
First, the \lya\ forest
region of the spectrum is divided into $N$ chunks of 5~\AA, and \HI\
bins which contain less than 25 pixels or contributions from fewer than
five chunks are discarded. Second, a new realization of the sample of
pixel redshifts 
is constructed by bootstrap resampling the spectra; i.e., $N$ chunks
are picked at random \emph{with replacement}
and the median $\tau_{\rm CIV}$ is computed for each \HI\ bin. This
procedure is repeated 100 times. Finally, for each \HI\ bin we take
the standard deviations of the median $\log \tau_{\rm CIV}$ as our best 
estimate of the errors in the medians (the distribution of the medians
from the bootstrap realizations is approximately lognormal, although
with somewhat more extended tails).

\begin{figure*}
\resizebox{0.9\textwidth}{!}{\includegraphics[60,186][544,606]{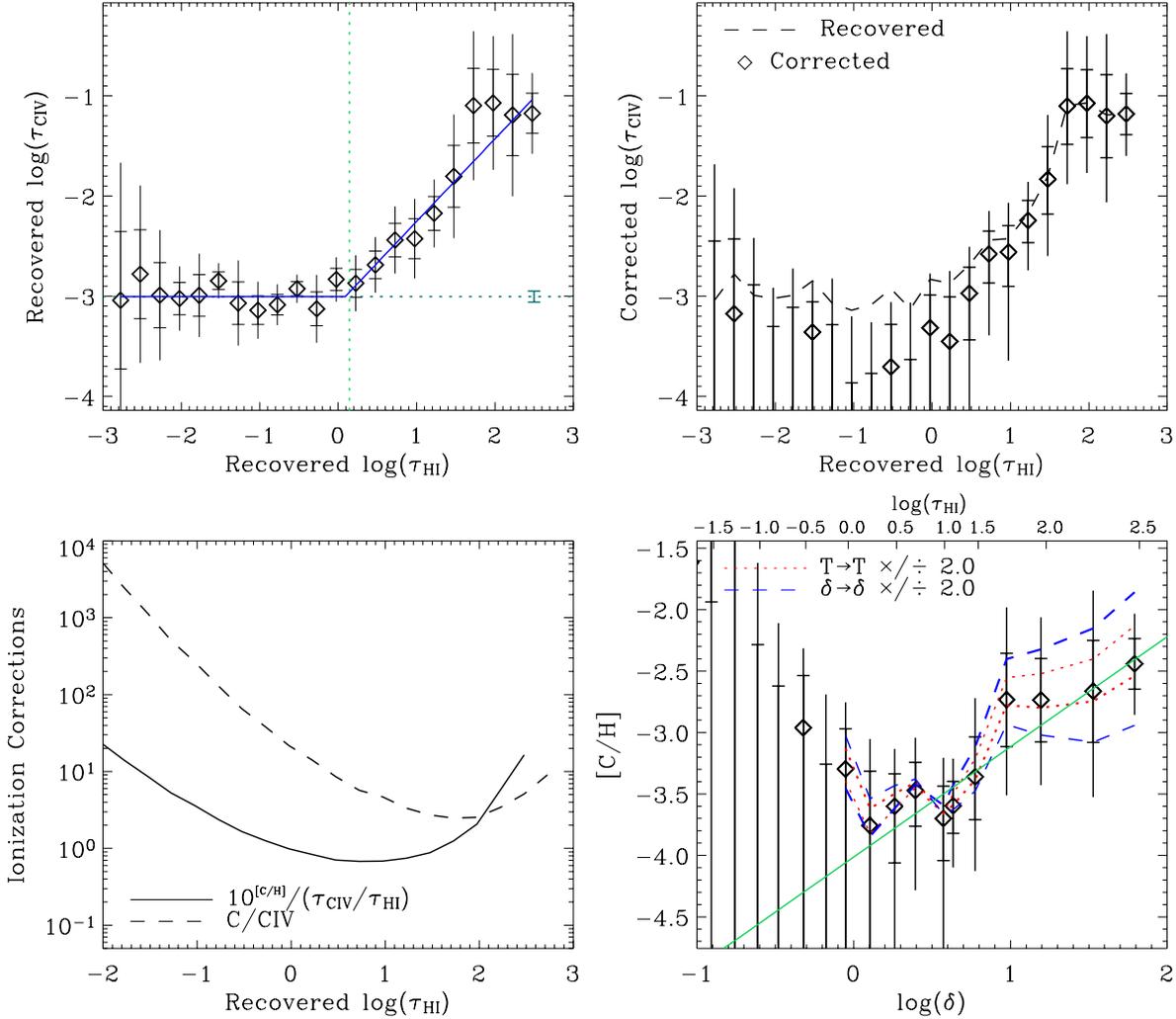}}
\caption{Measuring the carbon abundance in
Q1422+230. Data points are plotted with 1 and 2\,$\sigma$ error bars.
\emph{Top left:} Recovered $\tau_{\rm CIV 1548}$ plotted as a 
function of the recovered $\tau_{\rm HI 1216}$. For $\tau_{\rm HI} <
\tau_c$ (\emph{vertical dotted line}) the signal is lost due to
contamination, noise, and/or continuum errors and $\tau_{\rm CIV}
\approx \tau_{\rm min}$ (\emph{horizontal dotted line}). \emph{Top right:}
Corrected $\tau_{\rm CIV}$ vs.\ recovered $\tau_{\rm HI}$. Data
points have been corrected by subtracting $\tau_{\rm min}$ from the
recovered $\tau_{\rm CIV}$ plotted in the top left-hand panel and indicated
here by the dashed curve. The correction affects points close to
$\tau_{\rm min}$, converting the lowest points into upper
limits. \emph{Bottom left:} Solid and dashed curves show the ionization
correction factor and the inverse of the carbon fraction,
respectively. The ionization correction is insensitive to the \lya\
optical depth in the regime where $\tau_{\rm CIV}$ is best measured.
\emph{Bottom right:} The carbon metallicity as a function of the
overdensity (\emph{bottom axis}) or $\tau_{\rm HI}$ (\emph{top
axis}). Data points were obtained by inserting the corrected $\tau_{\rm
CIV}$, the recovered $\tau_{\rm HI}$ (i.e., the data points shown in
the top right-hand panel), and the ionization correction factor (\emph{solid
curve in the bottom left-hand panel}) into
eq.\~(\protect\ref{eq:metallicity}). The solid line shows the
best-fit power law: $[{\rm C/H}] = -3.12_{-0.10}^{+0.09} +
0.90_{-0.18}^{+0.19} (\log\delta-1.0)$.
Dashed and dotted curves indicate the result of changing the
density and temperature, respectively, by a factor of 2. 
\label{fig:1422_fullinv}}
\end{figure*}

The top left-hand panel of Figure~\ref{fig:1422_fullinv} shows the results
for Q1422+230, one of our best quasar spectra. The recovered \CIV\ and \HI\
optical depths are correlated down to $\log \tau_{\rm HI} \approx
0.1$, below which $\log \tau_{\rm CIV} \approx -3.0$ independent of
$\tau_{\rm HI}$. In Paper I we demonstrated that the correlation
flattens at low optical depth due to the combined effects of noise,
contamination, and continuum fitting errors.

A similar analysis was applied to an even higher S/N spectrum of the
same quasar by Ellison et al.\ (2000). Our results agree with theirs
for $\log \tau_{\rm HI} > 0.3$, below which Ellison et al.'s correlation
flattens off at $\log \tau_{\rm CIV} \approx -2.8$. We thus detect the
$\tau_{\rm CIV}-\tau_{\rm HI}$
correlation down to smaller optical depths
despite the fact that our data has an S/N about a factor of 2
lower than the spectrum of Ellison et al.\ (2000). The improvement is mainly
due to our correction for (self-)contamination.

Note that there is additional information contained in the
distribution of $\tau_{\rm CIV}$ that is not used when we only
consider the median. Different from Paper I, we will also
consider other percentiles than the 50th (i.e., the median), which
will allow us to characterize the scatter in the metallicity at a
fixed density. Our method to measure this scatter is
discussed in \S\ref{sec:method:scatter}.


\emph{4. Correct the median $\tau_{\rm CIV}(\tau_{\rm HI})$.}
The correlation between the recovered $\tau_{\rm CIV}$ and $\tau_{\rm
HI}$ flattens off at low $\tau_{\rm HI}$ where noise, continuum fitting
errors, and contamination wash out the signal. In Paper I we used
simulations to calibrate the difference between the true and recovered
$\tau_{\rm CIV}$ and then used this information to correct the
$\tau_{\rm CIV}$. Because we would like to minimize our use of the
simulations and because the correction for \CIV\ does not show any
dependence on $\tau_{\rm HI}$, we choose to correct the recovered
\CIV\ optical depth using the observations themselves. 

Assuming that, as is the case in the simulations (see Paper I), the
asymptotic flat level $\tau_{\rm CIV}(\tau_{\rm HI}\rightarrow -\infty)$
is the median signal due to contamination, 
noise, and/or continuum fitting errors and that this spurious signal
is independent of $\tau_{\rm HI}$, we can correct for this component
to the signal by subtracting it from the data points. 

The asymptotic flat level is determined as follows. We first find
$\tau_c$, the \HI\ optical depth 
below which the correlation vanishes, by fitting a power law to the data
\begin{eqnarray}
\log(\tau_{\rm CIV}) = \left \{ \begin{array}{ll}
A & \quad \tau_{\rm HI} < \tau_c \\
A + B\log (\tau_{\rm HI}/\tau_c) 
& \quad \tau_{\rm HI} \ge \tau_c \end{array}
\right .
\label{eq:taufit}
\end{eqnarray}
Next, we define the asymptotic \CIV\ optical depth, $\tau_{\rm
min}$, as the median $\tau_{\rm CIV}$ of those pixels that have
$\tau_{\rm HI} < \tau_c$,
\begin{equation}
\tau_{\rm min} \equiv {\rm median}(\tau_{\rm CIV}|\tau_{\rm HI} <
\tau_c)
\end{equation}
(and similarly for other percentiles than the median).
The horizontal and vertical dotted lines in the top left-hand panel of
Figure~\ref{fig:1422_fullinv} indicate $\tau_{\rm min}$ and $\tau_c$,
respectively. As was the case for the data points, the error
on $\tau_{\rm min}$ was computed by bootstrap resampling the
spectrum. To be conservative, we add the error in
$\tau_{\rm min}$ linearly (as opposed to in quadrature) to the errors
on the recovered $\tau_{\rm CIV}$. Finally,
for display purposes only, we fit equation (\ref{eq:taufit}) again to
the data, but this time keeping $\tau_{\rm min}$ fixed at
the value obtained before. The result is shown as the solid curve in
the figure.

The corrected data points are shown in the top right-hand panel of
Figure~\ref{fig:1422_fullinv} and should be compared with the dashed
curve in the same panel, which connects the recovered data points
that were plotted in the top left-hand panel. The
correction is significant only for those 
points that are close to $\tau_{\rm min}$. Data points with $\tau_{\rm
HI} < \tau_c$ are effectively converted into upper limits.

\emph{5. Compute the carbon abundance as a function of density.}

Assuming ionization equilibrium, which should be a very good
approximation for gas photoionized by the UV background,
we can compute the ionization 
balance of hydrogen and carbon given the density, temperature, and a
model for the ionizing background radiation. As discussed in
\S\ref{sec:ionizcorr}, we use CLOUDY to compute
ionization balance interpolation tables as a function of redshift,
density, and temperature, and we use our hydrodynamical simulation to
create tables of the density and temperature as a function of redshift
and \HI\ optical depth. 

To convert the optical depth measurements into metallicity estimates, 
we first compute the median redshift and
$\tau_{\rm HI}$ of each \HI\ bin and then use the interpolation tables
to compute the corresponding density, temperature, and ionization
balance. For example, the dashed
curve in the bottom left panel of Figure~\ref{fig:1422_fullinv} shows
C/\CIV, the inverse of the \CIV\ fraction, as a
function of $\tau_{\rm HI}$. The solid curve shows the ionization
correction factor: $10^{[{\rm C/H}]}/(\tau_{\rm CIV}/\tau_{\rm HI})$.

Given the corrected \CIV\ optical depth $\tau_{\rm CIV}$, the
recovered \HI\ optical depth $\tau_{\rm HI}$, and the ionization
correction, we can compute the carbon metallicity: 
\begin{equation}
[{\rm C/H}] 
=  \log \left ({\tau_{\rm CIV} (f\lambda)_{\rm HI} \over \tau_{\rm HI}
(f\lambda)_{\rm CIV}}  {\rm C \over CIV}{\rm HI \over H}\right ) - ({\rm
C/H})_\odot,
\label{eq:metallicity} 
\end{equation}
where $f_i$ and $\lambda_i$ are the oscillator strength and rest
wavelength of transition $i$, respectively ($f_{\rm CIV} = 0.1908$,
$f_{\rm HI} = 0.4164$, $\lambda_{\rm CIV} = 1548.2041$~\AA,
$\lambda_{\rm HI} = 1215.6701$~\AA), and we use the solar abundance
$({\rm C/H})_\odot = -3.45$ (number density relative to hydrogen;
Anders \& Grevesse 1989). The resulting 
data points are plotted in the bottom right-hand panel of
Figure~\ref{fig:1422_fullinv}, which shows the carbon metallicity as a
function of the overdensity (\emph{bottom axis}) or \lya\ optical depth
(\emph{top axis}). 

Metals are detected over two decades in density. The median
metallicity increases from a few times $10^{-4}$ around the
mean density to $\sim 10^{-3}$ to $10^{-2}$ around an overdensity $\delta
\sim 10^2$.  The best-fit\footnote{All fits in this paper are obtained
by minimizing $\chi^2 = \sum_i\left ([Z_i - Z(\delta_i)]/
\sigma_i \right )^2$, where $Z_i = [{\rm C/H}]_i$ are the metallicity
data points and $Z(\delta)$ is the function being fitted. The sum is
over all data points for which $\tau_{\rm HI} > \tau_c$. The errors in
the recovered $\tau_{\rm CIV}$ are lognormal, but the subtraction of
$\tau_{\rm min}$ results in asymmetric error bars for $[{\rm
C/H}]$. We therefore compute a closely spaced grid of errors (e.g.,
$\pm 0.01\sigma, 0.02\sigma, \ldots$) in $\tau_{\rm CIV}$ and
$\tau_{\rm min}$, propagate these to obtain an equivalent (irregular)
grid of errors for $[{\rm C/H}]$, and then compute $\chi^2$ using the
latter grid.}
constant metallicity is $[{\rm C/H}] = -3.30_{-0.08}^{+0.08}$, but
this fit is ruled 
out at high confidence ($\chi^2/{\rm dof} \approx 30.6/9\approx 3.4$
for the data points with $\tau_{\rm HI}>\tau_c$,
corresponding to a probability 
$Q=3.5\times 10^{-4}$). The best-fit power-law metallicity is $[{\rm
C/H}] = -3.12_{-0.10}^{+0.09} + 0.90_{-0.18}^{+0.19}
(\log\delta-1.0)$, which has $\chi^2/{\rm
dof} \approx 3.42/8\approx 0.43$, or a probability $Q=0.91$.

The errors in 
this plot reflect only the errors in the corrected median $\tau_{\rm
CIV}$ (\emph{top right-hand panel}) and do not take uncertainties
in the ionization balance into account. The dotted (dashed) curves
illustrate how the metallicity changes if the temperature
(density) is changed by a factor of 2. The metallicity is most
sensitive to the assumed density for 
high $\delta$, where the ionization correction factor changes
rapidly with $\tau_{\rm HI}$. Fortunately, the ionization correction
is relatively insensitive to both the density and the temperature in
the regime where our data is best.

\subsection{The distribution of metals at a fixed density}
\label{sec:method:scatter}
\begin{figure*}[t]
\plottwo{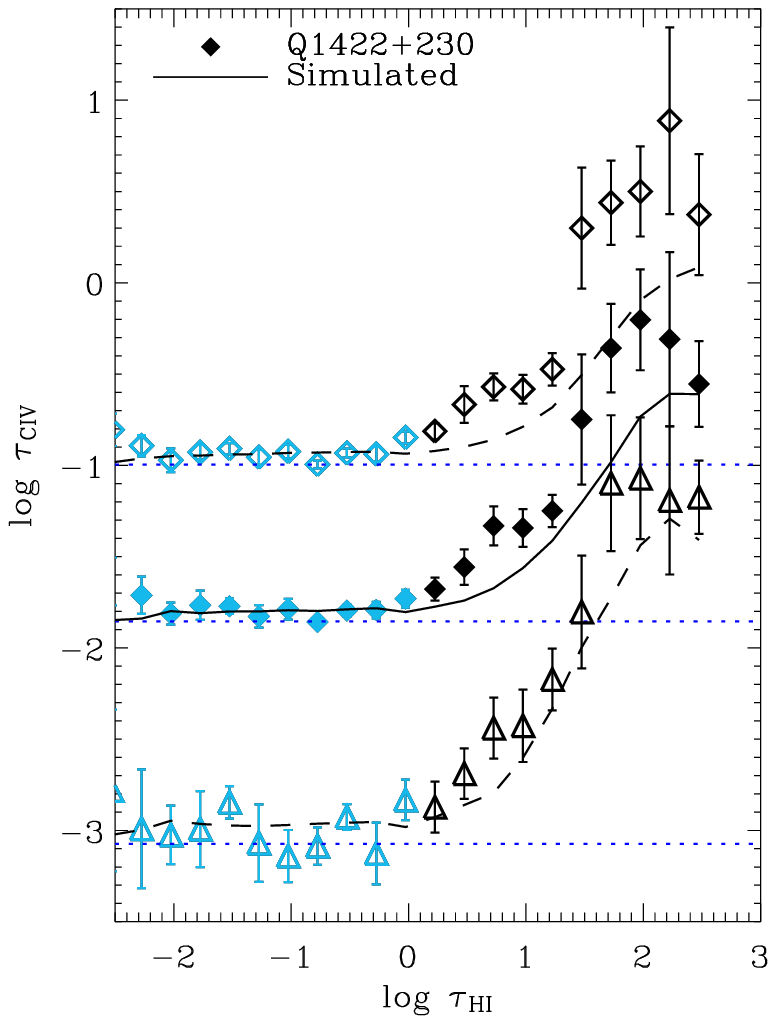}{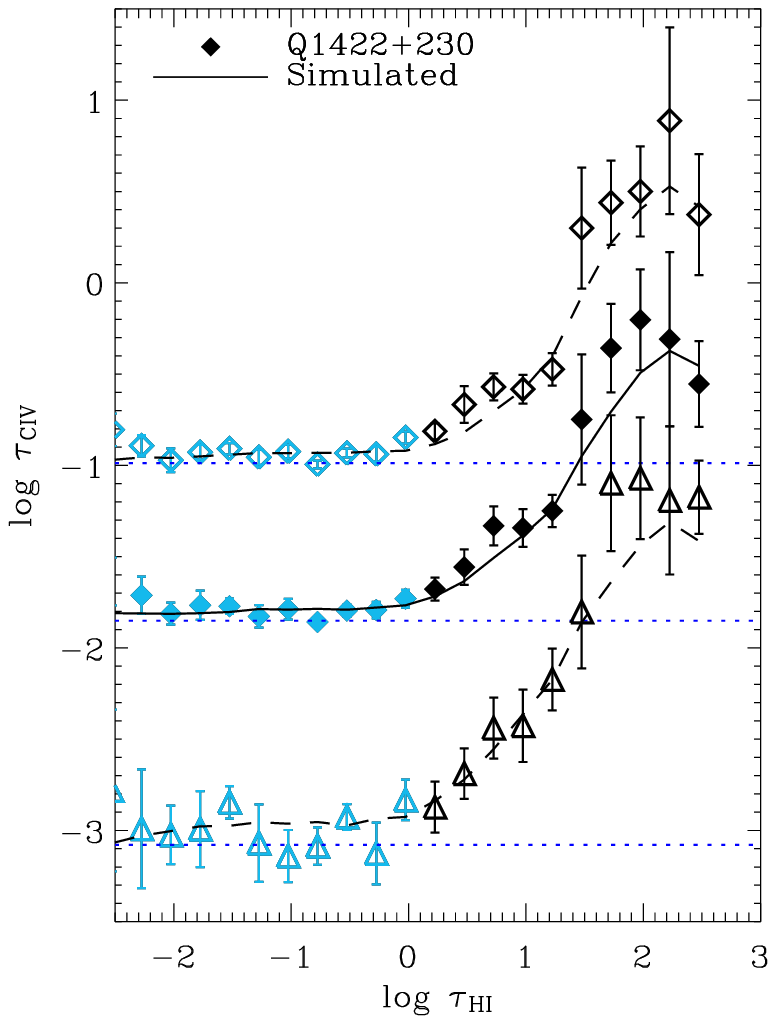}
\caption{Comparison of the optical depth statistics of observed
and simulated spectra using the metal distribution measured
from the observations. From top to bottom the three sets of data
points are the 84th (\emph{open diamonds}), 69th (\emph{solid
  diamonds}) and 50th 
(\emph{triangles}) percentiles of the recovered CIV
optical depth as a function of $\tau_{\rm HI}$ for Q1422+230. For
clarity, the 84th and 69th percentiles have been offset by +1.0 and +0.5
dex, respectively. The
curves in the left-hand panel are for a simulation in
which each particle was given the median metallicity measured from the
observations, $[{\rm C/H}] = -3.12 + 0.90(\log\delta-1.0)$. The
simulation can fit the observed median $\tau_{\rm CIV}$ ($\chi^2$
probability $Q=0.21$), but not the observed $\tau_{\rm CIV}(\tau_{\rm
  HI})$ for the other percentiles ($Q<10^{-4}$). The curves in the
right-hand panel are for a simulation that has 
the same median metallicity, but which includes scatter. The
simulation cube was divided into $10^3$ cubic sections, and all
particles in each section were given a metallicity of $[{\rm C/H}] =
-3.12 + s + 0.90(\log\delta-1.0)$, where $s$, which is the same for
all particles in the subvolume, is drawn at random from a lognormal
distribution with mean 0 and variance $\sigma = 0.81~{\rm dex}$ as
measured from the observations. The simulation provides an acceptable
fit to all percentiles (from top to bottom $Q=0.33$, 0.69, and 0.90).
\label{fig:1422_fwd}}
\end{figure*}

Having measured the median carbon abundance as a function of the
density, we use the simulation to check whether our measured
metallicity profile can reproduce the observed optical depth
statistics. The bottom set of data points in the left-panel of
Figure~\ref{fig:1422_fwd} shows the observed median \CIV\ 
optical depth as a 
function of $\tau_{\rm HI}$. The bottom solid curve shows
the median $\tau_{\rm CIV}$ in the simulation (averaged over 10
spectra) for the best fit metallicity profile measured from the
observations [each gas particle was given a metallicity $[{\rm C/H}] =
-3.12 + 0.90(\log\delta-1.0)$]. Clearly, the simulation curve is a
good fit of the observations ($\chi^2$ 
probability\footnote{Because we do not want to compare simulated and
observed noise, the $\chi^2$ is computed only for the data points
with $\tau_{\rm HI}>\tau_c$, where $\tau_c$ is computed for the
observed spectra. The light data points were excluded from the $\chi^2$
calculation. For the same reason, we have added $\tau_{\rm
  min,obs}-\tau_{\rm min,sim}$ to the simulated data points. The
horizontal dotted lines indicate the original $\tau_{\rm min,sim}$.} 
$Q=0.21$), which
gives us confidence that, given a model for the UV background, our
method for measuring the median carbon abundance as a function
of density is robust.


Although the median $\tau_{\rm CIV}$ is reproduced by the simulation,
this is not the case for other percentiles. From top to bottom the sets of
data points in Figure~\ref{fig:1422_fwd} correspond to the 84th, 69th,
and 50th  
percentiles. The simulation clearly cannot reproduce the width of the
observed metallicity distribution ($Q<10^{-4}$ for the higher
percentiles). This indicates 
that the scatter in the observations is greater than can be accounted
for by contamination and noise effects, both of which are present in
the simulations. The scatter is also unlikely to be due to continuum
fitting errors because those would result in a scatter $\sigma(\tau_{\rm
CIV})$ that is independent of $\tau_{\rm HI}$, contrary to what is
observed. Hence, to reproduce the observed optical depth statistics, we
need to extend the method of Paper I to allow for scatter in the
metallicity.  

\begin{figure*}[t]
\plotone{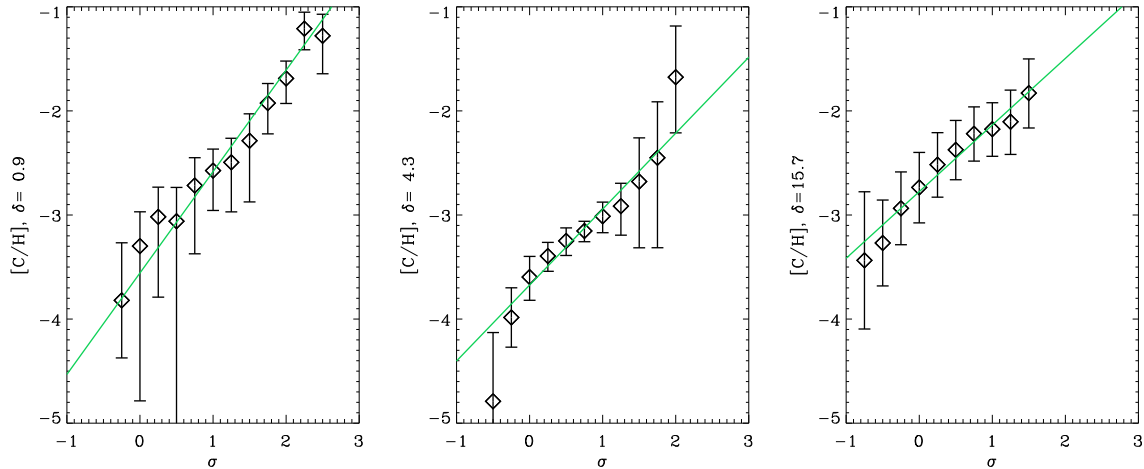}
\caption{Distribution of carbon metallicities for three gas
overdensities: $\delta \approx 1.6$ (\emph{left}), $\delta \approx 5.4$
(\emph{middle}), and $\delta \approx 19.4$ (\emph{right}). For each
density (i.e., 
HI bin) the CIV pixels were ranked according to $\tau_{\rm CIV}$
and a number of percentiles were computed. The metallicity 
corresponding to each percentile was computed as described for the
median in \protect\S\ref{sec:method:median}. The missing percentiles
correspond to negative optical depths (i.e., the normalized flux is
greater than unity). The curves are lognormal fits to the data points. The
best-fit slopes are $\sigma([{\rm C}/{\rm H}]) = 0.97$, 0.73, and
0.64 for the left, middle, and right panels, respectively.
\label{fig:lgnorm_sigma}}
\end{figure*}

Our method to reconstruct the median metallicity can of course also be
applied to other percentiles. As we will show below, for a fixed
density the metallicity distribution is approximately lognormal.
A lognormal distribution is parameterized by the mean
$\left < \log Z\right > = {\rm median}(\log Z)$ and the variance
$\sigma^2(\log Z)$, and the fraction $f$ of pixels that have a
value smaller than $x\sigma$, is given by the Gaussian integral
\begin{equation}
f(x) = {1 \over \sqrt{2\pi}} \int_{-\infty}^{x} e^{-t^2/2} dt.
\end{equation}
For a given \HI\ bin, we measure $\sigma$ by computing
$\log Z$ for a large number of percentiles (i.e., values of $f$)
corresponding to a regular grid of $x$-values, and fitting the
linear function $\log Z = \sigma x$ to the data points (we use a
spacing $\Delta x = 0.25$ and require a minimum of four data points). The
procedure 
is illustrated in Figure~\ref{fig:lgnorm_sigma}, which shows the
measured metallicity as a function of $x$ for three
different densities. The data
appear well fitted by lognormal distributions 
(\emph{solid curves}) with $\sigma\approx0.97$ for $\delta \approx 0.9$
($\tau_{\rm HI} = 0.95$; \emph{left-hand
panel}), $\sigma\approx0.73$ for $\delta \approx 4.3$ ($\tau_{\rm HI} =
16.8$; \emph{middle 
panel}), and  $\sigma\approx0.64$ for $\delta \approx 15.7$ ($\tau_{\rm HI}
= 94.4$, \emph{right-hand panel}). Note that we cannot get reliable errors on
$\sigma$ from the 
fits shown in the figure because the errors on the data points are all
correlated since the abscissa corresponds to a ranking. 

Figure~\ref{fig:lgnorm_sigma} demonstrates that the distribution of pixel
metallicities is well described by a lognormal function over at least
the range $-0.5\sigma$ to $+2\sigma$. To probe the distribution
further into the low-metallicity tail we would need to reduce the
noise, since the lowest percentile/$x$-value corresponds to the
minimum positive $\tau_{\rm CIV}$ (the rest of the pixels have a flux
greater than 
$e^{-\tau_{\rm min}}$). To probe further into the high-metallicity
tail we would 
need more pixels since for $N$ pixels one cannot measure a percentile
greater than $(N-1)/N$ [in practice we only make use of percentiles
for which $N(1-f) > 5$ in order to prevent a few anomalous pixels from
affecting the results].


The right-hand panel of Figure~\ref{fig:1422_fwd} compares the observed
percentiles with a simulation that uses 
a lognormal metallicity distribution, with the same median metallicity
as a function of density as before and with a constant scatter of 
$\sigma([{\rm C}/{\rm H}]) = 0.81~{\rm dex}$ (the median value of the
scatter measured in the observations for the different densities) on 
a comoving scale\footnote{The exact scale is unimportant as long as it
is at least as large as the effective smoothing scale of the absorbers
and smaller than the simulation box.} 
of $1.2~h^{-1}~\mpc$. Contrary to the
simulation without scatter (\emph{left-hand panel}), this simulation
is consistent 
with all the observed percentiles ($\chi^2$ probability $Q=0.33$,
0.69, and 0.90). To estimate the error in $\sigma([{\rm C}/{\rm H}])$
we compared the observed 
percentiles with those computed from simulated spectra using varying
amounts of scatter in the metallicity. We find that values of
$\sigma([{\rm C}/{\rm H}]) =  0.81 \pm 0.25$ are acceptable. 

Unlike the median metallicity, the scatter is independent of the
assumed spectral shape of the UV background radiation. This is because
the ionization correction 
consists of multiplying the observed optical depth ratio $\tau_{\rm
CIV}/\tau_{\rm HI}$ with a factor that is independent of $\tau_{\rm
CIV}$: $[{\rm C}/{\rm H}] = \log(\tau_{\rm CIV}/\tau_{\rm HI}) + X$,
where $X$ is the log of the ionization correction factor (see eq.\
[\ref{eq:metallicity}]) which depends only on $\tau_{\rm HI}$. Hence,
if $\log(\tau_{\rm CIV}/\tau_{\rm HI})$ is distributed lognormally
with mean $\mu$ and variance $\sigma^2$, then $[{\rm C}/{\rm H}]$ is
also distributed lognormally with variance $\sigma^2$, but with mean
$\mu + X$. Note, however, that we have
assumed the UV background to be uniform. If there are large fluctuations in
the UV radiation at a fixed \HI\ optical depth, then the true scatter
in the metallicity could be smaller.

Having found that the metallicity distribution is well fitted by a
lognormal function, we can estimate the mean metallicity as follows
\begin{eqnarray}
\log \left < Z\right > 
&=& \log \left ({1 \over \sqrt{2\pi}\sigma} \int_{-\infty}^{+\infty} Z
e^{- \left (\log Z - \left < \log Z\right >\right )^2/2\sigma^2} d\log Z
\right )\\
&=& \left < \log Z\right > + {\ln 10 \over
2}\sigma^2
\approx \left < \log Z\right > + 1.15\sigma^2,
\label{eq:meanz}
\end{eqnarray}
where $\left < \log Z\right > = {\rm median}(\log Z)$ and $\sigma$ are
obtained as explained above. For Q1422+230 this gives a mean
metallicity of $\log \left <Z\right > \approx -2.36 \pm 0.48$ at
$\delta=10$.  

Unlike estimates based on the
mean \CIV\ optical depth, this estimate of the mean metallicity is
based on a fit to the full metallicity distribution and is thus
relatively insensitive 
to anomalous pixels. On the other hand, since we have only a finite number of
pixels, this estimate of the mean does implicitly assume that the
metallicity distribution remains lognormal beyond where we can actually
measure it. While the mean metallicity is insensitive to the shape of the
low-metallicity tail of the distribution,
the contribution of the unmeasured high-metallicity tail to the mean 
metallicity depends strongly on the variance. For example, cutting of
the distribution at $+2\sigma$ reduces the mean by factors of 2.6 and
1.2 for $\sigma = 1.0$ and 0.5~dex, respectively. For Q1422+230 we find
$\sigma \approx 0.8$, which is small enough for the mean to be fairly
insensitive to the unmeasured high-metallicity tail.

\subsection{Summary of results for Q1422+230}
Using model QG for the UV background and the $\delta(\tau_{\rm
HI},z)$ and $T(\tau_{\rm HI},z)$ relations obtained from the
simulations, we have found the following from analyzing the statistics
of $\tau_{\rm  
CIV}(\tau_{\rm HI})$ in Q1422+230: 
\begin{enumerate}
\item The observed median optical depths are \emph{inconsistent} with a
  constant metallicity. The observations are well fitted by a median carbon
abundance that is a power law of the overdensity: $[{\rm C/H}] =
-3.12_{-0.10}^{+0.09} + 0.90_{-0.18}^{+0.19} (\log\delta-1.0)$.
\item The distribution of $\tau_{\rm CIV}$ at fixed $\tau_{\rm HI}$ is
\emph{inconsistent} with a metallicity that depends only on density. The
observations are well fitted by a lognormal metallicity distribution with
a scatter of $0.81\pm 0.25$~dex and a median that varies as above.
\end{enumerate}

\section{The CIII/CIV ratio}
\label{sec:ciii}

We compute the ionization corrections as a function of the \HI\ optical
depth and redshift using interpolation tables created from the
simulation. Because the simulation reproduces the observed evolution
of the mean absorption (Fig.~\ref{fig:taueff}) and the temperature of
the gas responsible for the low-column density \lya\ lines (Schaye et
al.\ 2000b), we are confident that the ionization balance of the gas
responsible for the \HI\ absorption is on average
well determined. However, it is possible in principle that the \CIV\
and \HI\ absorption at a given redshift is dominated by different
gas parcels. For example, Theuns et al.\ (2002b) predict that
many of the 
metals produced by dwarf galaxies reside in hot gas bubbles that do 
not contribute significantly to the \HI\ absorption. In their simulation
most of the carbon has $T\sim 10^6~\K$ and is too highly ionized to
give rise to \CIV\ absorption. In this scenario the observed \CIV\
absorption arises in the low-temperature tail of the
carbon temperature distribution.

Although the ionization corrections are insensitive to
small changes in the temperature (see Fig.~\ref{fig:1422_fullinv}),
they could be considerably in error if the temperature were
high enough for collisional ionization to be important ($T\ga 10^5~\K$
for \CIV). Furthermore, such a high temperature would 
change the relation between $\tau_{\rm HI}$ and $\delta$,
causing us to underestimate the density corresponding to a given \HI\
bin. 


\begin{figure}
\epsscale{1.2}
\plotone{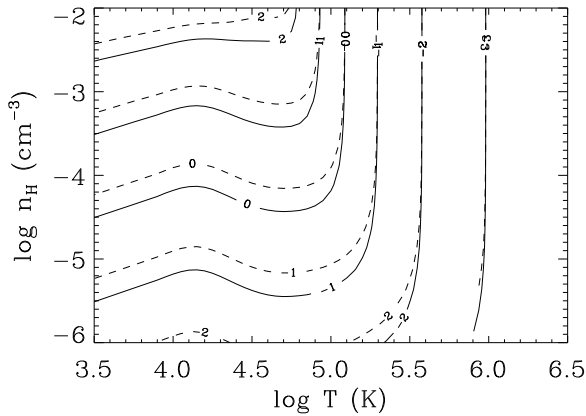}
\caption{Contour plot of $\log(\tau_{\rm CIII}/\tau_{\rm CIV})$ in
the density-temperature plane. Solid and dashed contours are for the
$z=3$ QG and Q UV backgrounds, respectively. For each density there is
a critical temperature below which the CIII/CIV ratio is nearly
constant and above which it decreases rapidly. For the densities of
interest here ($n_{\rm H} \sim 10^{-5}$ to $10^{-3}$), the ratio varies
from 0.1 to 10. 
\label{fig:c3c4pred}}
\end{figure}


Fortunately, there is a way to constrain the temperature of the gas
responsible for the carbon absorption by comparing the strengths of
the \CIII\ ($\lambda 977$) and \CIV\ ($\lambda\lambda 1548, 1551$)
transitions. Figure~\ref{fig:c3c4pred} shows a contour plot of
the predicted $\log \tau_{\rm CIII}/\tau_{\rm CIV}$ in the density-temperature
plane. The solid (dashed) contours are for the $z=3$ QG (Q) model of
the UV background radiation. From the figure it can be seen that a
measurement of the \CIII/\CIV\ ratio 
yields an upper limit on the temperature independent of the
density and the UV background.

\begin{figure*}[t]
\epsscale{1.15}
\plottwo{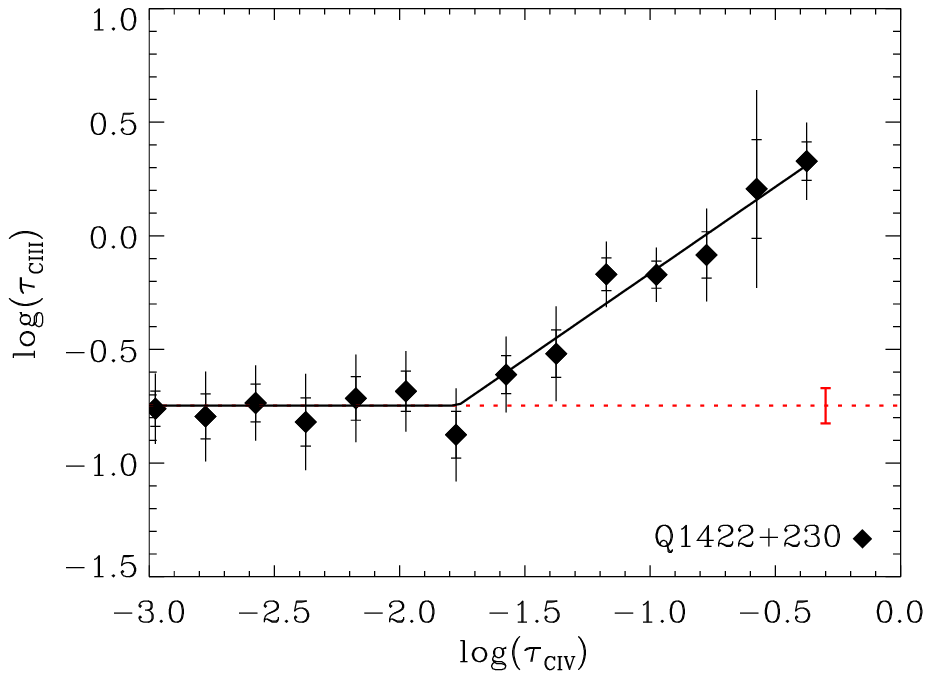}{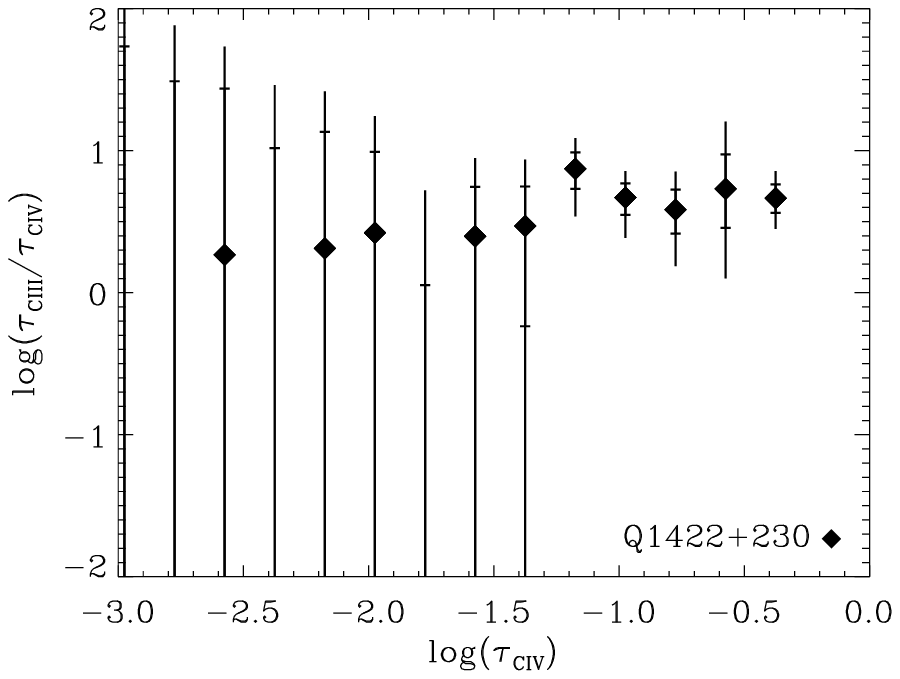}
\caption{$\tau_{\rm CIII}$ as a function of $\tau_{\rm CIV}$ for
  Q1422+230. Data points are 
plotted with 1 and $2\sigma$ error bars. \emph{Left:} Median
recovered $\tau_{\rm CIII}$ of pixels binned as a function of
recovered $\tau_{\rm CIV}$. CIII is detected down to $\log
\tau_{\rm CIV}\approx 
-1.7$. \emph{Right:} The $\tau_{\rm CIII}/\tau_{\rm CIV}$ ratio is
plotted as a function of 
$\tau_{\rm CIV}$. The median $\tau_{\rm CIII}$ were
corrected for noise, contamination, etc., as described for CIV in
\S\ref{sec:method:median} (step 4). The ratio $\tau_{\rm
CIII}/\tau_{\rm CIV}$ is much 
higher than it would be if the temperature were higher than
$10^5~\K$ (see Fig.~\protect\ref{fig:c3c4pred}). 
\label{fig:c3c41422}}
\end{figure*}

The left-hand panel of Figure~\ref{fig:c3c41422} shows the median \CIII\
optical depth as a function of $\tau_{\rm CIV}$ for Q1422+230. The
\CIV\ optical depths were recovered as described in
\S\ref{sec:method:median}, while the \CIII\ optical depths were
recovered as described in Paper I for \OVI, i.e., contaminating higher
order Lyman lines were removed. Absorption by \CIII\ is clearly
detected over 1.5 decades in $\tau_{\rm CIV}$. For $\log \tau_{\rm
CIV} < -1.7$ the apparent \CIII\ optical depth is constant, indicating
that for these low \CIV\ optical depths the true $\tau_{\rm CIII}$ is
smaller than the spurious signal resulting from contamination by other
absorption lines, noise, and continuum fitting errors.

Note that $\log \tau_{\rm CIV} =
-1.7$ corresponds to a typical \HI\ optical depth of a few tens (see
Fig.~\ref{fig:1422_fullinv}). Although there is
considerable scatter in the $\tau_{\rm CIV}$ corresponding to a fixed
$\tau_{\rm HI}$, this does suggest that we are detecting \CIII\ mostly
in gas with density $\delta \ga 10$. Indeed, the correlation between
$\tau_{\rm CIII}$ and $\tau_{\rm HI}$ (not plotted) is detectable only
for $\log \tau_{\rm HI} > 1$. 


We can see from
Figure~\ref{fig:c3c4pred} that if the carbon were purely photoionized ($T \ll
10^5~\K$), we would expect $\tau_{\rm CIII}/\tau_{\rm CIV}$
values between 1 and 10 at $z\approx 3$ and $\delta \ga 10$. The right-hand
panel of Figure~\ref{fig:c3c41422} shows the log of the ratio 
$\tau_{\rm CIII}/\tau_{\rm CIV}$ as a function of $\tau_{\rm CIV}$,
where the $\tau_{\rm CIII}$ are the median values plotted in the left-hand
panel but after subtraction of $\tau_{\rm min}$ (\emph{horizontal dashed
line}). The flat level $\tau_{\rm min}$ was determined in the same way
as we did for 
$\tau_{\rm CIV}(\tau_{\rm HI})$ in \S\ref{sec:method:median} (step
4). Percentiles other 
than the median (not plotted) give nearly identical $\tau_{\rm
  CIII}/\tau_{\rm CIV}$
ratios (after subtraction of the corresponding $\tau_{\rm min}$
percentiles), indicating that this ratio is rather uniform. 
For $\log \tau_{\rm
CIV} \ga -1.5$ we detect \CIII\ and $\log \tau_{\rm CIII}/\tau_{\rm
CIV} \sim 0.5$, as expected for photoionized gas at this
redshift. Crucially, optical depth ratios typical for collisionally
ionized \CIV\ ($T > 10^5~\K$) are ruled out by the data. Hence, the
temperature of the gas responsible for the carbon absorption is
smaller than $10^5~\K$, at least for densities $\delta \ga 10$. This
provides an important constraint for theories of the enrichment of the
IGM through galactic winds and gives us confidence that our assumption
that photoionization dominates is reasonable.

\begin{figure}
\epsscale{1.2}
\plotone{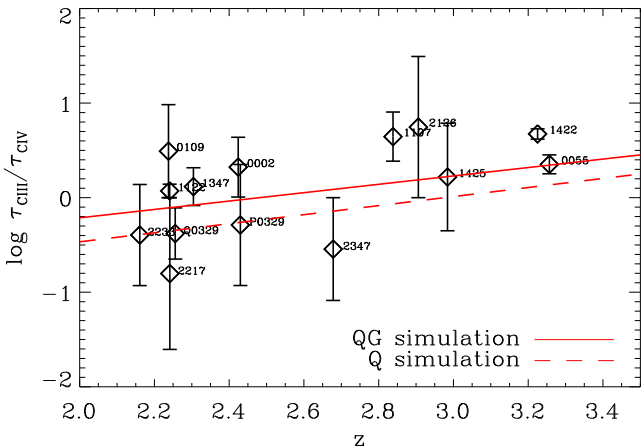}
\caption{The ratio $\tau_{\rm CIII}/\tau_{\rm CIV}$ is plotted as a
function of redshift for all $z_{\rm em}<4$ QSOs for which CIII is
covered (we do not detect CIII for QSOs above this redshift). For each 
QSO the median $\tau_{\rm CIII}$ were corrected for 
contamination, noise, etc., using the method described in
\S\protect\ref{sec:method:median} and the $\tau_{\rm CIII}/\tau_{\rm
  CIV}$ ratio was averaged 
over all bins with $\tau_{\rm CIV}>\tau_c$ (see
\S\protect\ref{sec:method:median} for the definition of
$\tau_c$). The solid line shows the best linear fit to the CIII/CIV
ratio measured from simulations of the same quasar spectra, which used
the QG UV background and the metallicity distributions measured from
the observations. The observed ratios agree well with the
simulations, 
and comparison with Fig.~\protect\ref{fig:c3c4pred} shows that they
are in the range expected for photoionized gas
with $n_{\rm H}\sim 10^{-4}~\cm^{-3}$ and $T < 10^5~\K$. 
\label{fig:c3c4}}
\end{figure}

We have computed the \CIII/\CIV\ ratio for all quasars in our sample
that cover the \CIII\ region. The results are summarized in
Figure~\ref{fig:c3c4}, in which $\tau_{\rm CIII}/\tau_{\rm CIV}$,
averaged over $\tau_{\rm 
CIV}$, is plotted as a function of redshift. While this plot uses
the median $\tau_{\rm CIII}$, the other percentiles give nearly 
identical results. For all QSOs the ratio is
in the range expected for photoionized gas with $n_{\rm H}\sim
10^{-4}~\cm^{-3}$ and $T< 10^5~\K$. 
The solid line shows the best linear least absolute
deviation fit to 
the simulated versions of all the quasar spectra, using the metallicity
distributions measured from the observed spectra and the QG
background. The dashed line shows a similar fit for the Q background.
Overall, the
simulations agree well with the observations, although they
may slightly underpredict the \CIII/\CIV\ ratios for $z>3$. The QG
background seems to fit the data slightly better than model Q, but the
difference is too small to be significant.
In the simulation the small decrease in the \CIII/\CIV\ ratio with
time is caused 
both by the decrease in the typical density of \CIV\ absorbers due to
the expansion of the universe and by the hardening of the
UV background radiation.

Thus, the observed
\CIII/\CIV\ ratios are in the range expected for photoionized gas, but
are \emph{inconsistent} with scenarios in which a large fraction of
the \CIV\ absorption takes place in gas with temperatures $T\ga
10^5~\K$. 


\section{Results for the full sample}
\label{sec:results}
In \S\ref{sec:method} we showed results for Q1422+230 to illustrate
our method for measuring the distribution of carbon as a function of
the gas density. In this section we will present the results from the
complete sample of quasars. The results from the individual quasars
are given in Appendix A.

All abundances are by number relative to the total hydrogen density,
in units of the solar abundance [$({\rm C/H})_\odot = -3.45$; Anders
\& Grevesse 1989].

Our goal is to measure the carbon abundance as a function of density
and redshift. For a lognormal distribution, which we found to provide
a good fit to the data (see \S\ref{sec:method:scatter}), the
distribution is determined by two parameters: the mean
$\left < [{\rm C}/{\rm H}]\right > = {\rm median}([{\rm C}/{\rm H}])$ and
the standard deviation $\sigma([{\rm C}/{\rm H}])$ (which we will
often refer to as the 
scatter). Thus, we can characterize the full distribution of carbon by 
fitting functions of the two variables
$\delta$ and $z$ to all the data points for $\left < [{\rm C}/{\rm
H}]\right >$ and $\sigma([{\rm C}/{\rm H}])$ obtained from the individual
quasars (see Fig.~\ref{fig:individual}). But before doing so, it is
instructive to bin the data in one variable and plot it against the
other. 

\begin{figure*}[t]
\epsscale{1.05}
\plotone{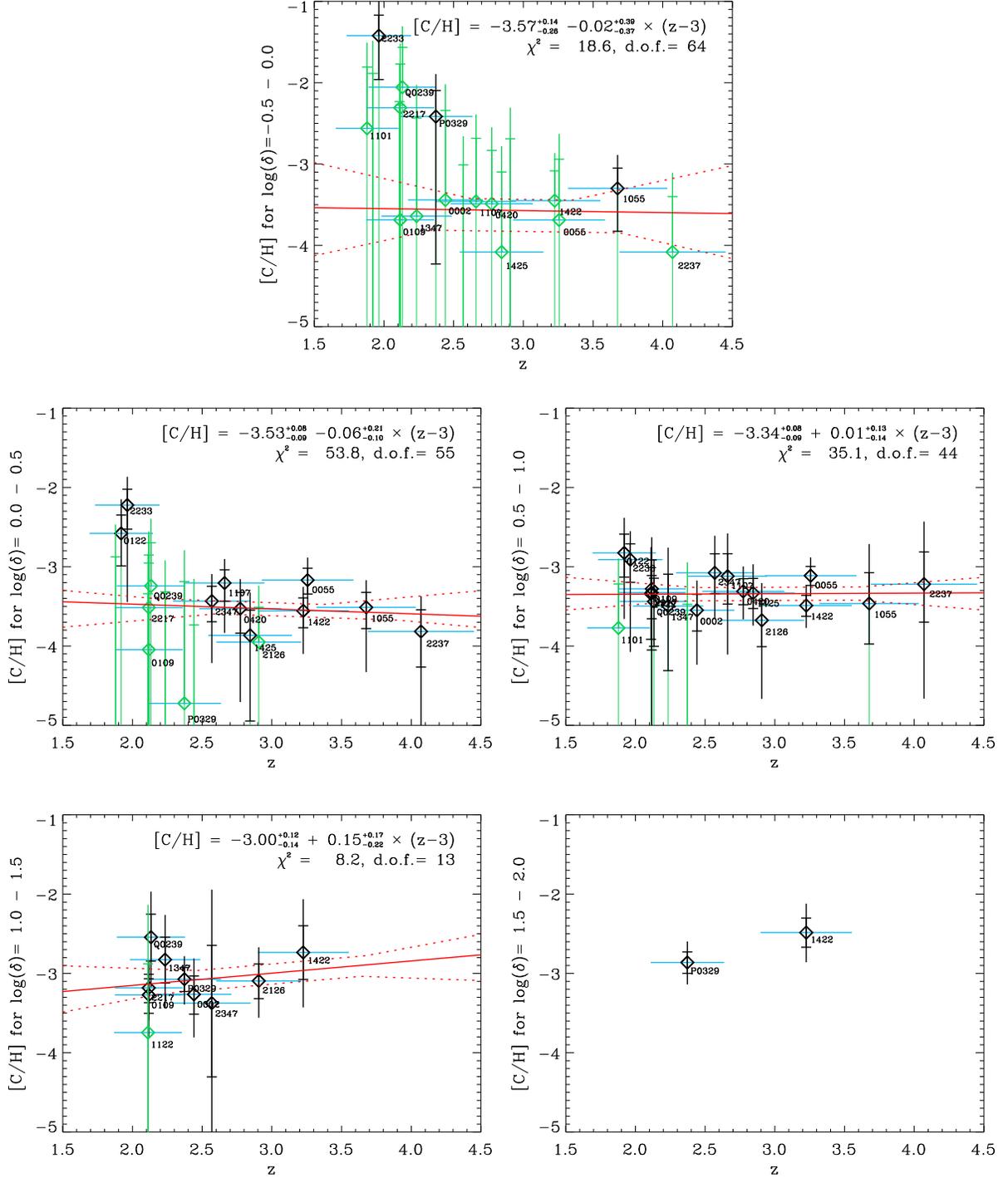}
\caption{Median metallicity as a function of redshift for different
overdensity cuts. From the top left to the bottom right, panels
correspond to the following density bins: $\log\delta = -0.5 - 0.0$,
$0.0 - 0.5$, $0.5 - 1.0$, $1.0-1.5$, and $1.5-2.0$. One data point per
QSO is plotted, labeled with the first four digits of the QSO
name. Horizontal error bars indicate redshift ranges, vertical error
bars indicate 1 and 2$\sigma$ errors. Light-colored data points have a lower
1$\sigma$ error of $-\infty$. For each QSO and density bin the plotted
data point was obtained by fitting 
a constant metallicity through all the HI bins of that QSO
spectrum (plotted in Fig.~\protect\ref{fig:individual}) that fall in
the density bin. This rebinning was done only for visualization
purposes, facilitating 
direct comparison with Fig.~\protect\ref{fig:individual}. The
least-squares fits, shown as the solid lines, were obtained by fitting
all data points (i.e., HI bins) from
Fig.~\protect\ref{fig:individual}. The errors on the parameters of
the fit were determined by bootstrap resampling the quasars. Dotted
curves indicate the 1$\sigma$ confidence limits. 
\label{fig:dbins}}
\end{figure*}

Figure~\ref{fig:dbins} shows the median metallicity versus redshift for five
different density bins of width 0.5~dex, centered about densities
increasing from $\log\delta = -0.25$ (\emph{top panel}) to $\log\delta =
1.75$ (\emph{bottom right-hand panel}). To facilitate comparison with
Figure~\ref{fig:individual} we have plotted only one data point per
quasar, obtained by fitting a constant metallicity to all the data points
of the quasar that fall in the given density bin. The points have been
labeled with the first four digits of the quasar name. The light data
points have $1\sigma$ lower limits that extend to minus infinity.  The
solid lines in each panel show the least-squares fits to the original
data points (i.e., the ones from Fig.~\ref{fig:individual}, not the
plotted rebinned data points) and the dotted curves indicate the
$1\sigma$ confidence limits. The errors on all fits were determined by
bootstrap resampling the quasars. Computing the errors using the
$\chi^2$ surface gives similar results, although the $\chi^2$ errors
tend to be somewhat larger (smaller) than the bootstrap errors if the
$\chi^2$ per degree of freedom is smaller (greater) than unity. 
The reduced $\chi^2$ are somewhat small, particularly at low
overdensities, indicating that we have
overestimated the errors. This is not unexpected as we have used 
conservative
estimates of the errors on data points for which the correction of the
``noise'' component is significant (i.e., $\tau_{\rm CIV} \approx
\tau_{\rm min}$; see \S\ref{sec:method:median}).

The first panel reveals that we have detected metals
in underdense gas: $[{\rm C}/{\rm H}] \approx -3.6$ at
$\log\delta=-0.25$, with a 2$\sigma$ lower limit of -4.12. We find
that $[{\rm C}/{\rm H}] > -5$ at the $2.4\sigma$ level (99.2\%
confidence). For percentiles higher than the 50th (i.e., the median)
the detection is even more significant. For example, for the 69th
percentile we find $[{\rm 
C}/{\rm H}] > -5$ at the $3.4\sigma$ level. Hence, barring errors in
our estimate of the density contrast, there is no question that a
large fraction of underdense gas has been enriched.

At the $1\sigma$ level all bins are consistent with no evolution. The
constraints are strongest for the bin $\log\delta = 0.5 - 1.0$, in
which the maximum allowed increase in the median metallicity per unit
redshift is 0.14~dex at the $1\sigma$ level and 0.25~dex at the
$2\sigma$ level.  
Comparing the metallicity at $z=3$ for the
different overdensity bins, we see that the metallicity increases with
overdensity from $[{\rm C}/{\rm H}] \approx -3.6$ at $\log\delta \approx -0.25$
to $[{\rm C}/{\rm H}] \approx -2.6$ at $\log\delta\approx 1.75$. This
trend may be 
more easily analyzed using the next figure.


\begin{figure*}[t]
\begin{center} 
\resizebox{0.9\textwidth}{!}{\includegraphics[52,205][560,592]{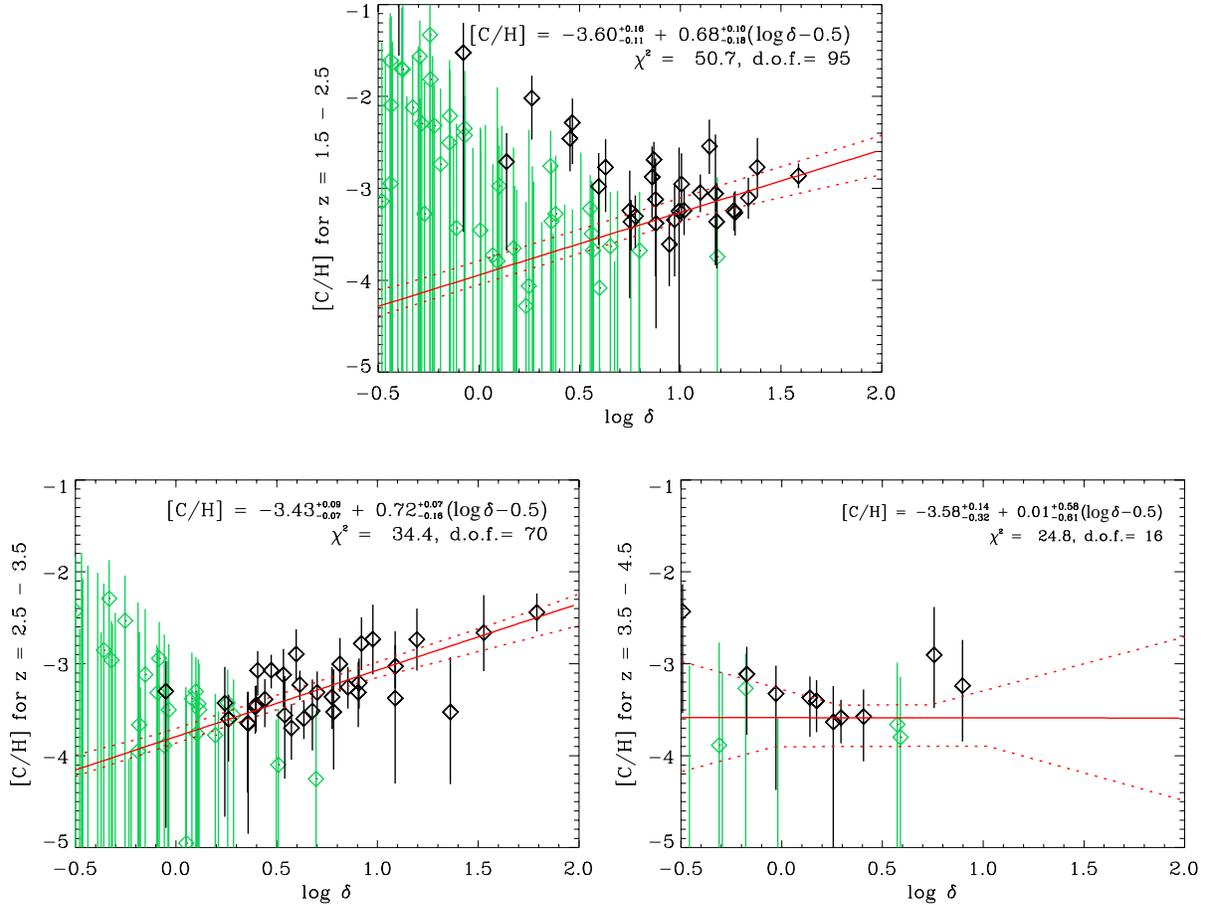}}
\caption{Median metallicity vs.\ overdensity $\delta$ for different
redshift bins. From the top left to the bottom right panels correspond
to the following redshift ranges: $z=1.5-2.5$, $2.5-3.5$, and
$3.5-4.5$. Data points with $1\sigma$ errors were taken from 
Fig.~\protect\ref{fig:individual}. Light-colored data
points have a lower 1$\sigma$ error of $-\infty$. Solid lines indicate
least-squares fits to the data points. Dotted curves indicate the
$1\sigma$ confidence limits, which were computed by bootstrap
resampling the quasars (for the $z=4$ bin, which contains only two
quasars, we resampled the data points).
\label{fig:zbins}}
\end{center}
\end{figure*}

Figure~\ref{fig:zbins} shows the median metallicity as a function of
overdensity for three different redshift bins of width $\Delta z =
1.0$ centered on $z=2$, 3, and 4. The data points were
taken directly from Figure~\ref{fig:individual}. As in the previous
figure, light-colored data points have 
$1\sigma$ lower limits of $-\infty$. Note that many of the
light-colored error bars correspond to data points that fall below the 
plotted range (some at $-\infty$). Note that the data are not uniformly 
distributed in these redshift bins: the median redshifts of the data
points in the three bins are $z\approx 2.12$, 2.84, and 4.07. Note also
that the last bin contains only two quasars: Q1055+461 and Q2237-061. 

The three redshift bins give similar results, confirming that there is
little evidence for evolution. We do, however, have a very significant
detection of a 
positive gradient of metallicity with overdensity in the $z=2$ and
$z=3$ bins (as seen also in the analysis of Q1422+230 in
\S\ref{sec:method:median}). The median metallicity increases from
$[{\rm C}/{\rm H}] \approx -4$ at $\log \delta=-0.5$ to $[{\rm C}/{\rm
    H}]\approx -3$ at $\log 
\delta=1.0$. The best-fit power law index $\alpha \equiv d[{\rm
    C}/{\rm H}]/d\log\delta$ is $0.68_{-0.18}^{+0.10}$ for $z=2$ and 
$0.72_{-0.16}^{+0.07}$ for $z=3$. 


In addition to measuring $\alpha$ from the combined quasar data, we
can also combine the $\alpha$ values measured in each quasar
individually, which can all be read off from
Figure~\ref{fig:individual}. A linear least-squares fit to the
data gives $\alpha =
0.71_{-0.17}^{+0.10}+0.56_{-0.46}^{+0.31}\times(z-3)$ (where the
errors were again computed by bootstrap resampling the quasars). As
expected, the two methods of measuring the gradient of the metallicity with
density give very similar results.

\begin{figure*}[t]
\begin{center} 
\epsscale{1.05}
\plotone{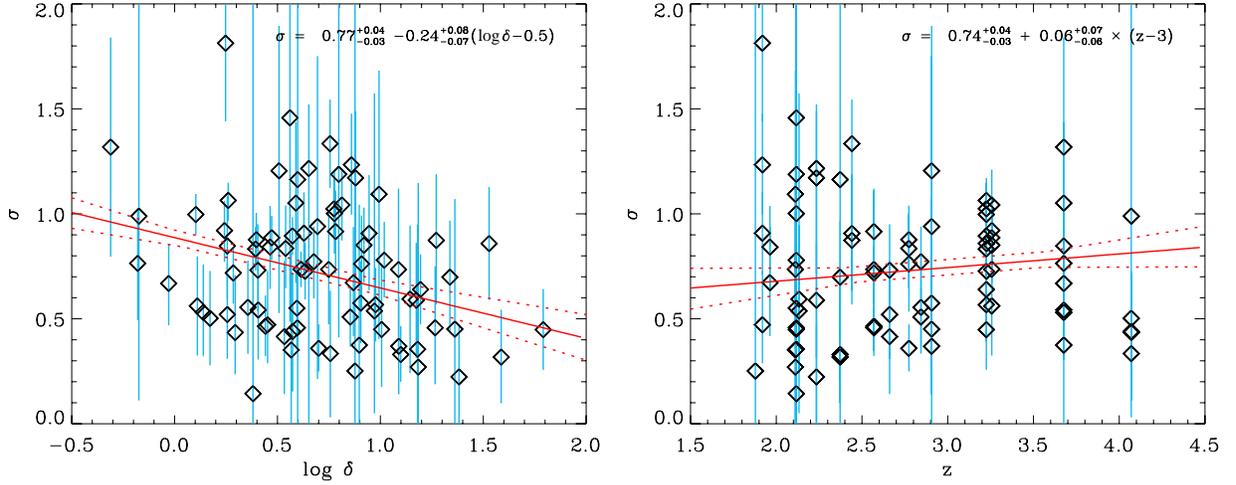}
\caption{Standard deviation of the lognormal
metallicity distribution as a function of overdensity (\emph{left})
and redshift 
(\emph{right}). Data points with $1\sigma$ errors correspond to the HI
bins shown in Fig.~\protect\ref{fig:individual}. The relative sizes of
the errors are reliable, but the absolute sizes are not. Solid lines are
least-squares fits to the data points. Dotted curves indicate the
$1\sigma$ confidence limits, which were computed by bootstrap
resampling the quasars.
\label{fig:sigma}}
\end{center}
\end{figure*}

Let us now turn to the width of the
lognormal fit to the 
metallicity distribution. Figure~\ref{fig:sigma} shows $\sigma([{\rm
    C}/{\rm H}])$ 
versus $\delta$ (\emph{left-hand panel}) and versus $z$
(\emph{right-hand panel}). The data points 
correspond directly to the points connected by the dashed lines in
Figure~\ref{fig:individual}. As discussed in \S\ref{sec:method:scatter},
our method for computing $\sigma([{\rm C}/{\rm H}])$ relies on fitting
different 
percentiles of $\tau_{\rm CIV}$ for a fixed $\tau_{\rm HI}$, and
because the percentiles are correlated, the derived error bars on
$\sigma([{\rm C}/{\rm H}])$ may 
not be robust. However, the relative errors among
the different points should still be reliable and hence the
least-squares fits to the data points --- shown as the solid lines in 
Figure~\ref{fig:sigma} --- should be accurate. The errors on the fits
were computed by bootstrap resampling the quasars and should
therefore also be robust estimates of the statistical errors. Note,
however, that the fact that the measurements of the scatter are mostly
one-sided (the low-metallicity tail is undetected), could result in
somewhat larger systematic errors.

We find a significant ($>2\sigma$) decrease in the scatter with
overdensity:
$\sigma([{\rm C}/{\rm
    H}])=0.76_{-0.04}^{+0.03}-0.24_{-0.07}^{+0.08}(\log\delta-0.5)$. 
There is a hint of evolution in the data: the best-fit scatter
decreases by about $0.14$~dex from 
$z=4$ to 2, but the a constant value is acceptable at about the
$1.2\sigma$ level.

The anticorrelation of $\sigma([{\rm C}/{\rm H}])$ with $\log\delta$
is somewhat 
difficult to see by eye, because the trend is not large
compared with the scatter between the data points. There are, however,
enough points to detect weak anticorrelations. The Spearman rank-order
correlation coefficient is -0.24, indicating that the 
anticorrelation exists at 97\% confidence. Note that this test is
based on a ranking of the data points and does not make use of the
errors.


We can summarize all of our data by fitting functions of redshift and
overdensity to the median metallicity and
the (lognormal) scatter, using all the $\log\delta >
-0.5$ data points shown in Figure~\ref{fig:individual}. The function 
\begin{eqnarray}
[{\rm C}/{\rm H}] &=& -3.47_{-0.06}^{+0.07}
+ 0.08_{-0.10}^{+0.09}\times(z-3) \nonumber \\ 
&& + 0.65_{-0.14}^{+0.10}\times(\log\delta-0.5)
\label{eq:3parsurface}
\end{eqnarray}
provides a very good fit to the data: $\chi^2=114.1$ for 184 degrees of
freedom. As for the projection fits, the reduced $\chi^2$ is low
because we overestimated the errors for the data points with
$\tau_{\rm CIV}\approx \tau_c$, i.e., points for which \CIV\ is barely
detected (see \S\ref{sec:method:median}). Excluding all data
points with $\tau_{\rm HI} < \tau_c$ does not change the fit
(the differences are smaller than the statistical errors), but does
result in a much more reasonable reduced $\chi^2$ of 0.89.

To test whether there is evidence for density-dependent evolution
or, equivalently, a redshift-dependent gradient with density, we have
also fitted a function including a $(z-3)(\log\delta-0.5)$ term to the
data. The best-fit coefficient of the cross term is consistent with
zero at the $\approx
0.8\sigma$ level, and including it improves the quality of the
fit only slightly, 
to $\chi^2=112.5$ for 183 dof. 

Fitting a three-parameter surface to the scatter in the lognormal
distribution yields
\begin{eqnarray}
\sigma([{\rm C}/{\rm H}]) &=& 0.76_{-0.08}^{+0.05} +
0.02_{-0.12}^{+0.08}\times(z-3) \nonumber \\
&& - 0.23_{-0.07}^{+0.09}\times(\log\delta-0.5).
\label{eq:sigma_3parsurface}
\end{eqnarray}
As was the case for the median, using the four-parameter function does
not improve the fit and yields no 
evidence for a nonzero coefficient of the $(z-3)(\log\delta-0.5)$ term.

The best-fit
parameters for all the surface fits are listed together with their 1
and $2\sigma$ errors in Tables~\ref{tbl:3parsurface} and
\ref{tbl:sigma_3parsurface}. 
Note that all these fits are based on data in the range
$\log\delta = -0.5 - 1.8$ ($\log\delta = -0.5 - 0.9$ for $z>3.5$),
$z=1.8 - 4.1$ ($z=2.1-3.3$ for 
$\log\delta>1$) and that extrapolations outside this range of
parameter space could be inaccurate. 


\begin{deluxetable}{lcccc} 
\tablecolumns{5}
\tablewidth{0pc} 
\tablecaption{Fits for $[{\rm C}/{\rm H}]=\alpha + \beta (z-3) + \gamma (\log
\delta - 0.5)$ \label{tbl:3parsurface}} 
\tablehead{\colhead{Model} & 
\colhead{$\alpha_{-1\sigma,\,2\sigma}^{+1\sigma,\,2\sigma}$} &
\colhead{$\beta_{-1\sigma,\,2\sigma}^{+1\sigma,\,2\sigma}$} &
\colhead{$\gamma_{-1\sigma,\,2\sigma}^{+1\sigma,\,2\sigma}$} &
\colhead{$\chi^2,{\rm d.o.f.}$}
}
\startdata
QG & $-3.47_{-0.06,\,0.12}^{+0.07 ,\, 0.13}$ & $+0.08_{-0.10 ,\,0.20
}^{+0.09 ,\,0.19 }$ & $+0.65_{-0.14 ,\,0.27 }^{+0.10 ,\,0.18 }$ &
114.1, 184\\ 
Q & $-2.91_{-0.07,\,0.13 }^{+0.07 ,\,0.13 }$ & $-0.06_{-0.09 ,\,0.19
}^{+0.09 ,\,0.21 }$ & $+0.17_{-0.08 ,\,0.21 }^{+0.08 ,\,0.15 }$ &
113.8, 184\\ 
QGS3.2 & $-3.78_{-0.09 ,\,0.16 }^{+0.14 ,\,0.21 }$ & $-0.12_{-0.12
  ,\,0.22 }^{+0.06 ,\,0.13 }$ & $+0.93_{-0.32 ,\,0.47 }^{+0.18 ,\,0.28
}$ & 124.2, 184 \\ 
QGS & $-4.14_{-0.05 ,\,0.11 }^{+0.06 ,\,0.11 }$ & $+0.54_{-0.07 ,\,0.16
}^{+0.10 ,\,0.21 }$ & $+1.31_{-0.07 ,\,0.18 }^{+0.07 ,\,0.17 }$ &
114.2, 184
\enddata
\end{deluxetable} 

\begin{deluxetable}{lccc} 
\tabletypesize{\tiny}
\tablecolumns{4}
\tablewidth{0pc} 
\tablecaption{Fits for $\sigma =\alpha + \beta (z-3) + \gamma (\log
\delta - 0.5)$ \label{tbl:sigma_3parsurface}} 
\tablehead{\colhead{Model} & 
\colhead{$\alpha_{-1\sigma,\,2\sigma}^{+1\sigma,\,2\sigma}$} &
\colhead{$\beta_{-1\sigma,\,2\sigma}^{+1\sigma,\,2\sigma}$} &
\colhead{$\gamma_{-1\sigma,\,2\sigma}^{+1\sigma,\,2\sigma}$}}
\startdata
QG & $+0.76_{-0.08 ,\,0.16 }^{+0.05 ,\,0.09 }$ & $+0.02_{-0.12 ,\,0.22 }^{+0.08 ,\,0.17 }$ & $-0.23_{-0.07 ,\,0.15 }^{+0.09 ,\,0.21 }$ \\
Q & As for QG \\
QGS3.2 & As for QG \\
QGS & As for QG
\enddata
\end{deluxetable}


The surface fits confirm the picture suggested by the
projection plots. There is very little room for evolution, but there
is strong evidence for both an increase in the  
median metallicity and a decrease in the lognormal scatter with
overdensity. 

\subsection{Varying the UV background}
\label{sec:varyinguv}
All ionization corrections were computed as discussed in
\S\ref{sec:ionizcorr}, using model QG (Haardt \& Madau 2001) for the
UV/X-ray background from galaxies and quasars (see
\S\ref{sec:uvmodels}), rescaled to match the evolution of the mean
\lya\ absorption as measured from our sample of observations (see
\S\ref{sec:taueff}). Since 
the spectral shape of the UV background is not well constrained, it is
important to investigate what effect changes in the spectrum have on
the derived metallicities.

In particular, since photons with energies greater
than 4~ryd can ionize \CIV\ but not \CIII, the ionization corrections
can be sensitive to the break at the \HeII\ Lyman limit, which is much
greater if the background is dominated by stars as opposed to quasars,
and can also be very large if \HeII\ is not fully reionized. 
As Figure~\ref{fig:ionizcorr} shows, for gas with densities $n_{\rm H} \la
10^{-5}~\cm^{-3}$ the $z=3$ ionization correction is much larger for
model Q --- which includes only contributions from quasars --- than
for model QG. Thus, as expected, 
for low gas densities the derived metallicity is sensitive to the
hardness of the UV background radiation, with harder spectra yielding
higher metallicities. 

To investigate how sensitive the metallicities are to changes in the
UV radiation field, we have recomputed the surface fit
$[{\rm C}/{\rm H}](\delta,z)$ (eq.\ [\ref{eq:3parsurface}]) using
the two additional models for the UV 
background radiation described in \S\ref{sec:uvmodels}. Model Q
(Haardt \& Madau 2001) is an updated version of the Haardt \& Madau
(1996) model for the background radiation from quasars only. Model QGS
is identical to model QG, except that above 4~ryd the flux has been
decreased by a factor of 10. The results are listed in
Table~\ref{tbl:3parsurface}.

Model QGS is much softer than model QG and therefore gives lower
metallicities and a much stronger gradient with overdensity.  In
addition, the metallicity is everywhere {\em decreasing} with
time, clearly an unphysical\footnote{Note that for fixed high
overdensities the metallicity could in principle decrease with time
because of infall of metal-poor gas (see \S\ref{sec:summary}).}
result. We get the same unphysical trend if we decrease the flux above
4~ryd by only a factor of 3 instead of a factor of 10 as in QGS. Although
using model  
QGS at all redshifts results in a metallicity evolution that does not
make physical sense, QGS could be a reasonable model for the UV
background at $z\sim 4$ if \HeII\ was reionized late. We have
therefore tested model QGS3.2, which is identical to model QG for
$z<3.2$ and identical to model QGS for $z>3.2$. This transition
redshift was chosen because various lines of evidence suggest that
\HeII\ reionization was incomplete at earlier times (e.g., Songaila
1998; Schaye et al.\ 2000b; Heap et al.\ 2000; Theuns et al.\ 
2002a). 

\begin{figure*}[t]
\epsscale{1.15}
\plotone{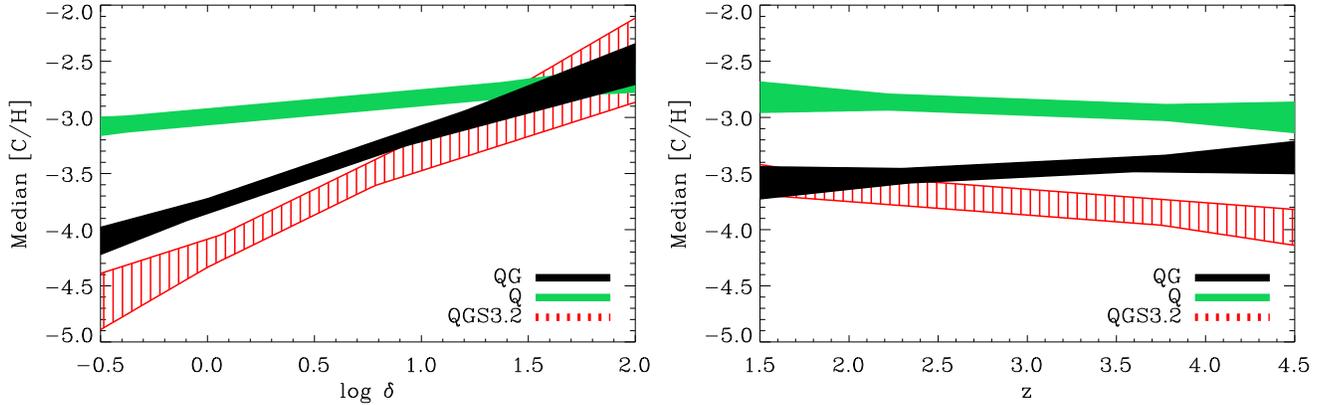}
\caption{Visualization of the fits to the median metallicity as a
function of overdensity and redshift for $z=3$ (\emph{left-hand panel}) and
$\log\delta = 0.5$ (\emph{right-hand panel}) for our fiducial UV
background model 
QG, as well as for models Q
and QGS3.2. The shaded regions enclose the $1\sigma$
confidence contours for the three-parameter surface fits listed in
Table~\protect\ref{tbl:3parsurface}. Harder UV backgrounds yield
higher metallicities for $\delta < 10$. Strong evolution is ruled out
for all models.
\label{fig:uvmodels}}
\end{figure*}

Figure~\ref{fig:uvmodels} shows the two projections of the three-parameter
surface fits (listed in Table~\ref{tbl:3parsurface}) for models QG, Q, and
QGS3.2. The left-hand panel shows the median metallicity versus $\log\delta$
for $z=3$, while the right-hand panel plots the median
metallicity versus $z$ for $\log\delta = 0.5$.
Although these plots cannot shed light on density-dependent
evolution, they do  
illustrate the key differences between the various UV backgrounds. 


As for model QG, model Q gives no evidence for evolution. However, for
model Q the results do differ from our fiducial model in 
other ways. Model Q yields much higher metallicities for $\log\delta <
1$ and gives a much smaller gradient of the
median metallicity with density. In the next section we will show that
this small gradient, combined with the observed decrease in the
scatter with density, results in a \emph{mean} metallicity that
decreases significantly with increasing density, probably an
unphysical result.  Note that model Q also does not fare as well as
model QG in fitting the observed \CIII/\CIV\ ratios (see
\S\ref{sec:ciii}).

Model QGS3.2 yields a significantly lower metallicity for $z>3.2$ than
our fiducial QG model. As a result, there is evidence for (weak)
evolution ($d[{\rm C}/{\rm H}]/dz = -0.12_{-0.12}^{+0.06}$). Because
many of our low-density 
detections have high redshifts, the gradient with overdensity is even
stronger than for model QG. Since it is only the high-redshift data
that are responsible for the larger gradient, there is evidence
for density-dependent evolution,
including a cross term in the fit improves it by $\Delta\chi^2= 9$
relative to the three-parameter fit.

Although uncertainties in the UV background are important for
estimates of the median metallicity, this is not the case for our
measurements of the lognormal scatter, which are independent of the
assumed UV background (see \s\ref{sec:method:scatter}). Our finding
that the metallicity distribution is lognormal 
and our measurements of the scatter in this distribution are therefore
much less model-dependent than our measurements of the median
metallicity. They are, however, still not completely model-independent
because we have assumed that the UV radiation is constant for fixed
$\tau_{\rm HI}$. If there were significant scatter in the UV radiation
field corresponding to a fixed overdensity, then this would generate
scatter in the $\tau_{\rm CIV}$ distribution, leading us to
overestimate the width of the metallicity distribution.

\section{Discussion}
\label{sec:discussion}

\subsection{Summary of measurements}
\label{sec:summary}
We have measured the distribution of carbon as a function of
overdensity and redshift using data in the range $\log\delta = -0.5 -
1.8$ and $z=1.8-4.1$. For a fixed overdensity and redshift the metallicity
distribution is close to lognormal, at least from about $-0.5$
to $+2\sigma$. We measure a lognormal scatter
$\sigma([{\rm C}/{\rm H}]) = 0.76_{-0.08}^{+0.05}+0.02_{-0.12}^{+0.08}(z-3) -
0.23_{-0.07}^{+0.09}(\log\delta-0.5)$. Thus, we find no evidence for
evolution, but we do find a significant decrease in the scatter with
overdensity. The measurements of the scatter are independent of the
spectral shape of the UV background radiation.

Unlike the scatter, the median metallicity does depend on the
model for the UV background, although it is insensitive to the
spectral shape for $\delta \ga 10$. For our fiducial model QG, which
includes contributions from both galaxies and quasars (see Haardt \&
Madau 2001), we find $[{\rm C}/{\rm H}] =
-3.47_{-0.06}^{+0.07}+0.08_{-0.10}^{+0.09}(z-3)+
0.65_{-0.14}^{+0.10}(\log\delta-0.5)$, i.e., no evidence for
evolution, but a strong gradient with overdensity. 

Harder UV backgrounds yield higher metallicities at low overdensities. Model Q,
which includes only UV radiation from quasars, gives a median
metallicity higher by about 0.6~dex for $\log\delta=0.5$, a much
weaker gradient with overdensity 
($d[{\rm C}/{\rm H}]/d\log\delta = 0.17_{-0.08}^{+0.08}$), but again no
evidence for evolution. As we will discuss 
below, this UV background is probably too hard since it yields a
\emph{mean} metallicity that decreases with overdensity. Note that
measurements of $N_{\rm HeII}/N_{\rm HI}$ also imply that the true UV
background is softer than model Q (Heap et al.\ 2000; Kriss et al.\
2001). Making the UV background too soft also gives unphysical
results: model QGS, which has a 10 times smaller flux above 4~ryd than
model QG,\footnote{Reducing the flux above 
4 ryd further has very little effect because the CIV ionization rate
is already small.}
yields a metallicity that strongly decreases with time for all
overdensities. We find the same unphysical trend if we decrease the
flux above 4~ryd by a factor of 3 instead of 10. 

Hence, it appears that the metal distribution (median and
scatter) evolves very little from $z=4$ to $z=2$, which suggests
that most of the enrichment of the low-density IGM took place at
higher redshifts. There is, however, a caveat: we do
find (weak) evolution in the median metallicity if we make the UV background
much softer at high redshift (or much harder at low redshift). For
example, model QGS3.2, which has a 10 times smaller flux above 4~ryd
for $z>3.2$ (as may be appropriate if \HeII\ had not yet reionized by
that time), gives $d[{\rm C}/{\rm H}]/dz = -0.12_{-0.12}^{+0.06}$ for
$\log\delta = 0.5$ and stronger evolution for lower overdensities.

Note that even if the enrichment of the IGM was completed at $z\gg 4$,
one would still expect some evolution in the gradient of metallicity with
overdensity, unless the initial gradient was zero. Because gravity
tends to increase density contrasts, any initial gradient of metallicity with
overdensity should become weaker with time. In fact, our measurements
favor a small increase in the gradient with redshift: for
model QG the coefficient of the $(z-3)(\log\delta - 0.5)$ term in our
four-parameter surface fit is 
$0.23_{-0.28}^{+0.15}$. Although a detailed 
comparison with models for the enrichment of the IGM is beyond the
scope of this paper, we note that the simulations we have performed show
that the expected ``passive'' evolution in the gradient of the
metallicity with overdensity from $z=4$ to $z=2$ is sufficiently small
to be easily compatible with our measurements.

\subsection{Mean metallicities}

Having measured the median and the scatter of the lognormal
distribution, we can also compute the mean (see eq.\
[\ref{eq:meanz}]).  For our fiducial model we find
$\log\left<Z\right>\approx -2.8$ at $\log\delta = 0.5$, where $Z$ is
short for $10^{[{\rm C}/{\rm H}]}$, with no evidence for 
evolution. At the 1$\sigma$ ($2\sigma$) level the data allow an
increase of about 0.1 dex (0.3 dex) per unit redshift. For model Q,
$\log\left<Z \right >$ is $\sim 0.6$\,dex higher 
at this overdensity and, as for model QG, there is no evidence for
evolution. Note that the limits on the allowed evolution of the mean
metallicity are weaker than for the median metallicity because there
is additional uncertainty in the redshift dependence of the lognormal
scatter. 

For our fiducial model the mean metallicity increases with overdensity
from $\log\left <Z \right > = -2.9$ at $\delta \sim 1$
to -2.3 at 
$\delta\sim 10^2$ for $z=3$. For model Q, however, it decreases from
-2.1 to -2.5 over the same density range. This negative gradient is
significant at the $1\sigma$ (but not $2\sigma$) level and may imply
that model Q is 
unphysical since models for the enrichment of the IGM generically predict
positive metallicity gradients with overdensity (e.g., Cen \&
Ostriker 1999; Aguirre et al.\ 2001a, 2001b).

\subsection{Global metallicities}
\label{sec:global}
By combining our measurements of the carbon distribution (equations
[\ref{eq:3parsurface}] and [\ref{eq:sigma_3parsurface}]) with the
mass-weighted probability density distribution for the gas density
obtained from our 
hydrodynamical simulation, we can compute the contribution of the
carbon that we see to the global, mass-weighted mean metallicity. For
$z=3$ and $\log\delta = -0.5 - 2.0$ we find a cosmic carbon
abundance $[{\rm C}/{\rm H}] = -2.80 \pm 0.13$, with no
evidence for evolution. Extrapolating our measurements to the full
density range of the simulation, we find $[{\rm C}/{\rm H}] = -2.56 \pm
0.16$ (the higher
overdensities being responsible for the difference). For model Q all
values are about 0.5~dex higher.

Relative to the critical density, our measurement of the contribution
of the forest to the global carbon abundance corresponds to a
global carbon density of 
\begin{eqnarray}
\Omega_{\rm C} &=& 10^{[{\rm C}/{\rm H}]+({\rm C/H})_\odot} m_{\rm C}
{\bar{n}_{\rm H} \over \rho_c},\\
&\approx& 2.3\times10^{-7} 10^{[{\rm C}/{\rm H}]+2.8}\left ({\Omega_b
\over 0.045}\right ),
\end{eqnarray}
where $m_{\rm C}$ is the atomic mass of carbon, $\bar{n}_{\rm H}$ is
the total, comoving number density of hydrogen, and $\rho_c$ is the
critical mass density at redshift zero.

\subsection{Filling factors}

\begin{figure} 
\epsscale{1.2}
\plotone{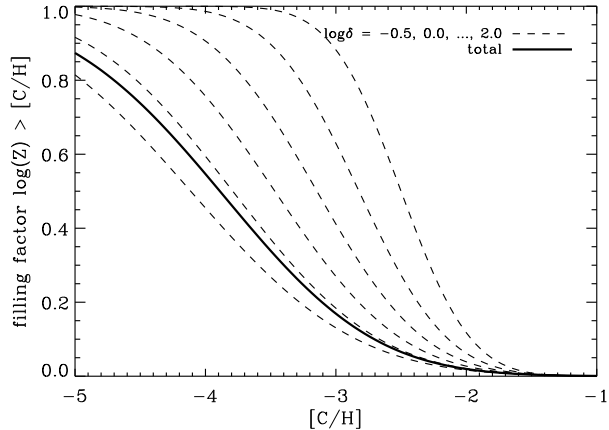}
\caption{Filling factor of
gas with a carbon abundance greater than $[{\rm C}/{\rm H}]$ plotted
as a function of $[{\rm C}/{\rm H}]$ for $z=3$ and UV background
QG. Dashed curves indicate the filling factors at overdensities of,
from left to right, $\log\delta = -0.5, 0.0, \ldots, 2.0$. The thick
solid curve indicates the total filling factor, i.e., the volume
fraction.
\label{fig:fillfactor}}
\end{figure}

The dashed curves in Figure~\ref{fig:fillfactor} show the fraction of
gas with a metallicity greater than $[{\rm C}/{\rm H}]$ as a function
of $[{\rm C}/{\rm H}]$ for various overdensities. All curves are for
model QG and $z=3$. The filling factors were computed using the
surface fits to the median metallicity and lognormal scatter as a
function of overdensity and redshift (equations [\ref{eq:3parsurface}]
and [\ref{eq:sigma_3parsurface}]). Note that for metallicities below
the median (i.e., filling factor 0.5), we detect \CIV\ only for
significantly overdense gas. Thus, the lognormal shape of the curves
in the upper half of the plot is a result of extrapolating the shape
measured for the bottom half. The thick, solid curve
shows the total filling factor, i.e., the volume fraction. It was
computed by combining our metallicity measurements with the volume
weighted probability density distribution for the gas density in our
hydrodynamical simulation at $z=3$ (using a physical smoothing scale of
75~kpc). The filling factors for $z=2$ and $z=4$ (not plotted) fall
nearly on top of this curve. 

Figure~\ref{fig:fillfactor} summarizes many of our results. The
intersection of the dashed curves with the filling factor 0.5 gives
the median metallicity at the various overdensities.  The fact that
the dashed curves become steeper going from low to high overdensities
(\emph{left to right}) reflects our finding that the scatter decreases with
overdensity. One consequence of this is that high-metallicity gas
($[{\rm C}/{\rm H}]\gg -2$) is rare\footnote{There could be somewhat
more high metallicity gas than suggested by this figure because we do
not have enough pixels to probe the metallicity beyond $+2\sigma$
(c.f.\ \S\ref{sec:method:scatter}).}. For all metallicities the filling
factor increases with overdensity. For model Q, which gives a median
metallicity that is nearly independent of overdensity, this is in fact
not the case for high metallicities. Since a filling factor that
decreases with overdensity seems implausible, this again
suggests that the UV radiation is too hard in model Q.

Essentially all collapsed gas clouds ($\delta \ga 10^2$) have a
metallicity greater than $5\times 10^{-4}~Z_\odot$. Metallicities of 1
and 5 times $10^{-4}$ have been claimed to be the 
maximum possible metallicities for the formation of supermassive stars
(Bromm et al.\ 2001; Schneider et al.\ 2002). Our results
therefore indicate that by $z=4$ the formation of such stars had
already come to an end. Some caution is
appropriate, however, because the top parts of the dashed curves are
based on extrapolations since we generally detect \CIV\ only if the
metallicity is similar or greater than the median value.


\subsection{Do we see all of the carbon?}

Throughout this paper we assumed that the \CIV\ gas has a temperature
$T\ll 10^5~\K$, so that collisional ionization is unimportant. Our
measurements of the \CIII/\CIV\ ratio strongly support this assumption
for $\log\delta \ga 1$, but cannot rule out high temperatures for
lower density gas. However, the fact that we detect \CIV\ generally
even for \HI\ optical depths as low as $\tau_{\rm HI} \sim 1$ implies
that hot gas would need to have a substantial filling factor to
invalidate the assumption that the gas that we observe is predominantly
photoionized. Note that Carswell et al.\ (2002) and Bergeron et al.\
(2002) could rule out collisional ionization for many of the \OVI\ absorbers
associated with low column density \HI\ systems on the basis of their line
widths. 

Gas with a temperature $T\gg 10^5~\K$ is too highly ionized to cause
detectable absorption in \CIV\ 
or \HI\ and will therefore not introduce errors in our measurements of
the metallicity of the warm, photoionized component. It is, however,
important to note that our analysis cannot reveal carbon hidden in
hot, X-ray--emitting gas, which may contain a large fraction of the
intergalactic metals (Theuns et al.\ 2002b). The same is true for 
cold, self-shielded gas for which most of the carbon is neutral or
singly ionized. Comparisons to models in
which a significant fraction of intergalactic carbon is in gas with $T\gg
10^5~\K$ or $T\ll 10^4~\K$ must therefore be made with care. 

\subsection{Comparison to previous work}

There have been no previous systematic attempts to measure the metal
distribution in the \lya\ forest as a function of overdensity and
redshift. Previous studies using \CIV\ pixel statistics (Cowie \&
Songaila 1998; Ellison et al.\ 1999, 2000) reported optical depth
measurements consistent with our results. However, these studies used
less sensitive methods, considered only the median $\tau_{\rm CIV}$,
and did not 
attempt to correct for ionization. Published measurements of the
carbon abundance have generally been based on a comparison of Voigt
profile statistics of \CIV\ lines with either a hydrodynamical
simulation assuming a uniform metallicity or simple photoionization
models for individual absorbers.

Both simulations (e.g., Haehnelt et al.\ 1996, Rauch et al.\ 1997;
Hellsten et al.\ 1997; Dav\'e et al.\ 1998) and simple photoionization
models (e.g., Cowie et al.\ 1995; Songaila \& Cowie 1996; Rollinde et
al.\ 2001; Carswell et al.\ 2002) find that a carbon abundance of
$10^{-3}$ to $10^{-2}$ solar 
reproduces the $z\sim3$ observations. Rauch et al.\ (1997), Hellsten
et al.\ (1997), and Dav\'e et al.\ (1998) all find $[{\rm C}/{\rm H}]
= -2.5$ and also report evidence for scatter of about 1.0 dex (Dav\'e
et al.\ 1998 find 0.5 dex). These studies used the relatively
hard Haardt \& Madau (1996) model for the UV background from quasars,
which closely resembles our model Q. For this model we find a median
metallicity at $z=3$ of $[{\rm C}/{\rm H}] =
-2.91_{-0.07}^{+0.07}+0.17_{-0.08}^{+0.08}(\log\delta-0.5)$
 and a lognormal scatter of $\sigma([{\rm C}/{\rm H}]) =
0.76_{-0.08}^{+0.05}-0.23_{-0.07}^{+0.09}(\log\delta-0.5)$, both
nearly independent of redshift.
However, compared with our work, the metallicity measurements of
previous studies were dominated by denser gas: $\log\delta \sim
1.0-1.5$. For these densities we find a scatter of
about 0.6 dex, a median metallicity $[{\rm C}/{\rm H}]\approx -2.8$, and a
\emph{mean} metallicity $\log\left <[{\rm C}/{\rm H}]\right 
>\approx -2.4$ (see eq.\ [\ref{eq:meanz}]; note that it is
unclear whether we should be comparing the mean or the median when
comparing with previous work). Hence, if we use a hard UV background
and focus on the high density gas ($N_{\rm HI}\sim
10^{15}-10^{16}~\cm^{-2}$), then our results for $z=3$ agree with the
metallicity measurements obtained by previous studies. However, our
fiducial UV background (QG) gives a smaller metallicity, particularly
at low overdensities. 

Songaila (2001) measured the \CIV\
column density distribution as a 
function of redshift from $z=1.5$ to$z= 5.5$ and found no evidence for
evolution. It is, however, important not to overinterpret this
finding. To draw conclusions about the evolution of the
metallicity, one must correct for ionization, and this correction is
time-dependent because the universe expands and 
the UV background evolves.

Songaila (2001; see also Pettini et al.\ 2003) also computed
$\Omega_{\rm CIV}$ as a function of 
redshift by summing all of the observed \CIV\ systems. This provides a
strict lower limit to $\Omega_{\rm C}$ because not all carbon is
triply ionized and because some \CIV\ systems may have been lost in
the noise and contamination. Finite sample size is
also important because the integrals of fits to the \CIV\ column density
distribution diverge at high $N_{\rm CIV}$. Songaila found
$\Omega_{\rm CIV} (10^{12}~\cm^{-2} \le N_{\rm CIV} \le
10^{15}~\cm^{-2}) \approx (3 \pm 2)\times 10^{-8}$ at all redshifts.\footnote{
Songaila (2001) measured $\Omega_{\rm CIV} \approx (5 \pm 4) \times
10^{-8}$ for $z=2-4$ assuming $\Omega_m=1$. Since her method is
based on measuring the total CIV column density per unit redshift, this
value scales as $\sqrt{\Omega_m + (1-\Omega_m)/(1+z)^3}$ for a
cosmologically flat universe. For our cosmology ($\Omega_m=0.3$) this
becomes $\Omega_{\rm CIV} \approx (3 \pm 2) \times 10^{-8}$.} Using our
measurements of the carbon 
distribution for model QG and the mass-weighted probability density
distribution for the gas density from our hydrodynamical simulation,
we find $\Omega_{\rm C} \approx (2.3\pm 0.7)\times 10^{-7}$ for $\log\delta =
-0.5 - 2.0$, with no evidence for
evolution (see \S\ref{sec:global}). For model Q the cosmic carbon
abundance is about a factor of 3 higher, again with no evidence
for evolution. Thus, we find that
$\Omega_{\rm C}$ is indeed much higher 
than $\Omega_{\rm CIV}$ measured by Songaila (2001). The large
difference implies that the observed evolution of $\Omega_{\rm
CIV}$ tells us very little about the evolution of $\Omega_{\rm C}$.

Nevertheless, Songaila's measurements of $\Omega_{\rm CIV}$ provide an
important consistency check. Using our 
measurements of the distribution of carbon, the density distribution
of our hydrodynamical simulation, and our ionization
correction as a function of overdensity and redshift, we can compute
$\Omega_{\rm CIV}$. For $-0.5\le \log\delta \le 2.0$ we find
$\Omega_{\rm CIV}/10^{-8}\approx 
1.3 \pm 0.8$, $4\pm 1$, and $8\pm 5$ for $z=2$, 3, and 4, respectively (as
expected, the different UV background models give similar
numbers). The values are nearly the same if we extrapolate to
higher or lower overdensities, which implies that we are seeing most
of the \CIV. 
The quoted errors are estimates of the statistical errors;
i.e., they are based on the errors on the parameters of the surface
fits. Note that the difference between $\Omega_{\rm CIV}$ and
$\Omega_{\rm C}$ decreases from $z=2$ to $z=4$. 
Thus, our results are in good agreement with those of Songaila
(2001). Given that our calculation of $\Omega_{\rm CIV}$ relies on
convolving our measured carbon distribution with the gas distribution
extracted from our simulation, this is a highly nontrivial
consistency check.

\subsection{Uncertainties}
\label{sec:uncertainties}

Our method contains several steps and parameters chosen to
minimize the errors in the metallicity measurements in simulated
spectra. To test the robustness of our method to these choices, we have
redone the surface fits after dividing each quasar spectrum in two,
doubling the \HI\ bin size, omitting our correction to the continuum
fit of the \CIV\ region, excluding all data with $\delta < 1$,
excluding all data with $\tau_{\rm HI} < \tau_c$, 
or excluding all \HI\ bins containing fewer than 50 pixels and/or
contributions from fewer than 10 different chunks (our fiducial values
for these parameters are 25 and 5, doubling them removes many of the
higher \HI\ bins). We find that the coefficients of the various
surface fits always agree within their errors with those obtained
using our standard method. Thus, our results are insensitive to the
details of our method and the surface fits are not determined by the
highest/lowest \HI\ bins.

Although our results are robust with respect to small changes in the
methodology, there could of course still be systematic errors. The
main uncertainty in the median metallicity comes undoubtedly from the
uncertainties in the spectral shape of the UV background. Although 
the lognormal scatter is independent of the mean spectral shape, it is
sensitive to fluctuations. If there were
significant fluctuations in the UV background, then this would
introduce scatter in the $\tau_{\rm CIV}(\tau_{\rm HI})$ relation even
for a uniform metallicity, particularly at low overdensities where the
ionization correction is large. If fluctuations are
important, then our measurements should be interpreted as upper limits
on the scatter in the metallicity.

Even for a uniform UV background there are likely to be some
systematic errors in our ionization corrections. For example, our
measurements of the \CIII/\CIV\ ratio cannot rule out the possibility
that collisional ionization is important for $\delta \ll 10$. At some
level, the ionization corrections must be incorrect, since
redshift-space distortions will result in multiple gas elements
contributing to the absorption at a fixed redshift. 

Fortunately, we can use our hydrodynamical simulation to estimate the
size of these and other systematic uncertainties: we can distribute
metals in our simulation according to our measurements, create
synthetic versions of our observations, and then compare the simulated
and observed optical depth statistics. In 
\S\ref{sec:method:scatter} we showed that our method passed this test
for Q1422+230. It is important to note that
interpreting the results of this test is not completely
straightforward, because there is some ambiguity in the interpretation
of our measurements. For example, we do not know the exact scale on
which our overdensities are smoothed (probably $\sim 50 -100~\kpc$;
see \S\ref{sec:ionizcorr}), and we do not 
know the exact scale(s) over which we are measuring the
scatter. Consequently, it is not completely clear how to distribute the
metals in the simulations.
Having said this, we expect the density smoothing scale to be similar to
the smoothing scale of the forest --- which depends on the gas density
and is not very different from the resolution of our simulation ---
and we find that the precise scale on which the scatter is imposed is
not critical provided it is greater than the smoothing scale of the
forest.

We have carried out the test described above (and in more detail in
\S\ref{sec:method:scatter}) for our entire
sample, distributing the metals in our simulations according to
equations~(\ref{eq:3parsurface}) and (\ref{eq:sigma_3parsurface}),
although we neglected the small, observed decrease in the scatter
with overdensity since we have no practical way of implementing such a
trend in the simulation (we imposed the scatter on a scale of
$1.2h^{-1}\,\mpc$ comoving, i.e., 1/10th of the simulation box
size). Averaging over 10 realizations, we find for the data points
with $\tau_{\rm HI} > \tau_c$ a reduced $\chi^2$ of 1.02, 0.97, and
1.35 for the 50th, 69th, and 84th percentiles, respectively. The total
number of data points included in this estimate is 92 for the 50th
percentile and 95 for the other two percentiles. Hence, we have a
model that can reproduce the observed optical depth statistics. It
appears that any 
systematic errors must be small and that the ambiguities discussed
above are not critical.

We also carried out the following, more stringent test. As before, we
first generate a set of 
synthetic observations using our three-parameter fits to the data
(eqs.\ [\ref{eq:3parsurface}] and
  [\ref{eq:sigma_3parsurface}], but with $d\sigma /d\log\delta =
  0$). Second, we measure the carbon 
abundance from the simulated spectra in the same way as we did for the
data, and we fit the same three-parameter functions to the
results. Finally, we compare the fits to the ones we started
with. The differences between the
fits give us estimates of the systematic errors, or rather upper
limits to the systematic errors given the ambiguities in the smoothing
scale discussed above. Averaging over 10 realizations we find
systematic errors for the coefficients of the median metallicity
(eq.\ [\ref{eq:3parsurface}]) of $\Delta\alpha = -0.10$ ($1.7\sigma$),
$\Delta\beta = -0.20$ ($2.0\sigma$), and $\Delta\gamma=-0.14$
($1.0\sigma$), where the numbers in brackets give the ratios of the
systematic to the statistical errors. The variance between the
different realizations is comparable to the statistical errors.
Note that the signs of the
systematic errors is such that 
we may have slightly overestimated the evolution (i.e., $d[{\rm
C}/{\rm H}]/dz$
too negative) and underestimated the gradient with overdensity. For
the coefficients of the 
scatter (eq.\ [\ref{eq:sigma_3parsurface}]) we find $\Delta\alpha
= -0.11$ ($0.69\sigma$), $\Delta\beta = 0.02$ ($0.25\sigma$), and
$\Delta\gamma=0.01$ ($0.11\sigma$). Thus, the systematic errors are small
for the scatter, but may not be insignificant for the normalization
($\alpha$) and the trend with redshift ($\beta$) of the
median. No systematic errors are greater than twice the corresponding
statisical errors. Although these tests give us confidence that our
results are robust, they do of course not rule out the existence of
other, unknown, sources of systematic errors.

\subsection{Future prospects}

How could our measurements be improved? Higher signal-to-noise ratios
(and better continuum fits, but these are signal-to-noise limited) are
required to measure the median metallicity at lower overdensities and
to verify that the metal distribution remains lognormal below the
median metallicity. Better coverage of the higher-order Lyman lines
would enable us to increase the number of data points for $\log\delta
> 1$.  To improve the constraints on the evolution more data at $z>3$
would be very helpful since only four out of our 19 quasar spectra have a
median redshift greater than 3. For $\delta < 10$ the median
metallicity becomes increasingly sensitive to the softness of the UV
radiation, and it will be necessary to improve the constraints on the
spectral shape of the UV background to tighten the constraints.

\section{Conclusions}
\label{sec:conclusions}

We have measured the statistical correlation between \CIV\ and \HI\
absorption in a sample of 19 high resolution VLT/UVES and Keck/HIRES spectra of
QSOs spanning a redshift range $\sim 1.5-4.5$.  Hydrodynamical
simulations (which can reproduce the observed statistics of the
Ly$\alpha$ forest) link the observed \HI\ absorption to the density
and temperature of the absorbing gas.  Given a model of the ionizing
background radiation field, this allows us to convert the median ratio of
\CIV\ to \HI\ pixel optical depths into a median carbon abundance
$[{\rm C}/{\rm 
H}]$ as a function of density and redshift. In Paper I we tested this
technique extensively on realistic simulated spectra, and here we have
generalized the method to recover the distribution of metals from
the full distribution of $\tau_{\rm CIV}$ as a function of $\tau_{\rm
HI}$. 

Our fiducial model for the UV background is that of Haardt \& Madau
(2001) --- which includes contributions from both galaxies and quasars ---
renormalized using our measurements of the evolution of the mean
\lya\ absorption.

Our primary conclusions are as follows:

\begin{itemize}
\item The best-fit power-law evolution of the effective \lya\ optical
depth in the spectra of 21 quasars is $\log\tau_{\rm eff} =
(-0.44 \pm 0.01) + (3.57 \pm 0.20)\log[(1+z)/4]$. Contamination from
metal lines --- which we removed before making these measurements --- is
significant, particularly at lower redshifts.

\item \CIII\ to \CIV\ optical depth ratios typical of collisionally
ionized \CIV\ ($T\ga 10^5~\K$) are ruled out by the data, at least for
overdensities $\delta \ga 10$. Simulations with $T\sim 10^4~\K$
naturally reproduce the observations. The existence of very hot ($T\gg
10^5~\K$) or very cold ($T \ll 10^4~\K$) carbon cannot be ruled out as it
would not give rise to significant \CIII\ or \CIV\ absorption.

\item For a fixed overdensity $\delta$ and redshift $z$ the
  metallicity distribution is 
close to lognormal, at least from about $-0.5$ to $+2\sigma$.  

\item Over the range $\log\delta = -0.5 - 1.8$ and $z=1.8-4.1$ the
median carbon abundance is well fitted by the following function: $[{\rm
C}/{\rm H}] =  -3.47_{-0.06}^{+0.07} + 0.08_{-0.10}^{+0.09}(z-3) +
0.65_{-0.14}^{+0.10}(\log\delta-0.5)$. The overdensities have
effectively been smoothed on the same scale as the \lya\ forest ($\sim
50 - 100~\kpc$). From tests using simulations we
find that the known systematic errors are at most twice as large as
the quoted statistical errors. There is no evidence for a nonzero
$(z-3)(\log\delta-0.5)$ term. Thus, we find no evidence for
evolution, but strong evidence for a positive gradient with
overdensity. 

\item Harder UV backgrounds give higher carbon abundances for
$\log\delta < 1$ and thus smaller gradients with overdensity. The
(rescaled) Haardt \& Madau (2001) quasar only UV 
background yields a mean metallicity that decreases with overdensity,
which is probably unphysical. UV backgrounds that are much softer than
our fiducial model result in a median (and mean) metallicity that
strongly increases with redshift, which is also unphysical. However, a
UV background that is softer only for $z>3.2$, as may be appropriate
if the reionization of \HeII\ was incomplete before this time, is
allowed and results in significant (but still weak) evolution in the
median metallicity.  

\item We find a nonzero median carbon abundance for underdense gas
($\log\delta=-0.5-0.0$) at the 99.2\% ($2.4\sigma$) confidence level.

\item The lognormal scatter is $\sigma([{\rm C}/{\rm H}]) =
  0.76_{-0.08}^{+0.05} + 
0.02_{-0.12}^{+0.08}(z-3) -
0.23_{-0.07}^{+0.09}(\log\delta-0.5)$. Our measurements of the
scatter are strictly independent of the spectral shape of the UV background
radiation. They are, however, sensitive to fluctuations in the UV
background, which we have assumed to be small. Tests using simulations show
that the (known) systematic errors are smaller than the quoted
statistical errors. Thus, we find no evidence for evolution, but we do
find strong evidence for a modest decrease in the scatter with
overdensity.

\item By combining our measurements of the distribution of metals with
the mass weighted probability density distribution in our
hydrodynamical simulation, we find that the gas between $\log\delta=-0.5$
and 2.0 accounts for a cosmic carbon abundance of $[{\rm C}/{\rm H}]=-2.80
\pm 0.13$, with no evidence for evolution. Expressed in terms of the
critical density, this becomes $\Omega_{\rm C} \approx 2\times 10^{-7}
(\Omega_b/0.045)$. 
We find that $\Omega_{\rm C} \gg \Omega_{\rm CIV}$ which implies that 
ionization corrections are important. Therefore, the finding that
$\Omega_{\rm CIV}$ is roughly constant between $z=2$ and $z=5$
(Songaila 2001) does
not by itself provide interesting constraints on the evolution of
$\Omega_{\rm C}$. 

\end{itemize}

We have presented the first measurements of the distribution of carbon
as a function of overdensity and redshift.  Published models for the
enrichment of the IGM by galactic winds and/or Population III stars predict a
wide variety of metal distributions (e.g., Cen \& Ostriker 1999;
Aguirre et al.\ 2001a, 2001b; Madau, Ferrara, \& Rees 2001; Theuns et
al.\ 2002b; Scannapieco, Ferrara,
\& Madau 2002; Thacker, Scannapieco, \& Davis 2002; Furlanetto \& Loeb
2002; Scannapieco, Schneider, \& Ferrara 2003), which is perhaps not 
surprising given the large uncertainties in the parameters of the
models. Our measurements will provide some much needed constraints and
it will be interesting to see how they narrow the parameter space for
the models. 

More generally, the measurements reported here should provide
a new and stringent set of targets for the next generation of
models for the history of the formation of stars and galaxies, and the
effects of feedback on the intergalactic medium. 

\acknowledgments 
We are grateful to the ESO Archive for their efficient work. Without
their help this work would not have been possible. J.~S. and A.~A.
gratefully acknowledge support from the W.~M.~Keck foundation. T.~T.
thanks PPARC for the award of an Advanced Fellowship. W.~L.~W.~S
acknowledges support from NSF grant AST-0206067. This work has been
conducted with partial support by the Research Training Network ``The
Physics of the Intergalactic Medium'' set up by the European Community
under the contract HPRN-CT2000-00126 RG29185 and by ASI through
contract ARS-98-226. Research conducted in cooperation with Silicon
Graphics/Cray Research utilizing the Origin 2000 supercomputer at
DAMTP, Cambridge.

\appendix
\setcounter{table}{0}
\setcounter{figure}{0}

\section{A. Results from individual quasars}
\label{sec:individual}
Figure~\ref{fig:individual} shows the results for all 19 quasars listed
in Table~\ref{tbl:sample}. For each quasar two panels are plotted. The
left-hand panels show the median recovered \CIV\ optical depth as a
function of the recovered $\tau_{\rm HI}$, while the right-hand panels show
the median metallicity as a function of overdensity.\footnote{Recall
that all overdensities have effectively been smoothed on the same
scale as the \lya\ forest spectra. As discussed in
\S\ref{sec:ionizcorr}, the smoothing is due to a combination of
thermal broadening and Jeans smoothing and varies from $\sim 30$ to
$100$ kpc (physical) depending on density and redshift.} The dashed
lines in the right-hand panels indicate 
the $\pm 1\sigma$ scatter measured from lognormal fits to the
distribution of \CIV\ pixel optical depths. Our methods for measuring
the median and variance of the metallicity distribution are explained
in \S\ref{sec:method}.

To facilitate future comparisons with models for the distribution of
carbon, we have tabulated the optical
depths for several computed percentiles in Table~\ref{tbl:recod}.  


\begin{figure*}
\begin{center}
\epsscale{0.8}
\includegraphics{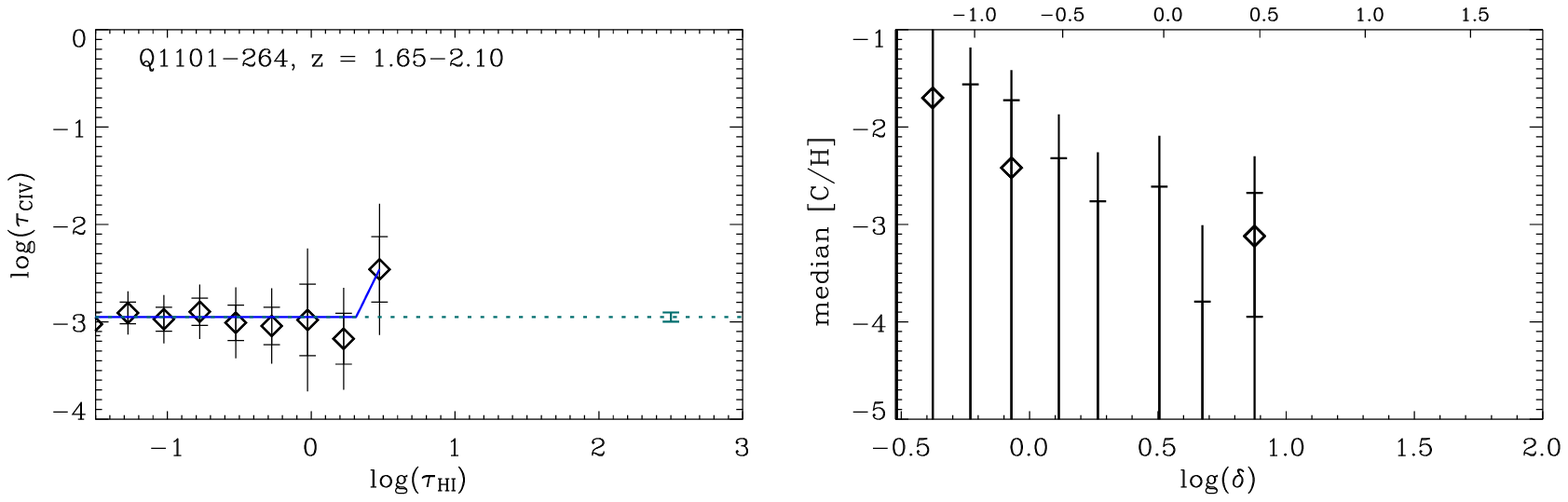}\\
\includegraphics{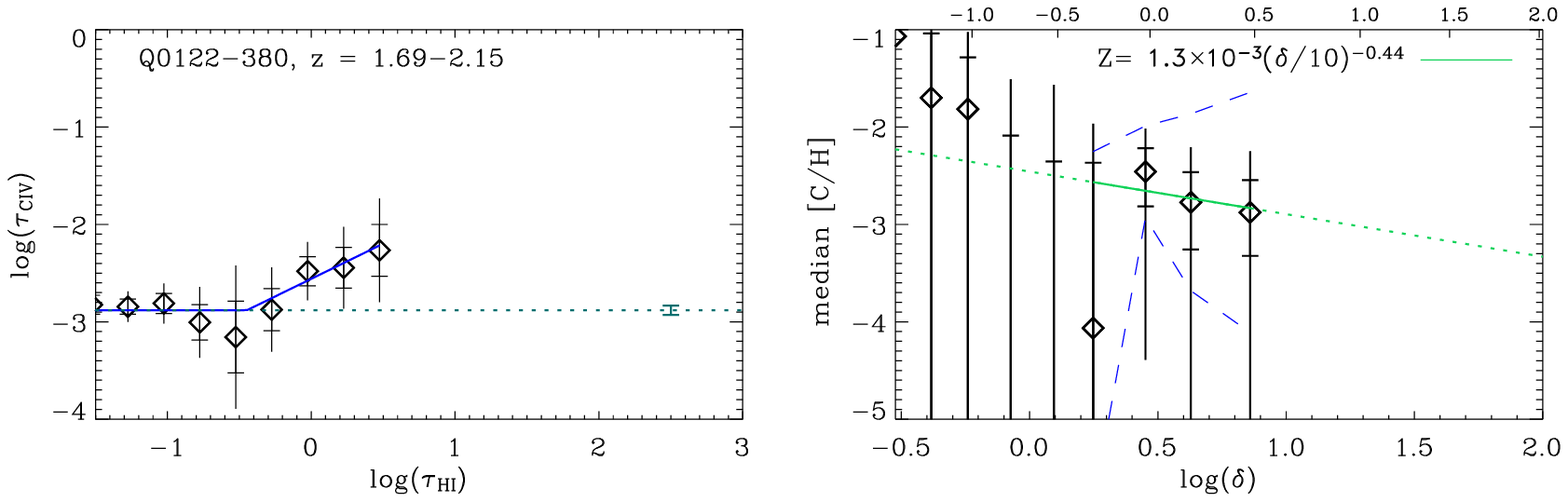}\\
\includegraphics{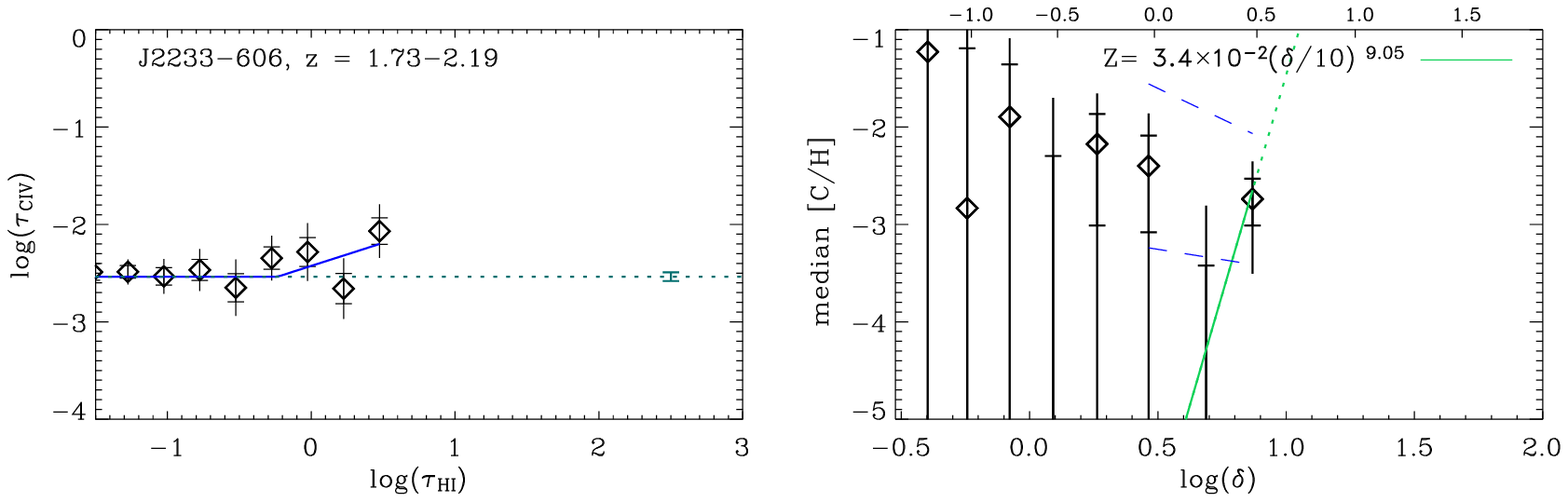}\\
\includegraphics{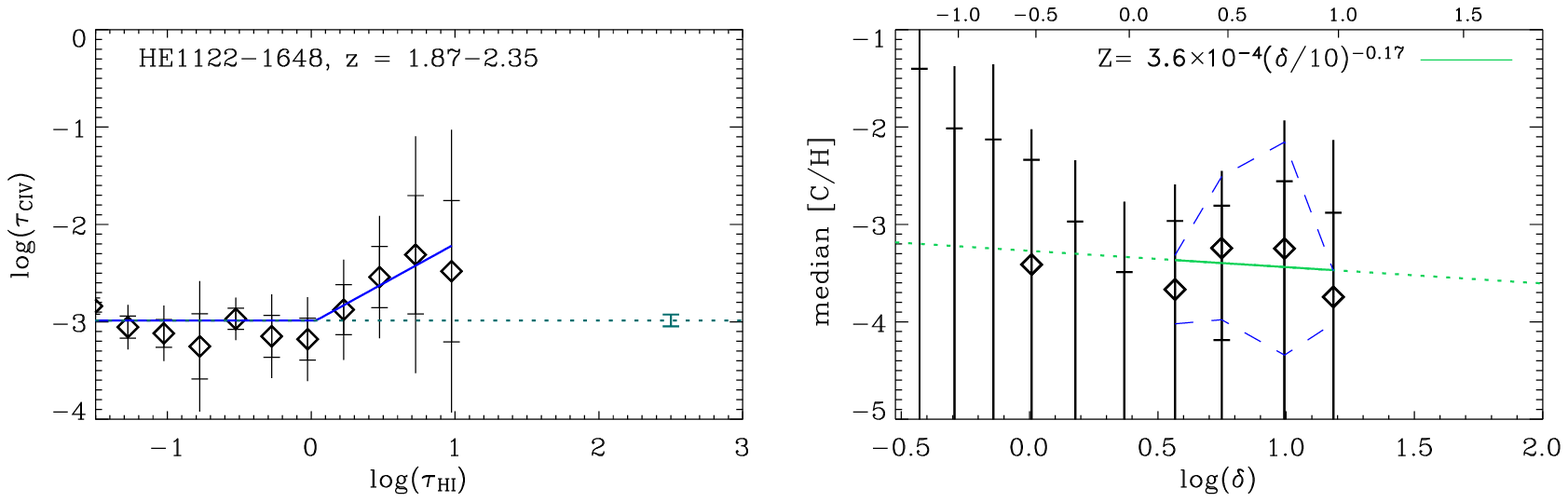}
\caption{Results for individual quasars. For each
quasar two panels are plotted. \emph{Left:} 
Median recovered CIV optical depth as a 
function of the recovered HI optical depth. The horizontal dotted line
indicates $\tau_{\rm min}$; the $1\sigma$ error in this level is indicated on
the right-hand side of the panel. The solid line is the best-fit
power law, eq.\ (\protect\ref{eq:taufit}). \emph{Right:} Median
carbon abundance $[{\rm C}/{\rm H}]$ as a function of log
overdensity. All metallicities were computed as described in
\S\protect\ref{sec:method:median} using the QG UV background.
The best-fit power law to the
$\tau_{\rm HI}>\tau_c$ data points is shown as the solid line
(dotted where extrapolated). The dashed lines show the $\pm 1\sigma$
lognormal scatter as measured from the full distribution of pixel
optical depths (see \S\protect\ref{sec:method:scatter}).
All data points are shown with both 1 and
$2\,\sigma$ error bars.
\label{fig:individual}}
\end{center}
\end{figure*}

\begin{figure*}
\begin{center}
\epsscale{0.8}
\includegraphics{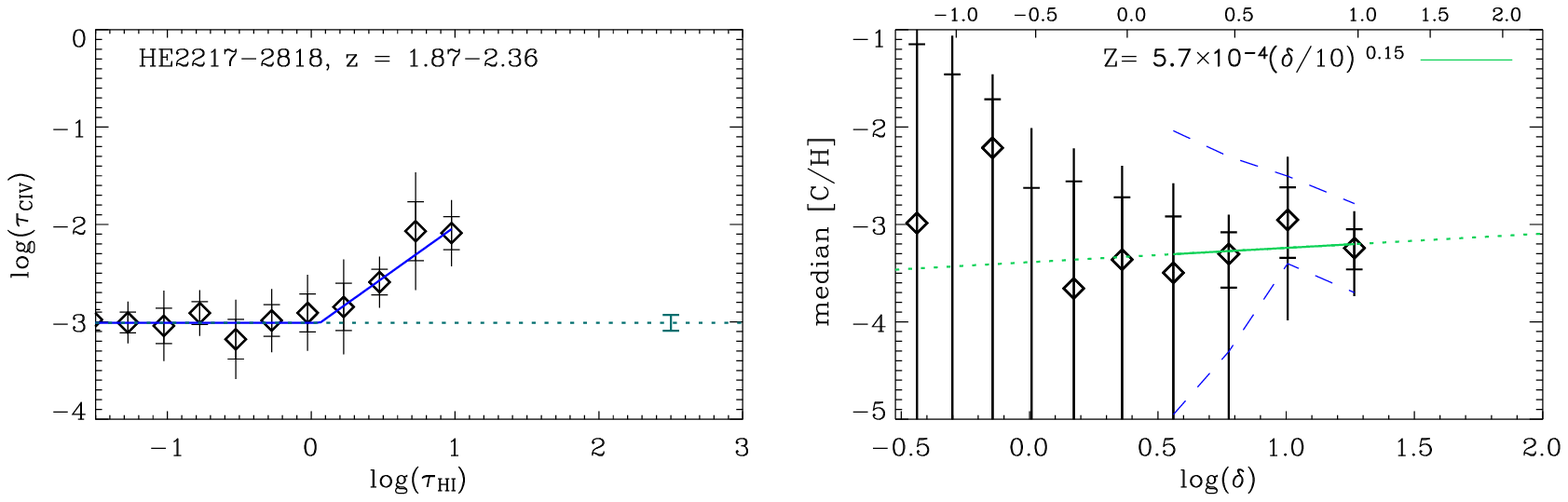}\\
\includegraphics{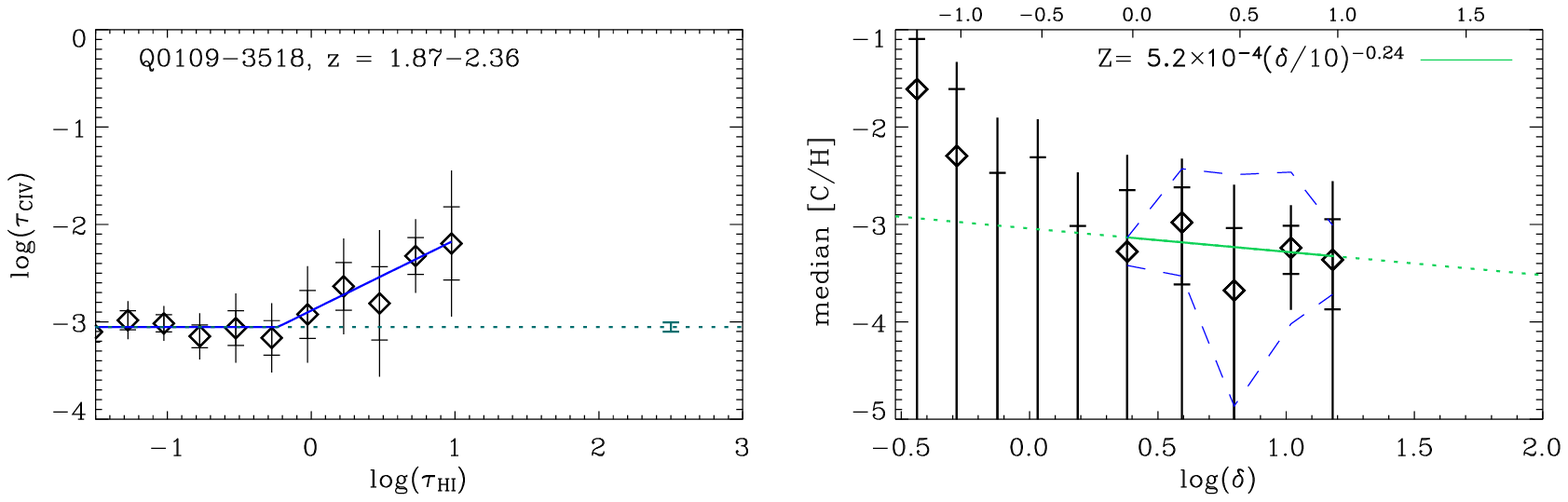}\\
\includegraphics{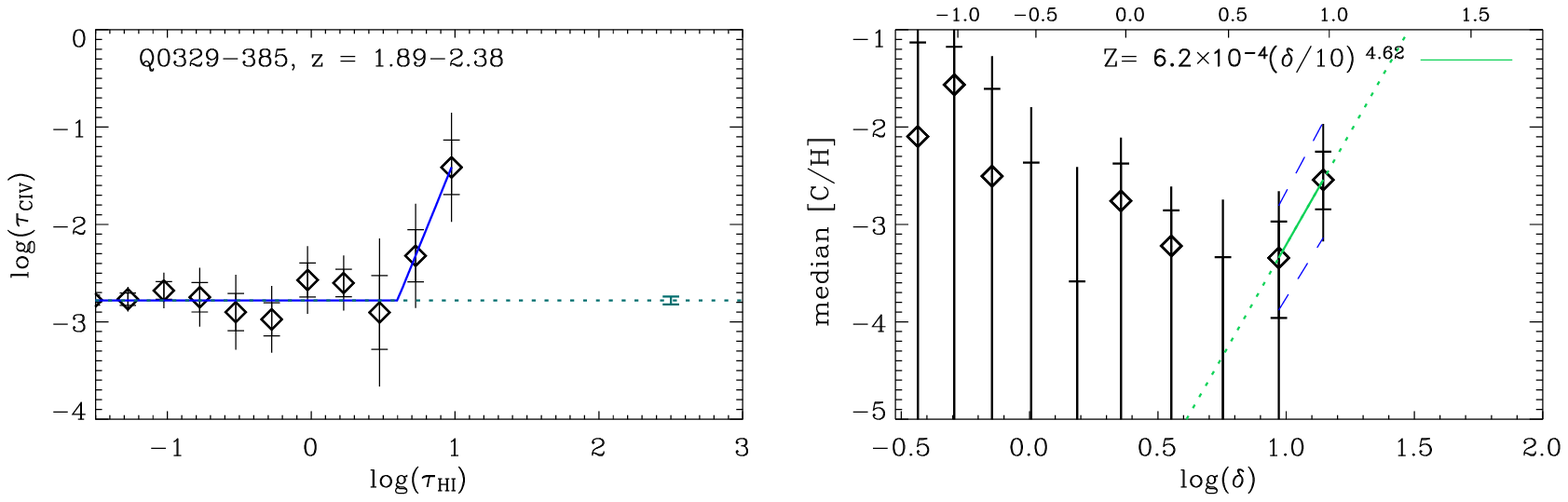}\\
\includegraphics{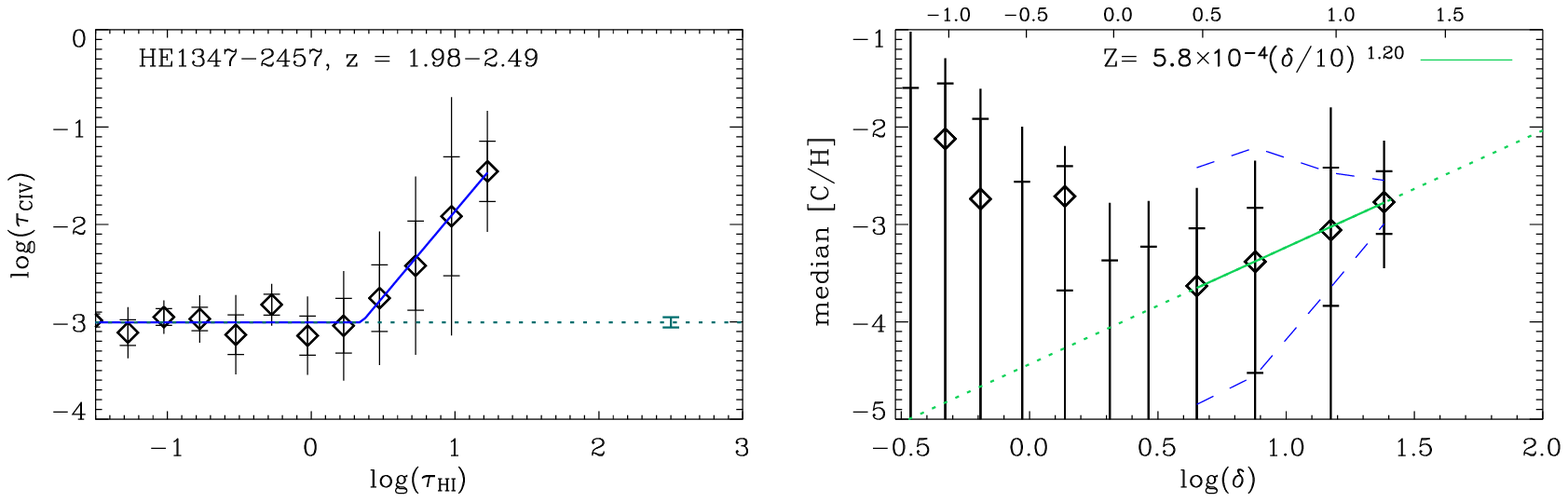}
\addtocounter{figure}{-1}
\figcaption[]{Continued.}
\end{center}
\end{figure*}

\begin{figure*}
\begin{center}
\includegraphics{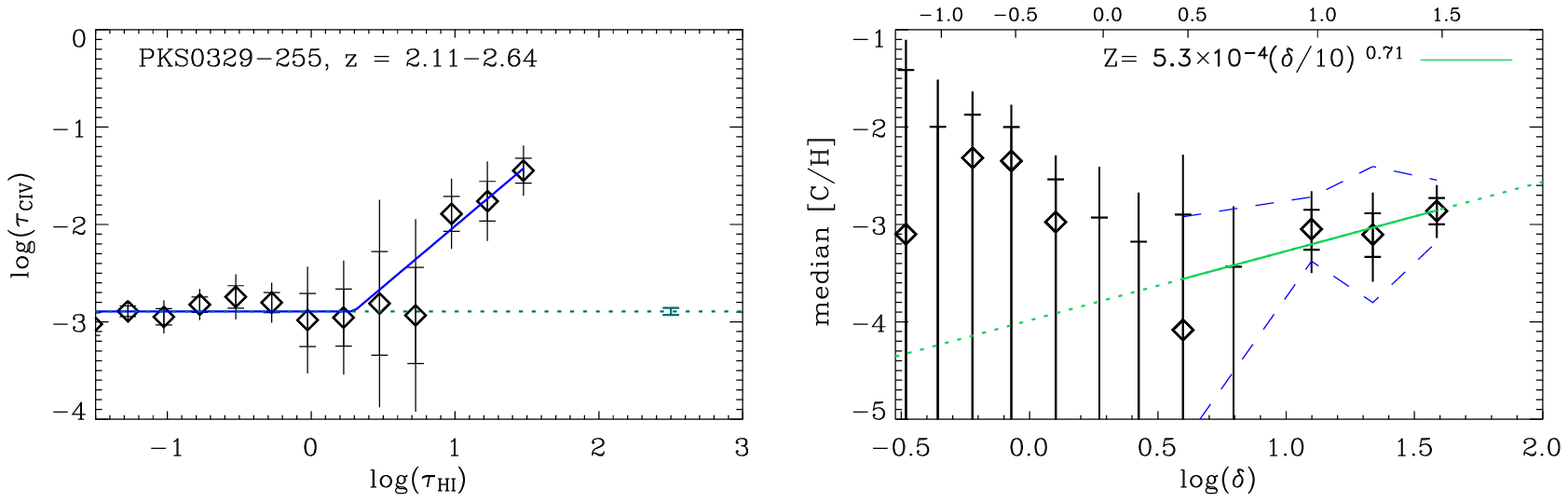}\\
\includegraphics{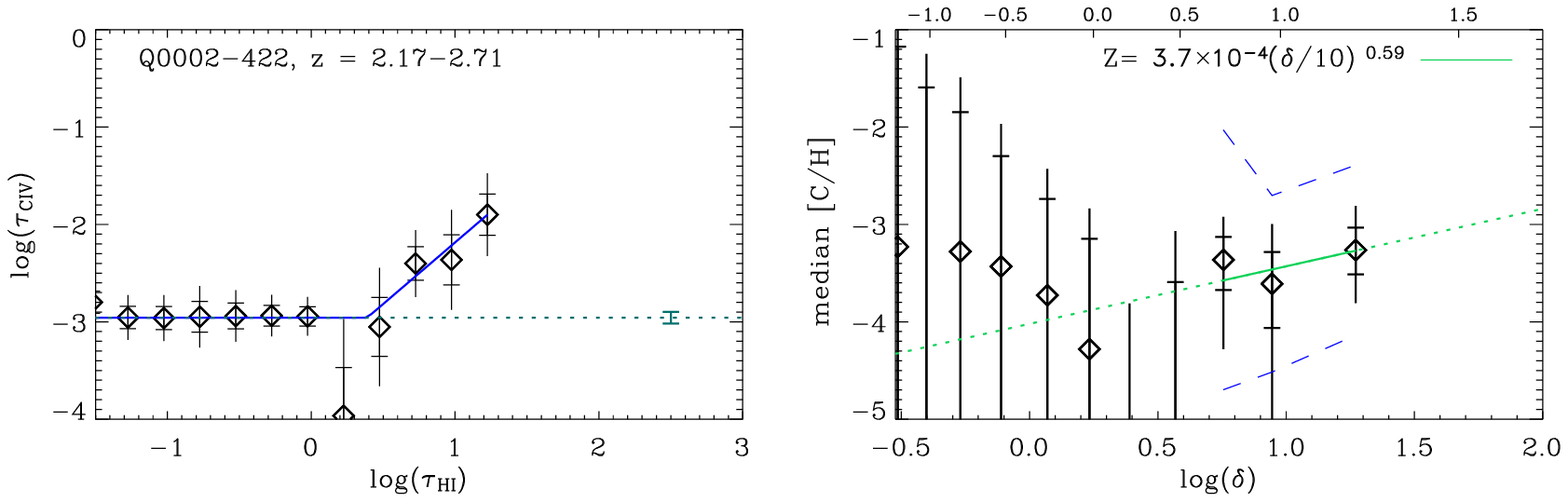}\\
\includegraphics{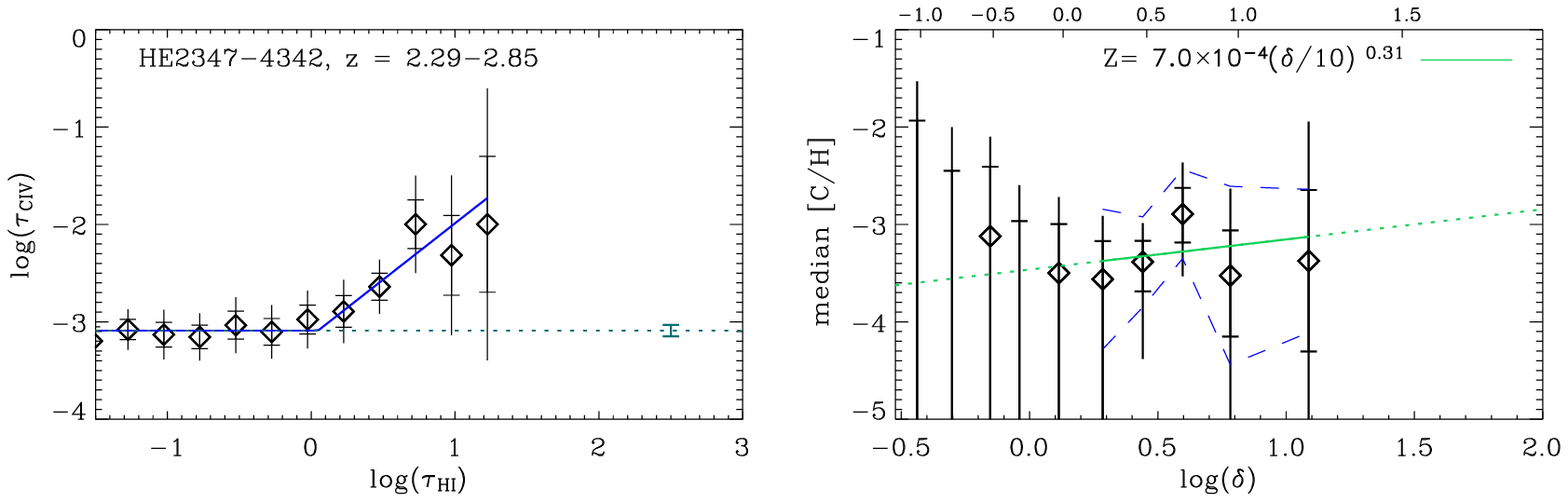}
\includegraphics{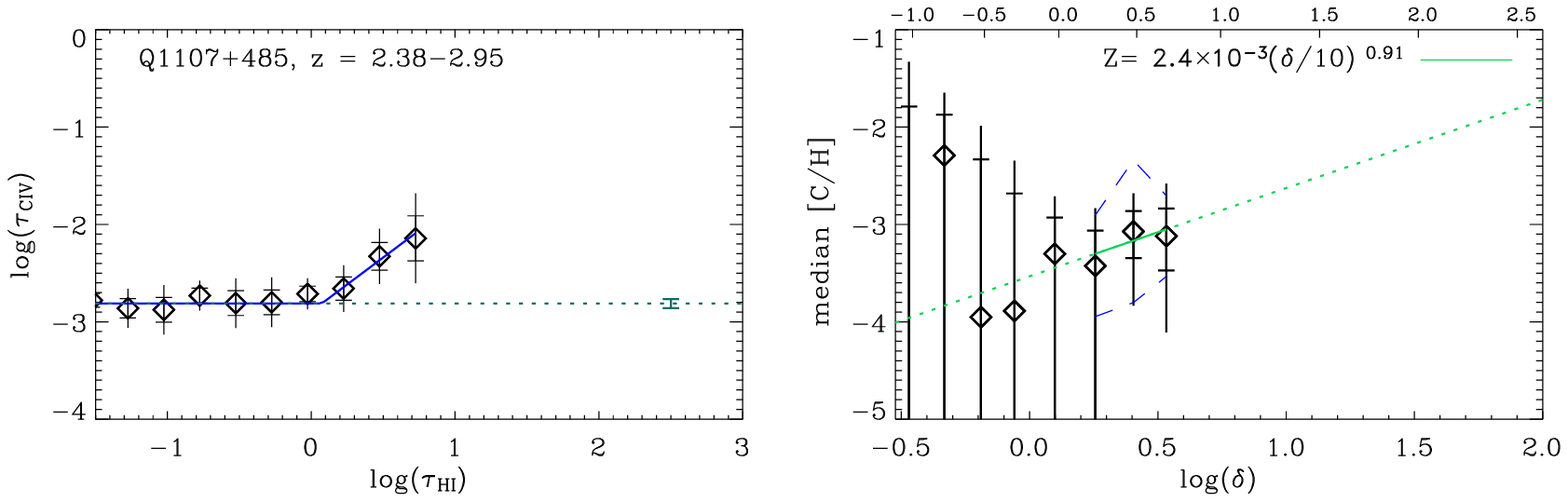}\\
\addtocounter{figure}{-1}
\figcaption[]{Continued.}
\end{center}
\end{figure*}

\begin{figure*}
\begin{center}
\includegraphics{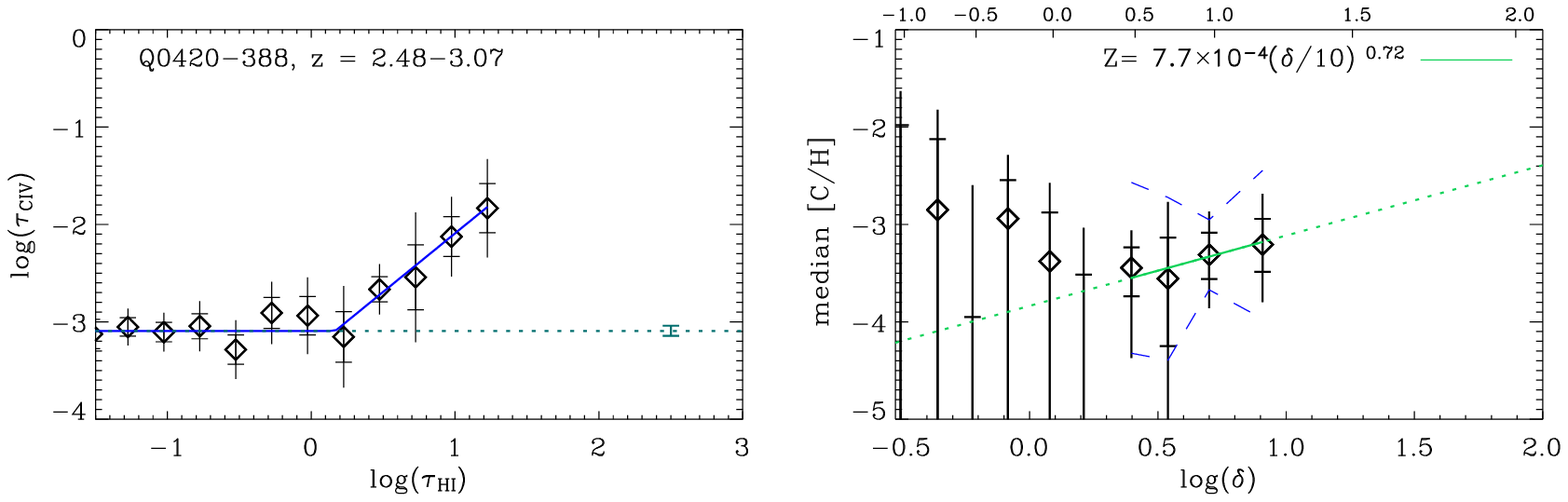}\\
\includegraphics{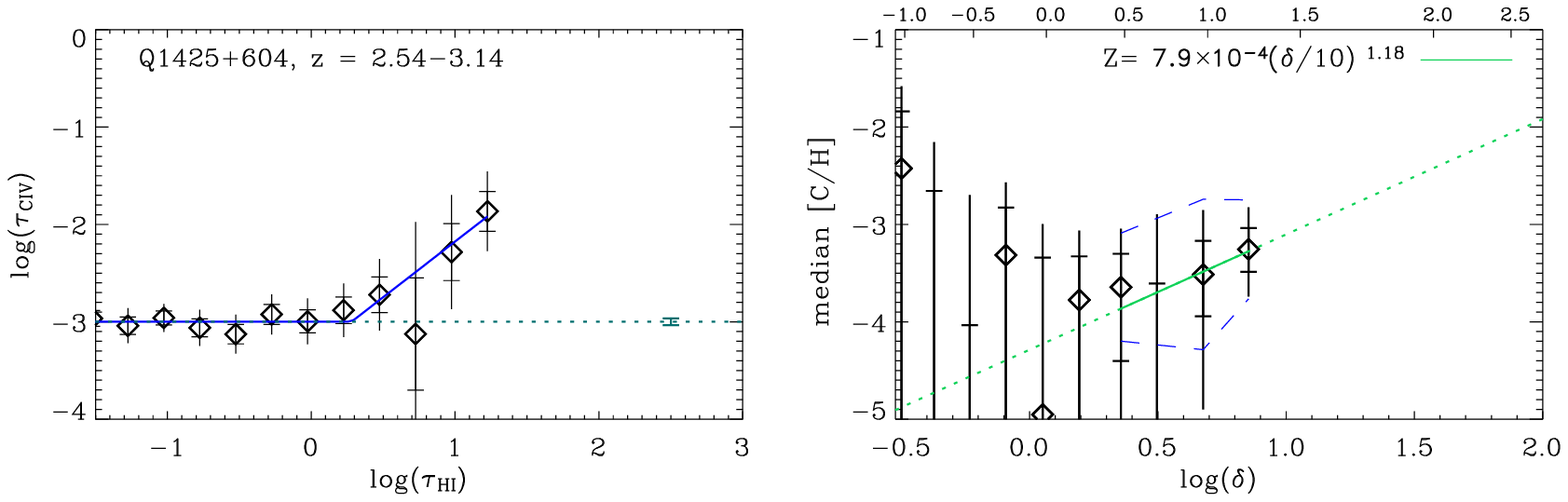}\\
\includegraphics{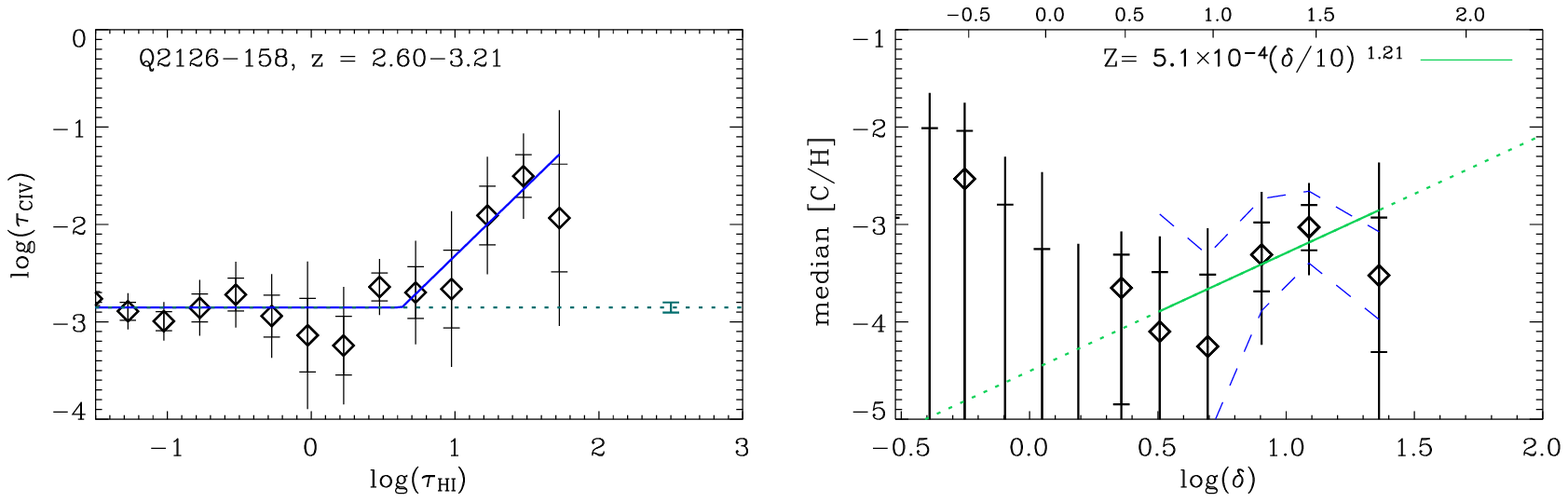}
\includegraphics{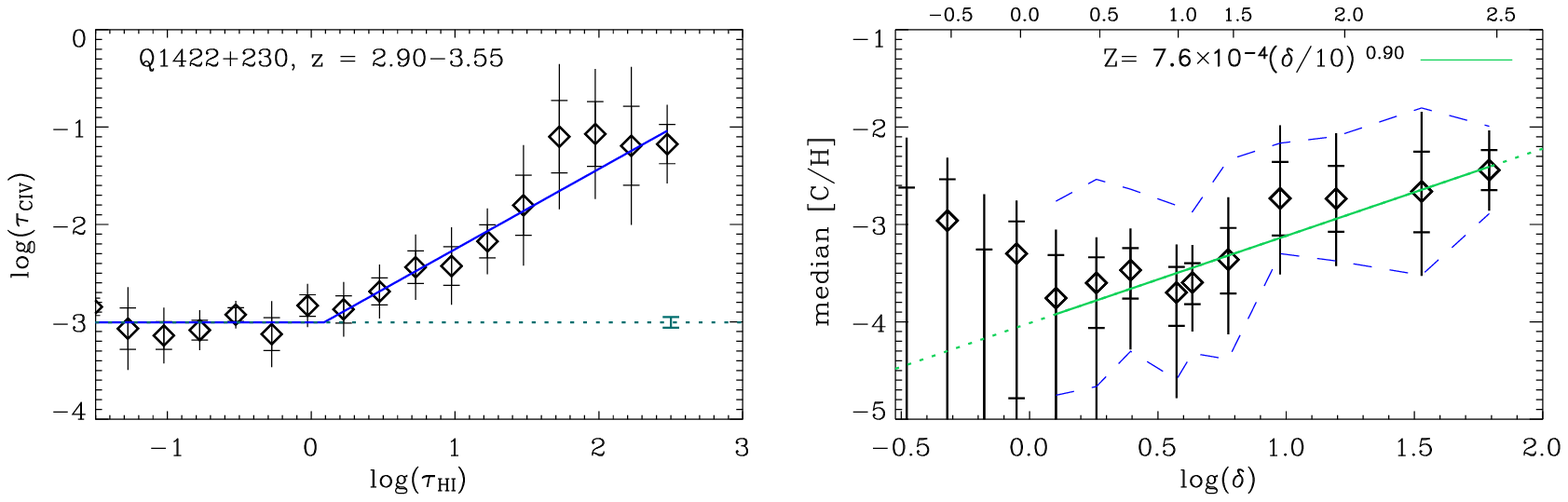}\\
\addtocounter{figure}{-1}
\figcaption[]{Continued.}
\end{center}
\end{figure*}

\begin{figure*}
\begin{center}
\includegraphics{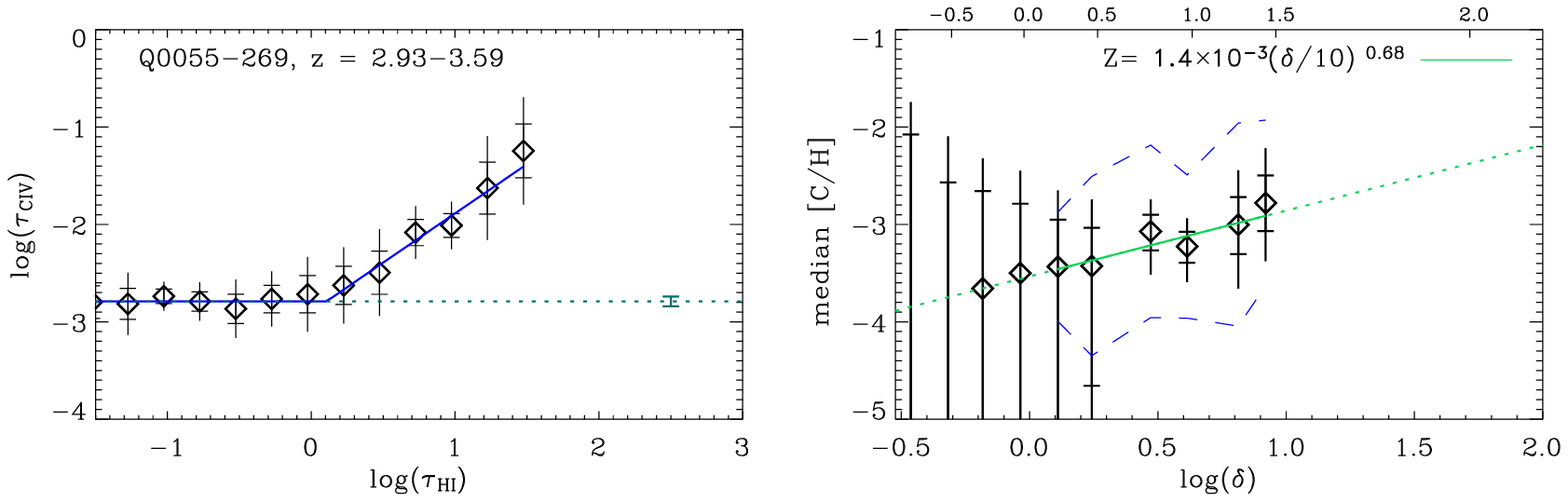}\\
\includegraphics{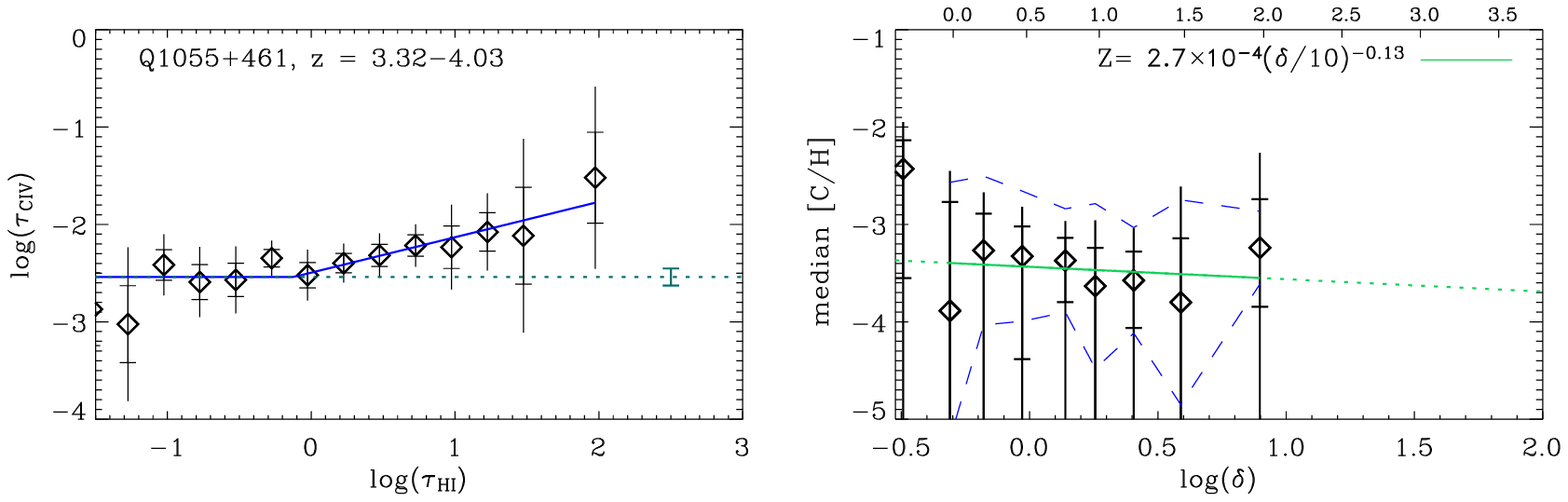}\\
\includegraphics{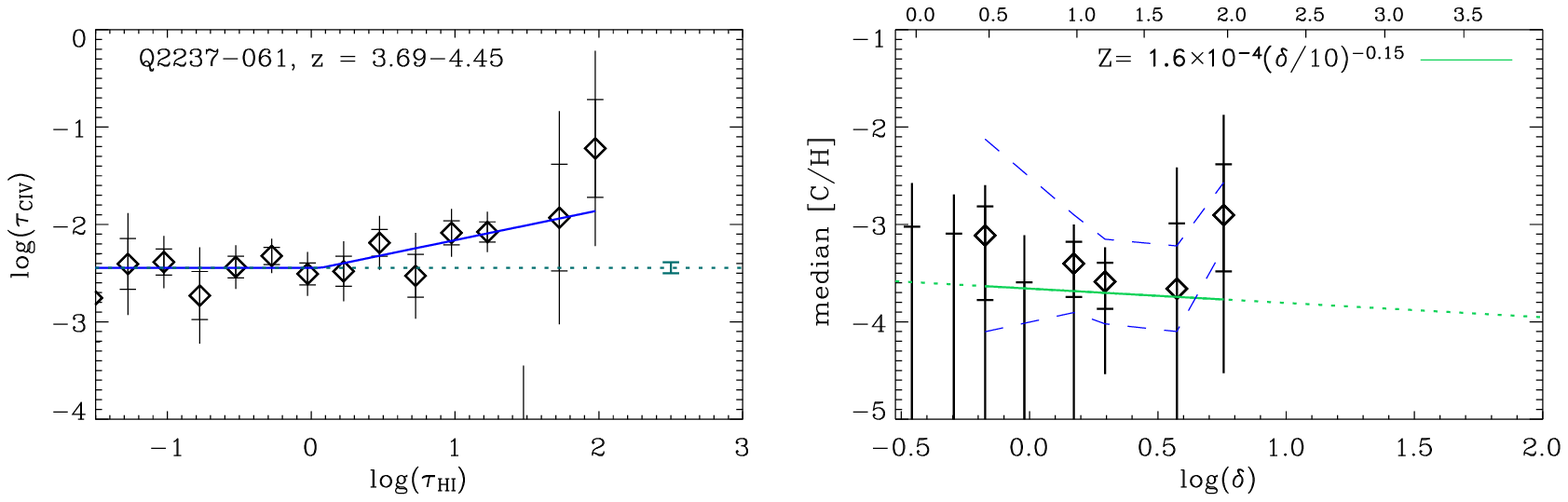}
\addtocounter{figure}{-1}
\figcaption[]{Continued.}
\end{center}
\end{figure*}


\begin{deluxetable*}{lrrrrrrrrrrrrr}
\tabletypesize{\tiny}
\tablecolumns{16}
\tablewidth{0pc} 
\tablecaption{Recovered CIV optical depths
\label{tbl:recod}}
\tablecomments{Cols.\ (1) and (2) contain the quasar name and the recovered HI
optical depth, respectively. Cols. (3) and (4) contain the 31st
percentile of the recovered CIV optical 
depth and the $1\sigma$ error on this value. Cols.\ (5)--(14)
show the same quantities for higher
percentiles (50th, 69th, 84th, 93th, and 98th). An HI optical depth of
$\log\tau_{\rm HI}=-9$ indicates that the corresponding $\tau_{\rm
CIV}$ are the $\tau_{\rm min}$ values. Negative errors indicate that
$\tau_{\rm HI} < \tau_c$ (see \S\protect\ref{sec:method:median} for
the definitions of $\tau_{\rm min}$ and $\tau_c$). Table~4
is published in its entirety in the electronic edition of  
the \emph{Astrophysical Journal}. A portion is shown here for guidance
regarding its form and content.}  
\tablehead{
\colhead{QSO} & 
\colhead{$\log\tau_{{\rm HI}}$} &	
\multicolumn{2}{c}{$\log\tau_{{\rm CIV}}, -0.5\sigma$} &	
\multicolumn{2}{c}{$\log\tau_{\rm CIV}, {\rm median}$} &	
\multicolumn{2}{c}{$\log\tau_{\rm CIV}, +0.5\sigma$} &	
\multicolumn{2}{c}{$\log\tau_{\rm CIV}, +1\sigma$}&
\multicolumn{2}{c}{$\log\tau_{\rm CIV}, +1.5\sigma$}&
\multicolumn{2}{c}{$\log\tau_{\rm CIV}, +2\sigma$}
}
\startdata
Q1101-264 & -9.000& \nodata & \nodata & -2.950 &  0.048 & -2.298 &
0.021 & -1. 67 &  0.022 & -1.767 &  0.015 & -1.619 &  0.028 \\
Q1101-264 & -2.775 & \nodata & \nodata & -2.816 & -0.481 & -2.424 & -0.094 & -2.092 & -0.074 & -1.833 & -0.058 & \nodata & \nodata \\
Q1101-264 & -2.525 & \nodata & \nodata & -2.856 & -0.211 & -2.240 & -0.062 & -1.979 & -0.049 & -1.787 & -0.036 & -1.656 & -0.070 \\
Q1101-264 & -2.275 & \nodata & \nodata & -2.820 & -0.119 & -2.232 & -0.076 & -1.929 & -0.031 & -1.796 & -0.029 & -1.665 & -0.069 \\
Q1101-264 & -2.025 & \nodata & \nodata & -2.959 & -0.092 & -2.334 & -0.045 & -1.999 & -0.029 & -1.753 & -0.031 & -1.615 & -0.130 \\
Q1101-264 & -1.775 & \nodata & \nodata & -2.887 & -0.083 & -2.271 & -0.035 & -1.933 & -0.032 & -1.758 & -0.019 & -1.620 & -0.027 \\
Q1101-264 & -1.525 & \nodata & \nodata & -3.025 & -0.117 & -2.305 & -0.042 & -1.961 & -0.030 & -1.768 & -0.029 & -1.593 &  0.029 \\
Q1101-264 & -1.275 & \nodata & \nodata & -2.908 & -0.111 & -2.301 & -0.053 & -1.966 & -0.047 & -1.739 & -0.030 & -1.561 &  0.078
\enddata
\end{deluxetable*} 

\clearpage
\section{B. The observed effective optical depth}

\setcounter{table}{0}

\begin{deluxetable*}{lcccccccc}[b]
\tablecolumns{9}
\tablewidth{0pc} 
\tablecaption{Observed effective optical depth\label{tbl:taueff}} 
\tablehead{ 
& \multicolumn{4}{c}{All pixels} &
\multicolumn{4}{c}{After removal of contamination}\\
\colhead{QSO} & \colhead{med$(z)$} & \colhead{$z_{\rm min}$} &
\colhead{$z_{\rm max}$} & \colhead{$\tau_{\rm eff}\pm 1\sigma$} &
\colhead{med$(z)$} & \colhead{$z_{\rm min}$} &
\colhead{$z_{\rm max}$} & \colhead{$\tau_{\rm eff}\pm 1\sigma$}}
\startdata 
Q1101-264 & 1.766 & 1.654 & 1.878 & $0.167 \pm 0.025$ & 1.756 & 1.654 & 1.901 & $0.099 \pm 0.016$ \\
Q1101-264 & 1.991 & 1.878 & 2.103 & $0.120 \pm 0.022$ & 2.000 & 1.901 & 2.102 & $0.093 \pm 0.016$ \\
Q0122-380 & 1.806 & 1.692 & 1.919 & $0.134 \pm 0.021$ & 1.799 & 1.692 & 1.917 & $0.120 \pm 0.026$ \\
Q0122-380 & 2.033 & 1.920 & 2.147 & $0.179 \pm 0.024$ & 2.034 & 1.917 & 2.147 & $0.162 \pm 0.026$ \\
J2233-606 & 1.848 & 1.732 & 1.963 & $0.236 \pm 0.036$ & 1.843 & 1.732 & 1.964 & $0.224 \pm 0.033$ \\
J2233-606 & 2.079 & 1.963 & 2.195 & $0.128 \pm 0.017$ & 2.080 & 1.964 & 2.194 & $0.124 \pm 0.019$ \\
HE1122-1648 & 1.990 & 1.869 & 2.112 & $0.134 \pm 0.019$ & 1.986 & 1.869 & 2.108 & $0.128 \pm 0.019$ \\
HE1122-1648 & 2.233 & 2.112 & 2.355 & $0.171 \pm 0.025$ & 2.224 & 2.108 & 2.355 & $0.164 \pm 0.026$ \\
Q0109-3518 & 1.996 & 1.874 & 2.117 & $0.199 \pm 0.025$ & 2.003 & 1.874 & 2.120 & $0.180 \pm 0.027$ \\
Q0109-3518 & 2.239 & 2.117 & 2.361 & $0.124 \pm 0.016$ & 2.239 & 2.120 & 2.361 & $0.113 \pm 0.018$ \\
HE2217-2818 & 1.996 & 1.874 & 2.117 & $0.168 \pm 0.024$ & 1.998 & 1.874 & 2.118 & $0.139 \pm 0.021$ \\
HE2217-2818 & 2.239 & 2.117 & 2.361 & $0.160 \pm 0.019$ & 2.242 & 2.118 & 2.361 & $0.149 \pm 0.018$ \\
Q0329-385 & 2.010 & 1.888 & 2.133 & $0.117 \pm 0.015$ & 2.010 & 1.888 & 2.133 & $0.115 \pm 0.014$ \\
Q0329-385 & 2.255 & 2.133 & 2.377 & $0.156 \pm 0.022$ & 2.256 & 2.133 & 2.377 & $0.156 \pm 0.021$ \\
HE1347-2457 & 2.108 & 1.982 & 2.234 & $0.141 \pm 0.016$ & 2.103 & 1.982 & 2.243 & $0.131 \pm 0.019$ \\
HE1347-2457 & 2.361 & 2.234 & 2.487 & $0.192 \pm 0.024$ & 2.366 & 2.243 & 2.487 & $0.175 \pm 0.022$ \\
Q1442+293 & 2.228 & 2.097 & 2.359 & $0.134 \pm 0.017$ & 2.217 & 2.097 & 2.338 & $0.137 \pm 0.020$ \\
Q1442+293 & 2.490 & 2.359 & 2.621 & $0.315 \pm 0.041$ & 2.496 & 2.338 & 2.621 & $0.214 \pm 0.029$ \\
PKS0329-255 & 2.241 & 2.109 & 2.372 & $0.186 \pm 0.027$ & 2.243 & 2.109 & 2.372 & $0.180 \pm 0.025$ \\
PKS0329-255 & 2.504 & 2.373 & 2.636 & $0.206 \pm 0.025$ & 2.506 & 2.372 & 2.636 & $0.205 \pm 0.022$ \\
Q0002-422 & 2.307 & 2.173 & 2.441 & $0.275 \pm 0.033$ & 2.308 & 2.173 & 2.434 & $0.234 \pm 0.029$ \\
Q0002-422 & 2.576 & 2.441 & 2.710 & $0.337 \pm 0.038$ & 2.572 & 2.434 & 2.710 & $0.283 \pm 0.032$ \\
HE2347-4342 & 2.430 & 2.291 & 2.569 & $0.226 \pm 0.029$ & 2.447 & 2.291 & 2.580 & $0.177 \pm 0.025$ \\
HE2347-4342 & 2.709 & 2.569 & 2.848 & $0.305 \pm 0.039$ & 2.715 & 2.580 & 2.848 & $0.308 \pm 0.041$ \\
Q1107+485 & 2.518 & 2.375 & 2.661 & $0.195 \pm 0.023$ & 2.509 & 2.375 & 2.651 & $0.182 \pm 0.021$ \\
Q1107+485 & 2.804 & 2.661 & 2.947 & $0.314 \pm 0.033$ & 2.805 & 2.651 & 2.947 & $0.273 \pm 0.030$ \\
Q0420-388 & 2.626 & 2.479 & 2.773 & $0.294 \pm 0.029$ & 2.626 & 2.479 & 2.778 & $0.279 \pm 0.033$ \\
Q0420-388 & 2.921 & 2.773 & 3.068 & $0.360 \pm 0.034$ & 2.920 & 2.779 & 3.068 & $0.343 \pm 0.037$ \\
Q1425+604 & 2.694 & 2.544 & 2.844 & $0.417 \pm 0.044$ & 2.670 & 2.544 & 2.882 & $0.232 \pm 0.023$ \\
Q1425+604 & 2.994 & 2.844 & 3.144 & $0.363 \pm 0.034$ & 3.009 & 2.882 & 3.144 & $0.360 \pm 0.040$ \\
Q2126-158 & 2.754 & 2.601 & 2.906 & $0.437 \pm 0.042$ & 2.752 & 2.601 & 2.943 & $0.329 \pm 0.043$ \\
Q2126-158 & 3.059 & 2.906 & 3.211 & $0.310 \pm 0.030$ & 3.078 & 2.944 & 3.211 & $0.271 \pm 0.030$ \\
Q1422+230 & 3.062 & 2.898 & 3.225 & $0.417 \pm 0.037$ & 3.058 & 2.898 & 3.218 & $0.423 \pm 0.034$ \\
Q1422+230 & 3.389 & 3.225 & 3.552 & $0.512 \pm 0.045$ & 3.382 & 3.218 & 3.552 & $0.496 \pm 0.048$ \\
Q0055-269 & 3.092 & 2.928 & 3.257 & $0.381 \pm 0.037$ & 3.088 & 2.928 & 3.249 & $0.366 \pm 0.039$ \\
Q0055-269 & 3.421 & 3.257 & 3.586 & $0.426 \pm 0.033$ & 3.411 & 3.249 & 3.586 & $0.445 \pm 0.041$ \\
Q0000-262 & 3.489 & 3.312 & 3.667 & $1.107 \pm 0.086$ & 3.708 & 3.607 & 3.808 & $0.705 \pm 0.074$ \\
Q0000-262 & 3.845 & 3.667 & 4.023 & $0.758 \pm 0.049$ & 3.912 & 3.808 & 4.023 & $0.811 \pm 0.065$ \\
Q1055+461 & 3.498 & 3.320 & 3.676 & $0.595 \pm 0.041$ & 3.517 & 3.338 & 3.695 & $0.644 \pm 0.056$ \\
Q1055+461 & 3.854 & 3.676 & 4.033 & $0.774 \pm 0.050$ & 3.862 & 3.695 & 4.033 & $0.843 \pm 0.075$ \\
Q2237-061 & 3.880 & 3.690 & 4.070 & $0.828 \pm 0.059$ & 3.862 & 3.696 & 4.032 & $0.839 \pm 0.069$ \\
Q2237-061 & 4.261 & 4.070 & 4.451 & $0.890 \pm 0.063$ & 4.287 & 4.032 & 4.448 & $0.827 \pm 0.060$
\enddata 
\end{deluxetable*} 


\begin{thebibliography}{}

\bibitem[Aguirre, Schaye, \& Theuns(2002)]{2002ApJ...576....1A} Aguirre, 
A., Schaye, J., \& Theuns, T.\ 2002, \apj, 576, 1  (Paper I)

\bibitem[Aguirre et al.(2001)]{2001ApJ...560..599A} Aguirre, A., Hernquist, 
L., Schaye, J., Weinberg, D.~H., Katz, N., \& Gardner, J.\ 2001, \apj, 560, 
599 

\bibitem[Aguirre et al.(2001)]{2001ApJ...561..521A} Aguirre, A., Hernquist, 
L., Schaye, J., Katz, N., Weinberg, D.~H., \& Gardner, J.\ 2001, \apj, 561, 
52

\bibitem[Anders \& Grevesse(1989)]{1989GeCoA..53..197A} Anders, E.~\& 
Grevesse, N.\ 1989, \gca, 53, 197 

\bibitem[Barlow \& Sargent(1997)]{1997AJ....113..136B} Barlow, T.~A.~\& 
Sargent, W.~L.~W.\ 1997, \aj, 113, 136 

\bibitem[Bergeron, Aracil, Petitjean, \& Pichon(2002)]{2002A&A...396L..11B} 
Bergeron, J., Aracil, B., Petitjean, P., \& Pichon, C.\ 2002, \aap, 396, 
L11 

\bibitem[Bernardi et al.(2003)]{2003AJ....125...32B} Bernardi, M.~et al.\ 
2003, \aj, 125, 32 

\bibitem[Bi \& Davidsen(1997)]{1997ApJ...479..523B} Bi, H.~\& Davidsen, 
A.~F.\ 1997, \apj, 479, 523 

\bibitem[Boksenberg, Sargent, \& Rauch(2003)]{2513} Boksenberg, A.,
  Sargent, W. L. W., \& Rauch, M. 2003, ApJS, submitted (astro-ph/0307557)

\bibitem[Bromm, Ferrara, Coppi, \& Larson(2001)]{2001MNRAS.328..969B} 
Bromm, V., Ferrara, A., Coppi, P.~S., \& Larson, R.~B.\ 2001, \mnras, 328, 
969 

\bibitem[Carswell et al.(1984)]{1984ApJ...278..486C} Carswell, R.~F., 
Morton, D.~C., Smith, M.~G., Stockton, A.~N., Turnshek, D.~A., \& Weymann, 
R.~J.\ 1984, \apj, 278, 486 

\bibitem[Carswell, Schaye, \& Kim(2002)]{2002ApJ...578...43C} Carswell, B., 
Schaye, J., \& Kim, T.\ 2002, \apj, 578, 43 

\bibitem[Cen \& Ostriker(1999)]{1999ApJ...519L.109C} Cen, R.~\& Ostriker, 
J.~P.\ 1999, \apjl, 519, L109 

\bibitem[Cowie \& Songaila (1998)]{1998Natur.394...44C} Cowie, L. L. \& 
Songaila, A. 1998, \nat, 394, 44 

\bibitem[Cowie, Songaila, Kim, \& Hu(1995)]{1995AJ....109.1522C} Cowie, 
L.~L., Songaila, A., Kim, T., \& Hu, E.~M.\ 1995, \aj, 109, 1522 

\bibitem[Croft, Weinberg, Katz, \& Hernquist(1997)]{1997ApJ...488..532C} 
Croft, R.~A.~C., Weinberg, D.~H., Katz, N., \& Hernquist, L.\ 1997, \apj, 
488, 532 

\bibitem[Dav{\' e} et al.(1998)]{1998ApJ...509..661D} Dav{\' e}, R., 
Hellsten, U., Hernquist, L., Katz, N., \& Weinberg, D.~H.\ 1998, \apj, 509, 
661 

\bibitem[D'Odorico et al.(2000)]{2000SPIE.4005..121D} D'Odorico, S., 
Cristiani, S., Dekker, H., Hill, V., Kaufer, A., Kim, T., \& Primas, F.\ 
2000, \procspie, 4005, 121 

\bibitem[Ellison et al.(1999)]{1999ApJ...520..456E} Ellison, S.~L., Lewis, 
G.~F., Pettini, M., Chaffee, F.~H., \& Irwin, M.~J.\ 1999, \apj, 520, 456 

\bibitem[Ellison, Songaila, Schaye, \& Pettini(2000)]{2000AJ....120.1175E} 
Ellison, S.~L., Songaila, A., Schaye, J., \& Pettini, M.\ 2000, \aj, 120, 
1175 

\bibitem[Ferland(2000)]{2000RMxAC...9..153F} 
Ferland, G.~J.\ 2000, Revista Mexicana de Astronomia y Astrofisica
Conference Series, 9, 153

\bibitem[Ferland et al.(1998)]{1998PASP..110..761F} 
Ferland, G.~J., Korista, K.~T., Verner, D.~A., Ferguson, J.~W.,
Kingdon, J.~B., \& Verner, E.~M.\ 1998, \pasp, 110, 761

\bibitem[Furlanetto \& Loeb(2003)]{2003ApJ...588...18F} Furlanetto, 
S.~R.~\& Loeb, A.\ 2003, \apj, 588, 18 

\bibitem[Haardt \& Madau(1996)]{1996ApJ...461...20H} Haardt, F.~\& Madau, 
P.\ 1996, \apj, 461, 20 

\bibitem[Haardt \& Madau(2001)]{haardt01:cuba}
Haardt, F., \& Madau, P. 2001, to be published in the proceedings of
XXXVI Rencontres de Moriond, astro-ph/0106018

\bibitem[Haehnelt, Steinmetz, \& Rauch(1996)]{1996ApJ...465L..95H} 
Haehnelt, M.~G., Steinmetz, M., \& Rauch, M.\ 1996, \apjl, 465, L95 

\bibitem[Heap et al.(2000)]{2000ApJ...534...69H} Heap, S.~R., Williger, 
G.~M., Smette, A., Hubeny, I., Sahu, M.~S., Jenkins, E.~B., Tripp, T.~M., 
\& Winkler, J.~N.\ 2000, \apj, 534, 69 

\bibitem[Hellsten et al.(1997)]{1997ApJ...487..482H} Hellsten, U., Dave, 
R., Hernquist, L., Weinberg, D.~H., \& Katz, N.\ 1997, \apj, 487, 482 

\bibitem[Kim, Cristiani, \& D'Odorico(2001)]{2001A&A...373..757K} Kim, 
T.-S., Cristiani, S., \& D'Odorico, S.\ 2001, \aap, 373, 757 

\bibitem[Kim et al.(2002)]{2002MNRAS.335..555K} Kim, T.-S., Carswell, 
R.~F., Cristiani, S., D'Odorico, S., \& Giallongo, E.\ 2002, \mnras, 335, 
555 

\bibitem[Kim et al.(2002)]{2590} Kim, 
T.-S., Viel, M., \& Haehnelt, M., Carswell, R. F., \& Cristiani,
S. 2003, MNRAS, (astro-ph/0308103) 

\bibitem[Kriss et al.(2001)]{2001Sci...293.1112K} Kriss, G.~A.~et al.\ 
2001, Science, 293, 1112 

\bibitem[Madau, Ferrara, \& Rees(2001)]{2001ApJ...555...92M} Madau, P., 
Ferrara, A., \& Rees, M.~J.\ 2001, \apj, 555, 92 

\bibitem[Pettini et al.(2003)]{2602} Pettini, M., Madau, P., Bolte, M.,
Prochaska, J. X., Ellison, S. L., \& Fan, X. 2003, ApJ, in press 
(astro-ph/0305413) 

\bibitem[Rauch, Sargent, Womble, \& Barlow(1996)]{1996ApJ...467L...5R} 
Rauch, M., Sargent, W.~L.~W., Womble, D.~S., \& Barlow, T.~A.\ 1996, \apjl, 
467, L5 

\bibitem[Rauch, Haehnelt, \& Steinmetz(1997)]{1997ApJ...481..601R}
Rauch, M., Haehnelt, M.~G., \& Steinmetz, M.\ 1997, \apj, 481, 601

\bibitem[Rauch et al.(1997)]{1997ApJ...489....7R} Rauch, M.~et al.\ 1997, 
\apj, 489, 7 

\bibitem[Rollinde, Petitjean, \& Pichon(2001)]{2001A&A...376...28R} 
Rollinde, E., Petitjean, P., \& Pichon, C.\ 2001, \aap, 376, 28 

\bibitem[Scannapieco, Ferrara, \& Madau(2002)]{2002ApJ...574..590S} 
Scannapieco, E., Ferrara, A., \& Madau, P.\ 2002, \apj, 574, 590 

\bibitem[Scannapieco, Schneider, \& Ferrara(2003)]{2003ApJ...589...35S} 
Scannapieco, E., Schneider, R., \& Ferrara, A.\ 2003, \apj, 589, 35 

\bibitem[Schaye(2001)]{2001ApJ...559..507S} Schaye, J.\ 2001, \apj, 559, 
507 

\bibitem[Schaye, Theuns, Leonard, \& Efstathiou(1999)]{1999MNRAS.310...57S} 
Schaye, J., Theuns, T., Leonard, A., \& Efstathiou, G.\ 1999, \mnras, 310, 
57 

\bibitem[Schaye, Rauch, Sargent, \& Kim(2000)]{2000ApJ...541L...1S} Schaye, 
J., Rauch, M., Sargent, W.~L.~W., \& Kim, T.\ 2000a, \apjl, 541, L1 

\bibitem[Schaye et al.(2000)]{2000MNRAS.318..817S} Schaye, J., Theuns, T., 
Rauch, M., Efstathiou, G., \& Sargent, W.~L.~W.\ 2000b, \mnras, 318, 817 

\bibitem[Schneider, Ferrara, Natarajan, \& 
Omukai(2002)]{2002ApJ...571...30S} Schneider, R., Ferrara, A., Natarajan, 
P., \& Omukai, K.\ 2002, \apj, 571, 30 

\bibitem[Simcoe, Sargent, \& Rauch(2002)]{2002ApJ...578..737S} Simcoe, 
R.~A., Sargent, W.~L.~W., \& Rauch, M.\ 2002, \apj, 578, 737 

\bibitem[Songaila(1998)]{1998AJ....115.2184S} Songaila, A.\ 1998, \aj, 115, 
2184 

\bibitem[Songaila(2001)]{2001ApJ...561L.153S} Songaila, A.\ 2001, \apjl, 
561, L153 

\bibitem[Songaila & Cowie (1996)]{1996AJ....112..335S} Songaila, A.  \& 
Cowie, L. L. 1996, \aj, 112, 335 

\bibitem[Spergel et al.(2003)]{2654} Spergel, D. N., et al.\ 2003, ApJ,
in press (astro-ph/0302209)

\bibitem[Telfer et al.(2002)]{2002ApJ...579..500T} Telfer, R.~C., Kriss, 
G.~A., Zheng, W., Davidsen, A.~F., \& Tytler, D.\ 2002, \apj, 579, 500 

\bibitem[Thacker, Scannapieco, \& Davis(2002)]{2002ApJ...581..836T} 
Thacker, R.~J., Scannapieco, E., \& Davis, M.\ 2002, \apj, 581, 836 

\bibitem[Theuns, Schaye, \& Haehnelt(2000)]{2000MNRAS.315..600T} Theuns, 
T., Schaye, J., \& Haehnelt, M.~G.\ 2000, \mnras, 315, 600 

\bibitem[Theuns et al.(2002a)]{2002ApJ...567L.103T} Theuns, T., Schaye, J., 
Zaroubi, S., Kim, T., Tzanavaris, P., \& Carswell, B.\ 2002a, \apjl, 567, 
L103 

\bibitem[Theuns et al.(2002b)]{2002ApJ...578L...5T} Theuns, T., Viel, M., 
Kay, S., Schaye, J., Carswell, R.~F., \& Tzanavaris, P.\ 2002b, \apjl, 578, 
L5 

\bibitem[Vogt et al.(1994)]{1994SPIE.2198..362V} Vogt, S.~S.~et al.\ 1994, 
\procspie, 2198, 362 

\bibitem[Zhang, Meiksin, Anninos, \& Norman(1998)]{1998ApJ...495...63Z} 
Zhang, Y., Meiksin, A., Anninos, P., \& Norman, M.~L.\ 1998, \apj, 495, 63 

\end{thebibliography}
\end{document}